\documentclass[11pt]{article}

\usepackage{scalerel}
\usepackage{graphicx}
\usepackage{amsmath}
\usepackage{amsfonts}
\usepackage{amssymb}
\usepackage{dsfont}
\usepackage{physics}
\usepackage{microtype}
\usepackage{bbold}
\usepackage{bm}
\usepackage{mathtools}
\usepackage{cite}
\usepackage{cancel}
\usepackage{xcolor}
\usepackage{mdframed}
\usepackage{xpatch}
\usepackage{etoolbox}
\usepackage{comment}
\usepackage[normalem]{ulem}
\usepackage[colorlinks=true, linkcolor=purple, filecolor=blue, urlcolor=blue, citecolor=blue, anchorcolor=blue, menucolor=blue , linktocpage=true]{hyperref}
\usepackage[font=footnotesize,labelfont=bf,width=.9\textwidth]{caption}

\allowdisplaybreaks
\numberwithin{equation}{section}

\renewcommand{\title}[1]{\vbox{\center\LARGE{#1}}\vspace{5mm}}
\renewcommand{\author}[1]{\vbox{\center#1}\vspace{5mm}}
\newcommand{\address}[1]{\vbox{\center\footnotesize\em#1}}
\newcommand{\email}[1]{\vbox{\center\footnotesize\tt#1}\vspace{5mm}}

\begin{document}

\begin{titlepage}
\vfill

\begin{minipage}{\textwidth}
\vspace{2cm}
\end{minipage}

\begin{center} 

\title{Near-extremal dynamics away from the horizon}

\author{Alejandra Castro$^{a}$, Robinson Mancilla$^{b}$ and Ioannis Papadimitriou$^{c}$}

\address{$^{a}$Department of Applied Mathematics and Theoretical Physics, University of Cambridge\\ Cambridge CB3 0WA, United Kingdom\\ 
\vspace{0.7em}
$^{b}$Department of Physics, University of California, Santa Barbara, CA 93101, United States\\
\vspace{0.7em}
$^{c}$Division of Nuclear and Particle Physics, Department of Physics,
National and Kapodistrian University of Athens, Athens GR-157 84, Greece
}

\email{ac2553@cam.ac.uk, rhmancilla@ucsb.edu, ioannis.papadimitriou@phys.uoa.gr}
\end{center}


\begin{abstract}
Near-extremal black holes are usually studied by zooming into the throat that describes their near-horizon geometry. Within this throat, one can argue that two-dimensional JT gravity is the appropriate effective theory that dominates at low temperature. Here, we discuss how to capture this effective description by standing far away from the horizon. Our strategy is to construct a phase space within gravitational theories in AdS$_{d+1}$ that fixes the radial dependence while keeping the transverse dependence arbitrary. This allows us to implement a decoupling limit directly on the phase space while keeping the coordinates fixed. With this, we can relate the effective description in JT gravity to the CFT$_d$ description at the boundary of AdS$_{d+1}$, which we do explicitly in AdS$_3$ and non-rotating configurations in AdS$_4$. From the perspective of the dual CFT, our decoupling limit should be understood as a flow between a CFT$_{d}$ and a near-CFT$_1$. Our analysis shows that local counterterms can be constructed in the near-CFT$_1$, which arise from the  anomalies (or absence of them) in the  CFT$_{d}$. We show that one of these counterterms is the Schwarzian effective action, making this sector a scheme-dependent choice. This illustrates the delicate interplay between a far and a near analysis of near-extremal black holes. 
\end{abstract}


\end{titlepage}


\setcounter{tocdepth}{2}
 \tableofcontents

 \newpage

\section{Introduction} \label{sec:intro}
Extremal black holes are renowned for the enhancement of symmetries in their near-horizon geometry, which made them a centrepiece to decode classical \cite{Kunduri:2013ana} and quantum \cite{Sen:2007qy} properties of gravity. Nowadays, near-extremal black holes have taken that spotlight for how they break these symmetries: starting with the analysis in \cite{Almheiri:2014cka,Maldacena:2016upp}, it has been demonstrated that simple two-dimensional models of dilaton gravity, in particular Jackiw-Teitelboim (JT) gravity, capture the leading temperature effect that controls the classical and quantum aspects of near-extremal black holes. This new development has been crucial in resolving some puzzling aspects of extremal black holes, such as those discussed in \cite{Preskill:1991tb,Maldacena:1998uz}; see \cite{Turiaci:2023wrh} for a review on these puzzles. 

More concretely, the claim is that JT gravity is a model that captures the backreaction for AdS$_2$ under an injection of energy, and AdS$_2$ is a portion of the near-horizon geometry of extremal and near-extremal black holes. The explicit and precise connection between this theory and near-extremal black holes in $d+1$ dimensions relies on controlling a dimensional reduction from $d+1$ to two dimensions that preserves the relevant modes in the near-horizon geometry of the black hole. This has been achieved in several cases, as done, for example, in \cite{Achucarro:1993fd,Cvetic:2016eiv,Castro:2018ffi,Iliesiu:2020qvm,Castro:2021wzn}, where it is carefully demonstrated that, upon a consistent dimensional reduction on the near-horizon geometry, a subsector of the theory is described by JT gravity.\footnote{Another approach is to study the linearised equations of motion around the near-horizon geometry, as done in \cite{Castro:2019crn,Castro:2021csm,Hadar:2020kry,Porfyriadis:2018jlw,Mariani:2025hee}. This also provides the necessary information to identify the JT sector in the vicinity of the near-horizon region.}

Performing a dimensional reduction is a luxury that relies on an abundance of symmetry in the internal directions, and for the majority of black holes, it is not a desirable route. For this reason, various methods have been developed to demonstrate the role of JT gravity in near-extremal black holes without using a dimensional reduction. One very successful approach advanced in \cite{Iliesiu:2022onk} evaluates the Euclidean path integral around the near-horizon geometry;  the appearance of the JT sector comes from regulating zero modes that appear in the path integral at zero temperature in the near horizon of extremal black holes. By introducing temperature effects in the throat, the resulting contribution from the regulated modes matches exactly the quantum predictions from JT gravity. Quantum corrections in JT are captured by a Schwarzian effective action, which leads to logarithmic terms that are distinctive and tractable \cite{Charles:2019tiu,Iliesiu:2020qvm}. Another approach is to stay away from the black hole horizon and analyse the whole geometry at finite temperature and track the behaviour of quantum corrections as the temperature goes to zero. Recent work along these lines includes \cite{Kapec:2024zdj,Kolanowski:2024zrq,Arnaudo:2024bbd}. 
One aspect shared by this body of work is that the analysis is catered for Euclidean signature, with the aim of seeing the imprint of the Schwarzian effective action on the gravitational path integral. 

Our aim here is to capture the dynamics of near extremal solutions without zooming into the horizon, and relate it to JT gravity, but via a different approach from the one described above. We will be working in the context of AdS$_{d+1}$/CFT$_d$, since the CFT$_d$ will give an anchor on how to set up observables in the whole geometry. The idea is to give as input the explicit radial dependence of the fields in AdS$_{d+1}$ for some set of asymptotic boundary conditions, leaving the transverse dependence of the background fields off-shell. This transverse dependence will define a reduced phase space.\footnote{This phase space will include black holes, both extremal and non-extremal; however, it is richer than that, allowing for temporal and spatial dependence on the boundary coordinates.} In AdS/CFT, it is known that imposing the remaining bulk equations of motion, which determine the transverse dependence, is equivalent to Ward identities in the CFT; this will give us a useful characterisation of the gravitational dynamics in terms of one-point functions of conserved currents and their sources. Within this phase space, we introduce an \textit{off-shell decoupling limit}, which allows us to show explicitly how the reduced phase space of the AdS$_{d+1}$ connects to JT gravity.    

There are a few interesting highlights of our construction. First, the off-shell decoupling limit does not act on the radial direction of AdS$_{d+1}$, nor does it act on the transverse directions. This is an important deviation from the standard decoupling limits in the context of, for example, black holes and black branes. For our analysis, it is crucial that coordinates are not tampered with: this is key to providing a map from the variables in our phase space to the variables in JT gravity. Second, the analysis we perform is in Lorentzian signature. This makes it difficult to compare with the Euclidean analysis described above, but it has the advantage of directly discussing the physical configurations in the CFT$_d$ and JT gravity.

Thirdly, in our analysis, it is important how we characterise observables in JT gravity. This two-dimensional theory is exactly solvable, both at the classical and quantum level, where all of the information is encoded in boundary terms, topology, and boundary conditions. Here we will revisit the description of its dynamics, such that we can easily compare it with the nomenclature used in the higher dimensions. For this reason, we cast the dynamics of the theory in a Bondi gauge and construct in this gauge the holographic dictionary. We discuss properties of the reduced phase space, review in detail the local symmetries acting on it, and construct the appropriate functionals, Ward identities, and charges that parametrise this phase space. In this construction, a free constant parameter becomes prominent: $\omega$. From the perspective of the Ward identities, it seems to be controlling a trace anomaly; however, its value is not fixed and can be modified by a suitable boundary counterterm. When evaluating the JT action on this reduced phase space, a preferred value is chosen: $\omega=1$.  However, in connecting with the phase space of the CFT$_d$, it is crucial to keep this parameter arbitrary since its value is controlled by the UV theory. 
 
Lastly, our analysis will focus on AdS$_{d+1}$ gravity with $d+1=3,4$. Although it has been well studied how to connect the physics of AdS$_3$ to two-dimensional dilaton models, see for example \cite{Achucarro:1993fd,Almheiri:2016fws,Cvetic:2016eiv,Ghosh:2019rcj}, our approach adds a perspective that allows an understanding of near-extremality from the perspective of the CFT$_2$. Our construction of the reduced phase space will only include the gravitational sector, where the only current is the stress tensor, and we will include the case where the left and right-moving central charges in the CFT$_2$ are unequal. This will allow a more precise tracking of the relations between the phase spaces and the role of counterterms, i.e., $\omega$, in this situation. It will also provide a lamppost to the four-dimensional example. 

In AdS$_4$, we will focus on Einstein-Maxwell theory and consider situations that are not rotating. In contrast to AdS$_3$, local degrees of freedom in the gravitational theory prevent a full characterisation of the space of solutions. Still, there is a family of solutions where the radial evolution can be exactly solved. These are the Robinson-Trautman (RT) family of algebraically special solutions \cite{RT:1962}. RT solutions have gained attention in holography \cite{HR:2014RT,Skenderis:2014RT,Petropoulos:2015RT,Bakas:2016RT,Skenderis:2017RT,Petropoulos:2017RT}, particularly in studies of out-of-equilibrium physics in AdS/CFT correspondence, and its corresponding flat limit has been discussed in \cite{Fareghbal:2019RT}. Further recent investigations into different aspects of RT solutions include \cite{SUSY:2024RT,Charge:2024RT,Scalar:2024RT,Adak:2019RT}.
One drawback of these backgrounds is that their phase space already satisfies the trace Ward identity. For this reason, we construct a generalisation of the RT solutions such that the conservation and tracelessness conditions of the currents are both on the same footing at the level of equations of motion. It is important to say that this generalisation takes the radial component of Einstein's equation off-shell (hence we are not solving for the radial evolution). Still, the radial dependence of the fields will be fixed and imposing the equations of motion fixes the transverse dependence of the fields; these equations of motion are precisely the Ward identities of the CFT.  
Finally, the role of anomalies in the Ward identities is notable in the case of AdS$_4$/CFT$_3$. There are no anomalies in CFT$_3$, which makes it strange from a field theory perspective to recover the Ward identities appearing in JT gravity from a decoupling limit. Here is where $\omega$ will play a key role in reconciling this tension, and we will discuss the physical implications of this result. 

This paper is organised as follows. In Sec.\,\ref{Sec:JTgravity} we present a self-contained analysis of JT gravity in the Bondi gauge, which is useful for the subsequent comparison with the higher-dimensional theories. We revisit the analysis of Ward identities, making explicit the ambiguities that appear in renormalising the boundary theory due to local counterterms. From this perspective, we relate to the familiar Schwarzian effective action that describes boundary modes in JT. We also discuss the interplay of the Schwarzian action with thermodynamics and the counterterms. In Sec.\,\ref{sec:ads3} we present our first, and simplest, example that connects JT with black hole physics. The theory is pure AdS$_3$, and the relevant black hole in our phase space is the BTZ solution. Even in this familiar example, counterterms play a role, which we highlight and discuss. In Sec.\,\ref{sec:TMG} we generalise the discussion of AdS$_3$ by adding a gravitational Chern-Simons term. Although the analysis is more complex, it parallels the simplest case in Sec.\,\ref{sec:ads3}, with the advantage that anomalies can be tracked more explicitly. In Sec.\,\ref{sec:ads5} we discuss an example in AdS$_4$, where the flagship black hole is the static Reissner-Nordstr\"om black hole. Here, the main challenge is to build an appropriate phase space that contains the black hole. For this, we use and extend results known for the Robinson-Trautman solutions. We then implement our off-shell decoupling limit and discuss important differences of the flow to JT in this case, where the CFT$_3$ has no anomalies that resemble the physics of a Schwarzian action.  Finally, we conclude with a discussion of our results and future directions in Sec.\,\ref{sec:disc}.

\section{Aspects of \texorpdfstring{nAdS$_2$/nCFT$_1$}{nAdS2/nCFT1}} \label{Sec:JTgravity}

In this section, we revisit Jackiw-Teitelboim (JT) gravity \cite{Jackiw:1984je,Teitelboim:1983ux} as a prototype for near extremal dynamics. By decomposing the dynamics along and transverse to the radial coordinate using a suitable Bondi gauge, and explicitly solving the radial evolution, we obtain a description that is off-shell in the transverse direction. The transverse equations of motion can be identified with Ward identities in the holographic dual quantum mechanics and lead to Schwarzian dynamics. As we will see later on, the variables that describe the transverse dynamics of JT gravity naturally arise in a specific IR limit of higher-dimensional CFTs dual to higher-dimensional AdS black holes. The conventions and analysis in this section follow mainly \cite{Cvetic:2016eiv}; other useful references include \cite{Almheiri:2014cka,Maldacena:2016upp,Jensen2016}, as well as the reviews \cite{Sarosi:2017ykf,Mertens:2023rjt,Turiaci:2024ljt}.

JT gravity couples the two-dimensional metric $g_{ab}$ to a dilaton $\phi$ through the action \cite{Maldacena:2016upp}
\begin{equation}\label{eq:JTaction}
   S_{\rm JT}=\frac{1}{2\kappa_2^2}\int_{\mathcal{M}_2}\dd^2x\sqrt{-g}\,\phi \left(\mathcal{R}+\frac{2}{\ell_2^2}\right)+\frac{1}{\kappa_2^2}\int_{\partial \mathcal{M}_2} \dd x \sqrt{-h}\,\phi\left(K-\frac{1}{\ell_2}\right)\,.
\end{equation}
Here, $\kappa_2^2$ is the dimensionless Newton's constant in two dimensions and $\ell_2$ is the AdS$_2$ radius. The surface terms are the standard Gibbons-Hawking-York term that involves the trace, $K$, of the extrinsic curvature of the induced metric $h$ on the one-dimensional boundary $\partial\mathcal{M}_2$, and the local volume counterterm proportional to $1/\ell_2$. 

The field equations following from the action \eqref{eq:JTaction} are
\begin{equation}\label{eq:EoMJT}
\begin{aligned}
    \mathcal{R}+\frac{2}{\ell_2^2}&=0\,,\\
    (\nabla_a\nabla_b-g_{ab}\Box)\phi+\frac{g_{ab}}{\ell_2^2}\phi&=0\,.
\end{aligned}
\end{equation}
These equations do not describe any propagating degrees of freedom. In particular, the first equation fixes the two-dimensional metric to be locally AdS$_2$. 

\subsection{Reduced phase space and local symmetries}\label{sec:2Dphase-space}

The field equations \eqref{eq:EoMJT} can be completely integrated by selecting a suitable gauge \cite{Maldacena:2016upp,Engelsoy:2016xyb}. For our present analysis, it will be convenient to parametrise the solutions in the so-called Bondi gauge,
\begin{equation}\label{eq:local-AdS2}
    ds^2=-f(u,r)\dd u^2-2\alpha(u) \dd u \dd r\,,
\end{equation}
where $\alpha(u)$ is an arbitrary function of the time coordinate, $u$, unrestricted by the equations of motion. Demanding that the metric \eqref{eq:local-AdS2} is locally AdS$_2$ requires the radial profile of the lapse function $f(u,r)$ to be at most quadratic in $r$, i.e., 
\begin{equation}\label{eq:fJT}
    f(u,r)=\alpha(u)^2\frac{r^2 }{\ell_2^2}+f_1(u)\frac{r}{\ell_2}+f_0(u)\,,
\end{equation}
where $f_1(u)$ and $f_0(u)$ are arbitrary functions of $u$. Moreover, the $rr$ component of the second equation in \eqref{eq:EoMJT} determines that the dilaton $\phi$ must be linear in $r$, namely
\begin{equation}\label{eq:phiJT}
    \phi(u,r)= \nu(u)\frac{r}{\ell_2}+\phi_0(u)\,,
\end{equation}
where $\nu(u)$ and $\phi_0(u)$ are arbitrary functions.\footnote{\label{ChangeOfGauge}There exists a diffeomorphism that relates the Bondi gauge \eqref{eq:local-AdS2} with the Fefferman-Graham gauge $ds^2=d\bar r^2+\gamma_{tt}(\bar r,t)dt^2$ used in \cite{Cvetic:2016eiv}. The general solutions in these two gauges are related by an asymptotic diffeomorphism of the form $e^{\bar r/\ell_2}=r/\ell_2+\mathcal{O}(r^0)$, $t=u-\frac{\ell_2^2}{\alpha(u)r}+\mathcal{O}(r^{-2})$. Under this transformation, the functions $\alpha(u)$ and $\nu(u)$ here coincide respectively with the functions $\alpha(t)$ and $\nu(t)$ that parametrise the running dilaton solutions in \cite{Cvetic:2016eiv}.}

The relations \eqref{eq:fJT} and \eqref{eq:phiJT} fully determine the radial dependence of the Bondi gauge solution and, therefore, parametrise a reduced phase space. However, there exist three different local transformations that modify the functions $\alpha(u)$, $\nu(u)$, $f_1(u)$, $f_2(u)$ and $\phi_0(u)$, but preserve the form of the radial solution:
\begin{enumerate}
    \item[i)] Time reparametrisations, $u\to \bar{u}(u)$, under which:
\begin{equation}
    \begin{aligned}
    &\alpha(u)\to \bar\alpha(\bar u) = \Big(\frac{d\bar u}{du}\Big)^{-1}\alpha(u)\,,\qquad f_{1,0}(u)\to \bar{f}_{1,0}(\bar u) = \Big(\frac{d\bar u}{du}\Big)^{-2}f_{1,0}(u)\,,\\
    &\nu(u)\to \bar\nu(\bar u) = \nu(u)\,,\qquad \phi_0(u)\to \bar\phi_0(\bar u) = \phi_0(u)\,.   
    \end{aligned}
\end{equation}

    \item[ii)] Local rescalings of the radial coordinate, $r\to e^{\sigma(u)}r$, under which:
\begin{equation}
    \begin{aligned}
    &\alpha\to e^{\sigma}\alpha\,,\qquad f_{1}\to e^{\sigma}\big(f_{1}+2\ell_2\alpha\sigma'\big)\,,\qquad f_0\to f_0\,,\\
    &\nu\to e^{\sigma}\nu\,,\qquad \phi_0\to \phi_0\,.
    \end{aligned}
\end{equation}

\item[iii)] Local shifts of the radial coordinate, $r\to r+\zeta(u)$, under which:
\begin{equation}\label{eq:RadialShift}
    \begin{aligned}
    &\alpha\to \alpha\,,\qquad f_{1}\to f_1+2\alpha^2\zeta/\ell_2\,,\qquad 
    f_0\to f_0+\alpha^2\zeta^2/\ell_2^2+f_1\zeta/\ell_2+2\alpha\zeta'\,,\\
    &\nu\to \nu\,,\qquad \phi_0\to \phi_0+\nu\zeta/\ell_2\,.
    \end{aligned}
\end{equation}
The fact that $f_1(u)$ and $\phi_0(u)$ transform additively under this transformation implies that either of the two can be viewed as a compensator. In particular, taking $\phi_0(u)$ as the compensator, we can dress each of the variables with an appropriate function of $\phi_0(u)$ so that the resulting composite variable is invariant. Since $\alpha(u)$ and $\nu(u)$ are already invariant we need only consider $f_1(u)$ and $f_0(u)$. It is straightforward to verify that the variables
\begin{equation}\label{eq:tildef}
    \widetilde{f}_1=f_1-2\alpha^2\frac{\phi_0}{\nu}\,, \qquad \widetilde{f}_0=f_0+\alpha^2\frac{\phi_0^2}{\nu^2}-\frac{\phi_0}{\nu} f_1-2\ell_2\alpha\partial_u\Big(\frac{\phi_0}{\nu}\Big)\,,
\end{equation}
are invariant under the transformations \eqref{eq:RadialShift}.
\end{enumerate}
Using $\phi_0(u)$ as a compensator of the radial shift symmetry allows us to parametrise the reduced phase space in terms of the four shift-invariant functions $\alpha(u)$, $\nu(u)$, $\widetilde{f}_1(u)$ and $\widetilde{f}_0(u)$.\footnote{The Bondi gauge has been used in JT gravity to study boundary conditions and their asymptotic symmetries \cite{Godet:2020xpk,Ruzziconi:2021Int}. In our analysis, radial shifts do not generate additional charges.} As we will see next, it is precisely these variables that can be mapped to the sources and one-point functions of the dual quantum mechanics.

\subsection{Holographic dictionary and Ward identities}\label{sec:JT-Ward}

The remaining equations of motion in \eqref{eq:EoMJT} impose differential relations among the functions that parametrise the reduced phase space. These relations take the form 
\begin{equation}\label{eq:eom-f}
    \widetilde f_1=2\ell_2\alpha\frac{\partial_u\nu}{\nu}\,,\qquad \partial_u\left(\frac{\nu^2}{\alpha^2}\widetilde f_0\right)=0\,,
\end{equation}
and correspond to first-class constraints that generate the local symmetries i) and ii) above. In order to identify these constraints with Ward identities in the dual quantum mechanics, it is necessary to choose a suitable parametrisation of the reduced phase space. 

Examining the leading asymptotic form of the radial solutions \eqref{eq:fJT} and \eqref{eq:phiJT}, we identify $-\alpha(u)^2$ with the boundary metric and $\log\nu(u)$ with the source of the scalar operator dual to the dilaton \cite{Cvetic:2016eiv} (see footnote \ref{ChangeOfGauge}). We expect, therefore, that the shift-invariant functions $\widetilde{f}_1(u)$ and $\widetilde{f}_0(u)$ are related to the one-point functions of the stress tensor $\mathcal{O}_u$ and the scalar operator dual to the dilaton $\mathcal{O}_\phi$. The precise relation between the variables $\widetilde{f}_1(u)$, $\widetilde{f}_o(u)$ and $\mathcal{O}_u$, $\mathcal{O}_\phi$ is dictated by the requirement that the constraints imposed by the remaining equations of motion take the standard form of conformal Ward identities when expressed in terms of the operators $\mathcal{O}_u$, $\mathcal{O}_\phi$. This requirement determines that  
\begin{equation}
    \begin{aligned}\label{eq:JToperators}
    \widetilde{f}_1(u)=&\;2\kappa_2^2\gamma^2\frac{\alpha^2}{\nu}\int \dd u\,\alpha(\mathcal{O}_u+\mathcal{O}_\phi)+2(1-\omega^2\gamma^2)\ell_2\alpha\frac{\partial_u\nu}{\nu}\,,\\
    \widetilde{f}_0(u)=&\;2\kappa_2^2\frac{\alpha^2}{\nu}\ell_2 \mathcal{O}_u-\omega^2\ell_2^2\frac{(\partial_u\nu)^2}{\nu^2}\,, 
    \end{aligned}
\end{equation}   
where $\gamma$ and $\omega$ are arbitrary constants that reflect an ambiguity in the identification of the dual quantum mechanics operators from the reduced phase space variables. As we will see, $\omega$ is related to a local counterterm in the effective action of the dual quantum mechanics and will play a key role in the subsequent analysis.

In terms of the variables $\mathcal{O}_u$, $\mathcal{O}_\phi$ introduced via \eqref{eq:JToperators} the equations of motion in \eqref{eq:eom-f} become
\begin{subequations}\label{eq:ward2d}
   \begin{align}
    &{\cal O}_{u}+\mathcal{O}_{\phi}= \frac{\omega^2\ell_2}{\kappa_2^2\alpha}\partial_u\left(\frac{\partial_u\nu}{\alpha}\right)\,,\label{eq:traceJT}\\
     &\partial_u {\cal O}_u-{\cal O}_{\phi}\frac{\partial_u\nu}{\nu}=0\,.\label{eq:WardIdJT}
    \end{align} 
\end{subequations}
These match exactly the Ward identities in the dual conformal quantum mechanics \cite{Cvetic:2016eiv}, with the right-hand side of \eqref{eq:traceJT} identified with a ``conformal anomaly''. It is not a genuine conformal anomaly, since, as we mentioned above and we will review below, the value of the parameter $\omega$ can be modified by a local counterterm.

Unlike in higher dimensions, the Ward identities \eqref{eq:traceJT}, \eqref{eq:WardIdJT} can be integrated to obtain the form of the variables ${\cal O}_u$ and ${\cal O}_\phi$. In particular, the general solution of the Ward identities is \cite{Cvetic:2016eiv}
\begin{equation}\label{eq:solnsJT}
    \begin{aligned}
        {\cal O}_u&=-\frac{m}{\nu}+\frac{\omega^2\ell_2}{2\kappa_2^2\nu}\left(\frac{\partial_u\nu}{\alpha}\right)^2\,,\\
        {\cal O}_\phi&=\frac{m}{\nu}+ \frac{\omega^2\ell_2}{2\kappa_2^2}\left(\frac{2}{\alpha}\partial_u\left(\frac{\partial_u\nu}{\alpha}\right)-\frac{1}{\nu}\left(\frac{\partial_u\nu}{\alpha}\right)^2\right)\,,
    \end{aligned}
\end{equation}
where $m$ is an integration constant. These variables may be expressed as variations of the action 
\begin{equation}\label{eq:IJT}
    \begin{aligned}
        I_{\omega}= \int \dd u\,\alpha\left[-\frac{m}{\nu}-\frac{\omega^2\ell_2}{2\kappa_2^2}\frac{1}{\nu}\left(\frac{\partial_u\nu}{\alpha}\right)^2\right]\,,
    \end{aligned}
\end{equation}
with respect to $\alpha$ and $\nu$, respectively. Namely,
\begin{equation}\label{eq:deltaIJT}
    \delta I_{\omega} = \int \dd u \Big({\cal O}_u\, \delta \alpha + \frac{\alpha}{\nu}{\cal O}_\phi \,\delta \nu \Big)\,.
\end{equation}

The effective action $I_{\omega}[\alpha,\nu]$ allows us to exhibit the direct relation between the Ward identities \eqref{eq:traceJT}, \eqref{eq:WardIdJT} and the reduced phase space symmetries i) and ii), which we can now recognize, respectively, as boundary diffeomorphisms and Weyl transformations. Taking $\epsilon$ as the parameter of an infinitesimal boundary diffeomorphism and $\sigma$ of an infinitesimal boundary Weyl transformation, the sources $\alpha(u)$ and $\nu(u)$ transform as
\begin{equation}\label{eq:PBHJT}
    \delta_{\epsilon,\sigma} \alpha=\partial_u(\epsilon \alpha)+\alpha\sigma~, \quad \delta_{\epsilon,\sigma}\nu=\epsilon\partial_u\nu+\nu\sigma\,.
\end{equation}
Applying these transformations to the expressions \eqref{eq:solnsJT}, we can also determine the transformation of the variables ${\cal O}_u$ and ${\cal O}_\phi$, which we will need later on. A direct calculation gives
\begin{equation}
    \begin{aligned}\label{eq:PBHJT_operators}
\delta_{\epsilon,\sigma}\mathcal{O}_u=&\;\epsilon\, \partial_u\mathcal{O}_u-\sigma\mathcal{O}_u+\frac{\omega^2\ell_2}{\kappa_2^2}\frac{\partial_u\nu\partial_u\sigma}{\alpha^2}\,,\\
\delta_{\epsilon,\sigma}\mathcal{O}_{\phi}=&\;\epsilon\, \partial_u\mathcal{O}_{\phi}-\sigma\mathcal{O}_{\phi}+\frac{\omega^2\ell_2}{\kappa_2^2}\frac{\nu}{\alpha}\partial_u\left(\frac{\partial_u\sigma}{\alpha}\right)\,.
    \end{aligned}
\end{equation}
In particular, both operators transform as scalars under boundary diffeomorphisms, but possess an anomalous Weyl transformation when $\omega\neq 0$.\footnote{Note that the transformation of ${\cal O}_\phi$ in eq. (5.8b) of \cite{Cvetic:2016eiv} contains two typos.}   

Returning to the relation between the Ward identities \eqref{eq:traceJT}, \eqref{eq:WardIdJT} and the symmetry transformations \eqref{eq:PBHJT}, let us first consider boundary diffeomorphisms. The transformations \eqref{eq:PBHJT} lead us to identify $h_b=-\alpha^2$ with the metric on the one-dimensional boundary and $\nu$ with a boundary scalar. The effective action can then be expressed in the manifestly covariant form
\begin{equation}\label{eq:IJTcov}
    I_{\omega}=\int \dd u\sqrt{-h_b}\,\nu^{-1}\left(-m+\frac{\omega^2\ell_2}{2\kappa_2^2}h_b^{-1}(\partial_u\nu)^2\right)\,.
\end{equation}
It follows that $I_{\omega}$ is invariant under boundary diffeomorphisms. This means that specializing \eqref{eq:deltaIJT} to boundary diffeomorphisms we have
\begin{equation}
    \delta_\epsilon I_{\omega}=-\int \dd u\, \epsilon\, \alpha\left(\partial_u {\cal O}_u-{\cal O}_{\phi}\frac{\partial_u\nu}{\nu} \right)=0\,.
\end{equation}
Since this must hold for arbitrary $\epsilon(u)$, we recover the Ward identity \eqref{eq:WardIdJT}. 

This argument can be repeated for boundary Weyl transformations. However, the effective action is not invariant in this case. In particular, under an infinitesimal Weyl transformation
\begin{equation}
    \delta_\sigma I_{\omega}=\frac{\omega^2\ell_2}{\kappa_2^2}\int \dd u\,\sigma\partial_u\Big(\frac{\partial_u\nu}{\alpha}\Big)\,.
\end{equation}
From \eqref{eq:deltaIJT} follows that
\begin{equation}
    \delta_\sigma I_{\omega}=\int \dd u\, \sigma\, \alpha ({\cal O}_u+\mathcal{O}_{\phi}) =\frac{\omega^2\ell_2}{\kappa_2^2}\int \dd u\,\sigma\partial_u\Big(\frac{\partial_u\nu}{\alpha}\Big)\,,
\end{equation}
which implies the trace Ward identity \eqref{eq:traceJT}.

\subsection{Asymptotic symmetries}

In the previous subsection, we studied the effect of infinitesimal boundary diffeomorphism and Weyl transformations to obtain the appropriate Ward identities. Here, we focus our attention on a subset of those transformations; in particular, we will now consider a combined boundary diffeomorphism and Weyl transformation such that the boundary metric remains unchanged. This will define the notion of asymptotic symmetries of the system, and will be important in the next subsection when discussing the Schwarzian dynamics.

The parameters $\epsilon_o$ and $\sigma_o$ of a boundary diffeomorphism and Weyl transformation that leave the boundary metric unchanged satisfy 
\begin{equation}\label{eq:SchwarzianMode}
    \delta_{\epsilon_o,\sigma_o} \alpha=\partial_u(\epsilon_o \alpha)+\alpha\sigma_o=0\,.
\end{equation}
This condition reduces the local parameters from two to one. In particular, the two parameters satisfying the constraint \eqref{eq:SchwarzianMode} can be chosen as 
\begin{equation}\label{eq:xi-transformation-param}
    \epsilon_o=\frac{\xi}{\alpha}\,,\qquad \sigma_o=-\partial_{\tau}\xi\,, 
\end{equation}
where $\xi({\tau})$ is an arbitrary function and 
\begin{equation}\label{eq:proptime}
    {\tau}(u)=\int^u\dd u'\,\alpha(u')\,,
\end{equation}
is the proper time on the boundary. Under generic $\xi$-transformations, $\nu$ transforms as
\begin{equation}
    \delta_{\xi} \nu=\epsilon_o\partial_u\nu+\nu\sigma_o=\epsilon_o\partial_u\nu-\frac{\nu}{\alpha}\partial_u(\epsilon_o \alpha)=-\nu^2\partial_{\tau}(\nu^{-1}\xi)\,,
\end{equation}
or
\begin{equation}\label{eq:xi-beta}
    \delta_\xi\nu^{-1}=\partial_{\tau}(\nu^{-1}\xi)\,.
\end{equation}

Finite $\xi$-transformations play a central role in the dynamics of JT gravity and take the form
\begin{equation}\label{eq:t-beta}
   \nu(\tau)\to\nu_{\cal F}({\tau})=\frac{1}{\partial_{\tau}{\cal F}({\tau})}\nu({\cal F}({\tau}))\,,
\end{equation}
where ${\cal F}({\tau})$ is an arbitrary function with $\partial_{\tau} {\cal F}({\tau})\neq 0$. From now on we will be thinking of $\nu$ as a function of proper time $\tau$. Indeed, inserting ${\cal F}({\tau})={\tau}+\xi({\tau})$ in \eqref{eq:t-beta} and expanding for small $\xi({\tau})$ we recover the infinitesimal transformation \eqref{eq:xi-beta}. If $\omega\neq 0$, a specific finite $\xi$-transformation can be used to set $\omega=1$, at the expense of introducing an overall factor of $\omega$ in the effective action, $I_{\omega}$, and in the variables ${\cal O}_u$ and ${\cal O}_\phi$. In particular, a finite $\xi$-transformation with ${\cal F}({\tau})=\omega\, {\tau}$ amounts to the rescaling $\nu\to\nu/\omega$. Accordingly, $I_{\omega}\to \omega I_{\omega=1}$, ${\cal O}_u^\omega\to\omega\,{\cal O}_u^{\omega=1}$ and ${\cal O}^\omega_\phi\to\omega\, {\cal O}^{\omega=1}_\phi$. In particular, the Ward identities \eqref{eq:traceJT} and \eqref{eq:WardIdJT}, as well as the operator transformations \eqref{eq:PBHJT_operators} become independent of $\omega$.

Applying two successive infinitesimal $\xi$-transformations on $\nu$ we can determine the corresponding algebra. Using \eqref{eq:xi-beta}, we obtain
\begin{equation}
    [\delta_{\xi_1},\delta_{\xi_2}]\nu^{-1}=\delta_{\xi}\nu^{-1}\,,\qquad \xi=\xi_2\partial_{\tau}\xi_1-\xi_1\partial_{\tau}\xi_2\,.
\end{equation}
The representation of the algebra of $\xi$-transformations on the phase space of JT gravity solutions requires a Poisson bracket. The variation \eqref{eq:deltaIJT} of the effective action defines the symplectic form
\begin{equation}\label{eq:OmegaIJT}
    \Omega_{\rm JT} = \int \dd u \Big(\delta{\cal O}_u\wedge \delta \alpha + \frac{\alpha}{\nu}\delta{\cal O}_\phi \wedge\delta\nu \Big)\,.
\end{equation}
Inverting this leads to the Poisson bracket:
\begin{equation}\label{eq:PBJT}
   \{A,B\}=\int \dd u\left(\frac{\delta A}{\delta \alpha}\frac{\delta B}{\delta {\cal O}_u}+\frac{\nu}{\alpha}\frac{\delta A}{\delta \nu}\frac{\delta B}{\delta {\cal O}_\phi}-A\leftrightarrow B\right)\,, 
\end{equation}
where $A$, $B$ are arbitrary functions of the phase space variables, $\alpha$, $\nu$, ${\cal O}_u$, and ${\cal O}_\phi$.  

Boundary diffeomorphisms and Weyl transformations are generated through this Poisson bracket by the Ward identities \eqref{eq:traceJT}, \eqref{eq:WardIdJT}. In particular, introducing the phase space functional 
\begin{equation}\label{eq:GenLocSym}
    {\cal L}(\epsilon,\sigma)=\int \dd u\,\alpha\left[\epsilon\left(\partial_u {\cal O}_u-{\cal O}_{\phi}\frac{\partial_u\nu}{\nu}\right)-\sigma\left({\cal O}_{u}+\mathcal{O}_{\phi}-\frac{\omega^2\ell_2}{\kappa_2^2\alpha}\partial_u\left(\frac{\partial_u\nu}{\alpha}\right)\right)\right]\,,
\end{equation}
the transformation of any phase space variable $A$ is given by 
\begin{equation}\label{eq:local-trans}
    \delta_{\epsilon,\sigma} A=\{{\cal L}(\epsilon,\sigma),A\}\,.
\end{equation}
$\xi$-transformations are generated by \eqref{eq:GenLocSym} with the parameters given in \eqref{eq:xi-transformation-param}, namely
\begin{equation}\label{eq:xi-generator}
    {\cal L}(\xi)=\int \dd u\left[\xi\left(\partial_u {\cal O}_u-{\cal O}_{\phi}\frac{\partial_u\nu}{\nu}\right)+\partial_u\xi\left({\cal O}_{u}+\mathcal{O}_{\phi}-\frac{\omega^2\ell_2}{\kappa_2^2\alpha}\partial_u\left(\frac{\partial_u\nu}{\alpha}\right)\right)\right]\,.
\end{equation}
Dropping total derivative terms, this can be simplified to
\begin{equation}\label{eq:xi-generator-simple}
    {\cal L}(\xi)=-\int \dd{\tau}\, \xi\nu^{-1}\,\partial_{\tau}\left(\nu{\cal O}_\phi-\frac{\omega^2\ell_2}{\kappa_2^2}\Big(\nu\partial_{\tau}^2\nu-\frac12(\partial_{\tau}\nu)^2\Big)\right)\,.
\end{equation}

It is straightforward to verify that the $\xi$-transformations of $\alpha$ and $\nu$ we found above, namely $\delta_\xi\alpha=0$, $\delta_\xi\nu=-\nu^2\partial_{\tau}(\nu^{-1}\xi)$, are generated by \eqref{eq:xi-generator-simple} through the Poisson bracket \eqref{eq:PBJT}. Similarly, the $\xi$-transformation of the variables ${\cal O}_u$ and ${\cal O}_\phi$ may be obtained either by specializing the transformations \eqref{eq:PBHJT_operators} to $\xi$-transformations, or via the Poisson bracket. Either way one obtains
\begin{equation}
    \begin{aligned}\label{eq:ScwarzianJT_operators}
\delta_{\xi}\mathcal{O}_u=&\;\partial_{\tau}\left(\xi\mathcal{O}_u\right)-\frac{\omega^2\ell_2}{\kappa_2^2}\partial_{\tau}\nu\partial_u^2\xi\,,\\
\delta_{\xi}\mathcal{O}_{\phi}=&\;\partial_{\tau}\left(\xi\mathcal{O}_{\phi}\right)-\frac{\omega^2\ell_2}{\kappa_2^2}\nu\partial_{\tau}^3\xi\,.
    \end{aligned}
\end{equation}
In particular, the combination $\nu^{-1}{\cal O}_\phi$ transforms as \cite{Cvetic:2016eiv}
\begin{equation}\label{eq:VirasoroCurrent}
\delta_{\xi}(\nu^{-1}\mathcal{O}_{\phi})=\{{\cal L}(\xi),\nu^{-1}\mathcal{O}_{\phi}\}=2\nu^{-1}\mathcal{O}_{\phi}\partial_{\tau}\xi+\xi\partial_{\tau}\left(\nu^{-1}\mathcal{O}_{\phi}\right)-\frac{\omega^2\ell_2}{\kappa_2^2}\partial_{\tau}^3\xi\,,    
\end{equation}
which corresponds to the coadjoint representation of the Virasoro algebra with central charge proportional to $\omega^2\ell_2/\kappa_2^2$ \cite{Witten:1987ty}. However, the combination $\nu^{-1}\mathcal{O}_{\phi}-\frac{\omega^2\ell_2}{\kappa_2^2}\nu^{-2}\big(\nu\partial_{\tau}^2\nu-\frac12(\partial_{\tau}\nu)^2\big)$ that enters in the generating function \eqref{eq:xi-generator-simple} possesses a homogeneous $\xi$-transformation. As a result, the algebra of $\xi$-symmetry generators on the phase space closes without a central element, namely
\begin{equation}
    \{{\cal L}(\xi_1),{\cal L}(\xi_2)\}=\delta_{\xi_1}{\cal L}(\xi_2)={\cal L}(\xi_2\partial_{\tau}\xi_1-\xi_1\partial_{\tau}\xi_2)\,.
\end{equation}

Finally, let us consider the special case of $\xi$-transformations that preserve $\nu$, besides $\alpha$. From the $\xi$-transformation of $\nu$ in \eqref{eq:xi-beta} we see that $\nu$ is preserved provided $\xi=k\nu$, where $k$ is constant. In that case, $\epsilon_o$ in \eqref{eq:xi-transformation-param} corresponds to the unique conformal Killing vector (CKV) of JT gravity solutions. The existence of CKVs implies the existence of associated conserved charges, which can be constructed using the Ward identities \eqref{eq:WardIdJT} and \eqref{eq:traceJT}.\footnote{In higher dimensions, not all CKVs lead to a conserved charge in the presence of a conformal anomaly \cite{Papadimitriou:2005ii}. However, since the conformal anomaly in one dimension can always be eliminated by a local counterterm, a conserved charge always exists.} Multiplying the diffeomorphism Ward identity \eqref{eq:WardIdJT} with $\epsilon_o$ we obtain
\begin{equation}
   \frac{1}{\alpha}\big(\partial_u(\nu {\cal O}_u)-({\cal O}_u+{\cal O}_{\phi})\partial_u\nu\big)=0\,.
\end{equation}
Using the trace Ward identity \eqref{eq:traceJT}, this in turn can be written as
\begin{equation}
   \partial_{\tau}\left(\nu {\cal O}_u-\frac{\omega^2\ell_2}{2\kappa_2^2}(\partial_{\tau}\nu)^2\right)=0\,.
\end{equation}
It follows that the conserved charge associated with the unique CKV corresponding to constant $\xi$ is
\begin{equation}\label{eq:charge1D}
   {\cal Q}\equiv-\left(\nu {\cal O}_u-\frac{\omega^2\ell_2}{2\kappa_2^2}(\partial_{\tau}\nu)^2\right)=m\,,
\end{equation}
where in the second equality we have used the form of ${\cal O}_{u}$ in \eqref{eq:solnsJT}. An equivalent way to obtain ${\cal Q}$ is by evaluating \eqref{eq:xi-generator-simple} on the CKV $\xi = k \nu$, which yields ${\cal L}(k\nu) = k {\cal Q}$. It follows that the integration constant, $m$, parametrising the variables ${\cal O}_u$ and ${\cal O}_\phi$ is the conserved charge associated with the timelike CKV, i.e. the mass or energy of the solution.

\subsection{On-shell action and Schwarzian dynamics}\label{sec:Schw}

Let us next compute the on-shell value of the JT action \eqref{eq:JTaction}. The equations of motion \eqref{eq:EoMJT} imply that the bulk integral vanishes on-shell and so there is only a boundary contribution to the action
\begin{equation}\label{eq:on-shell-actionJT-cov}
S_{\rm JT}\Big|_{\rm on-shell}=\frac{1}{\kappa_2^2}\int_{\partial \mathcal{M}_2} \dd x \sqrt{-h}\,\phi\left(K-\frac{1}{\ell_2}\right)\,.
\end{equation}

Writing the Bondi gauge metric \eqref{eq:local-AdS2} in the ADM form 
\begin{equation}\label{eq:ADM}
    ds^2=(N^2+N_uN^u)\dd r^2+2N_u \dd u\dd r+h_{uu}\dd u^2\,,
\end{equation}
we identify $N_u=-\alpha$, $h_{uu}=-f$, and $N=\alpha f^{-\frac12}$. It follows that the extrinsic curvature is given by
\begin{equation}\label{eq:}
K_{uu}=\frac{1}{2N}(\partial_r h_{uu}-2D_u N_u)=\frac{1}{2N}(\partial_r h_{uu}-2\partial_uN_u+h_{uu}^{-1}\partial_u h_{uu}N_u)\,,
\end{equation}
where $D_u$ is the covariant derivative with respect to the induced metric $h_{uu}$. Evaluating the trace of the extrinsic curvature, $K=h_{uu}^{-1}K_{uu}$, inserting it in \eqref{eq:on-shell-actionJT-cov}, and taking the limit $r\to\infty$ we obtain (see also appendix B of \cite{Chaturvedi:2020jyy})
\begin{equation}\label{eq:on-shell-actionJT}
    \begin{aligned}
    S_{\rm JT}\Big|_{\rm on-shell}=&\;\int \dd u\, \alpha\left[\frac{m}{\nu}+\frac{\ell_2}{2\kappa_2^2}\left(\frac{2}{\alpha}\partial_u\left(\frac{\partial_u\nu}{\alpha}\right)-\frac{1}{\nu}\left(\frac{\partial_u\nu}{\alpha}\right)^2\right)\right]\,.
     \end{aligned}
\end{equation}
As expected, the on-shell action is independent of the parameters $\omega$ or $\gamma$ introduced in \eqref{eq:JToperators}, since the invariant functions $\widetilde f_{0,1}$ do not depend on these parameters once the explicit form of the operators in \eqref{eq:solnsJT} is taken into account. More importantly, the on-shell action is also independent of the function $\phi_0(u)$, which reflects the fact that the boundary theory is inert under the local symmetry iii).  

Comparing the on-shell action \eqref{eq:on-shell-actionJT} and the effective action $I_{\omega}$ in \eqref{eq:IJT}, we observe that they contain exactly the same terms, but with different coefficients. In particular, the two are related by
\begin{equation}\label{eq:relate-SJT-Iw}
S_{\rm JT}\Big|_{\rm on-shell}=I_{\omega}+2\int \dd u\sqrt{-h_b}\,\nu^{-1}{\cal Q}+(1-\omega^2)\,\frac{\ell_2}{2\kappa_2^2}\int \dd u\sqrt{-h_b}\,\nu^{-1}h_b^{-1}(\partial_u\nu)^2\,,
\end{equation}
where $h_b=-\alpha^2$ is the metric on the one-dimensional boundary and ${\cal Q}=m$ is the conserved charge in \eqref{eq:charge1D}. Each of the additional terms on the right-hand side of \eqref{eq:relate-SJT-Iw} plays an important role which is worth discussing separately:
\begin{enumerate}

\item The last term in \eqref{eq:relate-SJT-Iw} is a local counterterm that modifies the value of $\omega$. This is why $\omega$ is not a conformal anomaly as mentioned below \eqref{eq:ward2d}. To make this more manifest, we write 
\begin{equation}
\begin{aligned}\label{eq:counterterm}
    S_{\rm ct}&=\frac{\ell_2}{2\kappa_2^2}\int \dd u\sqrt{-h}\,\phi^{-1}h^{uu}(\partial_u\phi)^2\,,
\end{aligned}
\end{equation}
and 
\begin{equation}
    S_{\rm ct}\Big|_{\rm on-shell}=\frac{\ell_2}{2\kappa_2^2}\int \dd u\sqrt{-h_b}\,\nu^{-1}h_b^{-1}(\partial_u\nu)^2~.
\end{equation}
The addition of this boundary counterterm will play a crucial role in our subsequent analysis. As we will see in the coming sections, the UV completion of JT gravity determines $\omega$.

\item The term proportional to ${\cal Q}$ in \eqref{eq:relate-SJT-Iw} is simply the Gibbons-Hawking-York term evaluated on the surface of minimum radius \cite{Cabo-BizetKolPandoZayasEtAl2018,Castro:2018ffi}. This term is state-dependent, since it involves the mass parameter $m$, which places it in a different footing to $S_{\rm ct}$. Adding such contribution to  $I_{\omega}$ leads to some useful properties. We define 
\begin{equation}\label{eq:IJTmod}
\widetilde I_{\omega}=I_{\omega}+2\int \dd u\sqrt{-h_b}\,\nu^{-1}{\cal Q}= \int \dd u\,\alpha\,{\cal O}_\phi\,.
\end{equation}
Since the integrand of $\widetilde I_{\omega}$ is proportional to ${\cal O}_\phi$, its $\xi$-transformation is related to the transformation of a Virasoro current, and therefore to Schwarzian dynamics. To render this connection manifest, let us begin by writing the effective action \eqref{eq:IJTmod} in the form\footnote{The functional $\widetilde I_{\omega}$ looks very similar to the 1D Liouville theory discussed in \cite{Mertens:2017mtv,Engelsoy:2016xyb}, however, $\nu$ is not related to the Liouville field. Recall that $\nu$ is a source and its variations give correlation functions of ${\cal O}_\phi$. In the next subsection, starting from $\widetilde I_{\omega}$, we derive the Schwarzian effective action. This action has a direct relation to 1D Liouville.}
\begin{equation}\label{eq:IJTSchwarzian}
    \begin{aligned}
        \widetilde I_{\omega}= \frac{\ell_2}{\kappa_2^2}\int \dd {\tau}\,\nu\,{\cal S}_\omega\,,
    \end{aligned}
\end{equation}
where ${\cal S}_\omega$ is given by 
\begin{equation}\label{eq:Somega}
    {\cal S}_\omega[\nu;\tau]=\frac{\kappa_2^2}{\ell_2}\nu^{-1}{\cal O}_\phi=\frac{s_0}{2\nu^2}+\frac{\omega^2}{2}\left(\frac{2}{\nu}\partial_{\tau}^2\nu-\frac{1}{\nu^2}\left(\partial_{\tau}\nu\right)^2\right)\,,\qquad s_0=\frac{2\kappa_2^2}{\ell_2}m\,.
\end{equation}
The Ward identities in \eqref{eq:ward2d} imply that ${\cal S}_\omega$ satisfies the differential constraint
\begin{equation}\label{eq:SchwarzianWI}
\partial_{\tau}(\nu^2{\cal S}_\omega)-\omega^2\nu\partial_{\tau}^3\nu=0\,.
\end{equation}
Moreover, \eqref{eq:VirasoroCurrent} implies that ${\cal S}_\omega$ transforms as a Virasoro current under $\xi$-transformations, namely  
\begin{equation}\label{eq:VirasoroCurrentS}
\delta_{\xi}{\cal S}_\omega=2{\cal S}_\omega\partial_{\tau}\xi+\xi\partial_{\tau}{\cal S}_\omega-\omega^2\partial_{\tau}^3\xi\,.
\end{equation}
As we will see shortly, $\widetilde I_{\omega}$ and ${\cal S}_\omega$ will be central in decoding thermodynamic properties.
\end{enumerate}

Hence, we find that $\widetilde I_{\omega}$ coincides with the on-shell value of $S_{\rm JT}$ plus the local covariant counterterm, namely
\begin{equation}
S_{\rm JT}+(\omega^2-1) S_{\rm ct}\,.
\end{equation}
To facilitate comparison with subsequent sections, let us define $I_{\rm JT} \equiv I_{\omega=1}$ as the generating functional obtained via a canonical transformation from the on-shell action of JT gravity given in (\ref{eq:on-shell-actionJT}). This allows us to express the relation
\begin{equation}
\label{eq:MapI}
    I_{\rm JT}=I_{\omega}+(1-\omega^2)\frac{\ell_2}{2\kappa_2^2}\int \dd u \sqrt{-h_b}h_b^{-1}\nu^{-1}(\partial_u\nu)^2~,
\end{equation}
where $h_b = -\alpha^2$.
In parallel, let us define the operators generated by $I_{\rm JT}$ as ${\cal O}_{u}^{\scaleto{\rm JT}{4pt}} \equiv {\cal O}_{u}^{\omega=1}$ and ${\cal O}_{\phi}^{\scaleto{\rm JT}{4pt}} \equiv {\cal O}_{\phi}^{\omega=1}$. The boundary counterterm (\ref{eq:MapI}) then induces the following field redefinitions,
\begin{equation}
\label{eq:MapII}
\begin{aligned}
     {\cal O}_{u}^{\scaleto{\rm JT}{4pt}}&={\cal O}_{u}+\frac{(1-\omega^2)}{2}\frac{\ell_2}{\kappa_2^2}\frac{(\partial_u\nu)^2}{\nu\alpha^2}~,\\
     {\cal O}_{\phi}^{\scaleto{\rm JT}{4pt}}&={\cal O}_{\phi}-\frac{(1-\omega^2)}{2}\frac{\ell_2}{\kappa_2^2}\frac{(\partial_u\nu)^2}{\nu\alpha^2}+(1-\omega^2)\frac{\ell_2}{\kappa_2^2\alpha}\partial_u\left(\frac{\partial_u\nu}{\alpha}\right)~.\\
\end{aligned}
\end{equation}
This mapping preserves \eqref{eq:WardIdJT} while modifying \eqref{eq:traceJT}, thereby yielding 
\begin{equation}
\begin{aligned}
    {\cal O}_{u}^{\scaleto{\rm JT}{4pt}}+\mathcal{O}_{\phi}^{\scaleto{\rm JT}{4pt}}&= \frac{\ell_2}{\kappa_2^2\alpha}\partial_u\left(\frac{\partial_u\nu}{\alpha}\right)\,,\\
    \partial_u {\cal O}_u^{\scaleto{\rm JT}{4pt}}-{\cal O}_{\phi}^{\scaleto{\rm JT}{4pt}}\frac{\partial_u\nu}{\nu}&=0\,.
\end{aligned}    
\end{equation}

The key insight of this section is that, in one dimension, the concept of a ``conformal anomaly'' is inherently ambiguous because it can always be removed by a local counterterm. In sharp contrast, in higher dimensions, no local counterterm exists to alter the coefficient of the Weyl anomaly, making it a robust quantity.

In the following sections, we will make use of the operator mapping in~(\ref{eq:MapII}), derived from the counterterm in~(\ref{eq:MapI}), to relate a particular low-energy limit of higher-dimensional CFTs to the Ward identities of Schwarzian dynamics. In this context, we will focus primarily on the generating function $I_{\text{JT}}$, rather than its Lagrangian form $S_{\text{JT}}$, since they are related in a straightforward way by equation~\eqref{eq:relate-SJT-Iw} and \eqref{eq:IJTmod}.

\subsubsection{Schwarzian dynamics} \label{sec:schw-dyn}

The analysis so far has not revealed that the gravitational aspects of JT gravity can be described in terms of a Schwarzian effective action, as shown in \cite{Maldacena:2016upp}. In the following, we will show how this description is derived in our notation. The main idea is to characterize appropriately our reduced phase space under the boundary condition that $\nu$ is constant. 

We will start by being more explicit about the solutions to the equations of motion in the Bondi gauge, writing out explicitly \eqref{eq:local-AdS2}-\eqref{eq:phiJT}. With a convenient choice of the function $\phi_0$, the general solution to \eqref{eq:EoMJT} is
\begin{equation}\label{eq:sol}
    ds^2=-\left(\frac{r^2}{\ell_2^2}-2\ell_2^2\,{\cal S}_{\omega=1}\right)\dd {\tau}^2-2\,\dd {\tau} \dd r\,,\qquad \phi= \nu\frac{r}{\ell_2}-\ell_2\partial_{\tau}\nu\,,
\end{equation}
where ${\cal S}_{\omega=1}$ stands for \eqref{eq:Somega} with $\omega=1$ and the proper time $\tau$ is defined in \eqref{eq:proptime}; we also used \eqref{eq:tildef}, \eqref{eq:JToperators}, and \eqref{eq:solnsJT}. Recall that even though intermediate steps have $\omega$-dependence, all solutions are independent of the parameter $\omega$. 

Vacuum solutions correspond to configurations with ${\cal S}_{\omega=1}=0$. Via  \eqref{eq:SchwarzianWI}, this provides an equation for $\nu({\tau})$, whose general solution takes the form
\begin{equation}\label{eq:vacsol}
    \nu({\tau})=-a({\tau}-{\tau}_0)^2+b({\tau}-{\tau}_0)+\frac{s_0-b^2}{4a}\,,
\end{equation}
where $a$, $b$ and ${\tau}_0$ are integration constants. 

Analogously, a black hole solution with temperature $T$ corresponds to configurations with ${\cal S}_{\omega=1}=2\pi^2 T^2$. The general solution to \eqref{eq:SchwarzianWI} in this case that reduces to \eqref{eq:vacsol} as $T\to 0$ is  
\begin{equation}\label{eq:bhsol}
    \nu({\tau})=-\frac{a}{\pi^2T^2}\sinh^2\big(\pi T({\tau}-{\tau}_0)\big)+\frac{b}{2\pi T}\sinh\big(2\pi T({\tau}-{\tau}_0)\big)+\frac{\sqrt{a^2+(s_0-b^2)\pi^2T^2}-a}{2\pi^2 T^2}\,,
\end{equation}
where again $a$, $b$ and $\tau_0$ are integration constants. A static solution is obtained by setting $a=b=0$, in which case $\nu=\sqrt{s_0}/2\pi T$ or $s_0=(2\pi T\nu)^2$. This is in agreement with \eqref{eq:Somega}, which implies that ${\cal S}_{\omega=1}=s_0/2\nu^2$ for static solutions. 

Both \eqref{eq:vacsol} and \eqref{eq:bhsol} are solutions of the condition $\partial_{\tau}{\cal S}_{\omega=1}=0$. Importantly, this equation is not a consequence of the JT equations of motion, since \eqref{eq:sol} solves these for any function $\nu({\tau})$. However, $\partial_{\tau}{\cal S}_{\omega=1}=0$ enforces the regularity of JT gravity solutions -- that is smooth AdS$_2$ or existence of a horizon --  and can be derived from the on-shell action of JT gravity as follows \cite{Maldacena:2016upp}. Consider the finite $\xi$-transformation \eqref{eq:t-beta}, where we will reparametrise the proper time via $\tau\to t=t(\tau)$. It is convenient to cast this as an inverse transform, i.e., take $\tau(t)$, and hence we write   
\begin{equation}\label{eq:xitransScwarzian}
    \nu(t)\to \nu_{\tau}(t)=\frac{\nu({\tau}(t))}{{\tau}'(t)}\,, 
\end{equation}
under which the functional ${\cal S}_{\omega}[\nu;\tau]$ transforms as
\begin{equation}
   {\cal S}_{\omega}[\nu;\tau]\to {\cal S}_{\omega}[\nu_{\tau};t]={\tau}'^
    2 {\cal S}_{\omega}[\nu;\tau]-\omega^2\{{\tau},t\}\,.
\end{equation}
This is the finite version of \eqref{eq:VirasoroCurrentS}, where we have introduced the Schwarzian derivative,
\begin{equation}
    \{{\tau},t\}\,= -\frac{1}{2}\left(\frac{\tau''}{\tau'}\right)^2 + \left(\frac{\tau{''}}{\tau'}\right)'~,
\end{equation}
and primes here are derivatives with respect to $t$. Taking the initial $\nu(\tau)$ to be the vacuum solution \eqref{eq:vacsol}, this transformation reduces to 
\begin{equation}\label{eq:Somega-Schwarzian-vac}
   {\cal S}_{\omega=1}[\nu_{\tau};t]=-\{{\tau},t\}\,.
\end{equation}
The corresponding value of the effective action \eqref{eq:IJTSchwarzian} for $\omega=1$ is
\begin{equation}
    \begin{aligned}
        \widetilde I_{\omega=1}[\alpha,\nu_{\tau}]= \frac{\ell_2}{\kappa_2^2}\int \dd t\,\nu_{\tau}\, {\cal S}_{\omega=1}[\nu_{\tau};t]=-\frac{\ell_2}{\kappa_2^2}\int \dd t\,\nu_{\tau}\,\{{\tau},t\}\,.
    \end{aligned}
\end{equation}

The equation $\partial_{\tau}{\cal S}_{\omega=1}=-\partial_{\tau}\{{\tau},t\}=0$ follows by varying this action with respect to ${\tau}(t)$, keeping $\nu_{\tau}$ fixed and constant. In fact, keeping $\nu_{\tau}$ a fixed constant and the equation $\partial_{\tau}\{{\tau},t\}=0$ are not independent. Equations \eqref{eq:vacsol} and \eqref{eq:xitransScwarzian} imply that in order for $\nu_{\tau}$ to be constant, ${\tau}(t)$ must satisfy the first order equation
\begin{equation}
    \nu_{\tau}=\frac{-a\big({\tau}(t)-{\tau}_0\big)^2+b\big({\tau}-{\tau}_0\big)+\frac{s_0-b^2}{4a}}{{\tau}'(t)}\,.
\end{equation}
The general solution of this equation is 
\begin{equation}
    {\tau}(t)=\frac{\sqrt{s_0}}{2a}\tanh\left(\frac{\sqrt{s_0}}{2\nu_\tau}(t-t_0)\right)+\frac{b}{2a}+{\tau}_0\,,
\end{equation}
where $t_0$ is an integration constant. Moreover, the solution of the condition $\{{\tau},t\}=-2\pi^2T^2$ is
\begin{equation}
    {\tau}(t)=a_1\tanh\left(\pi T(t-t_0)\right)+a_2\,,
\end{equation}
where again $a_1$, $a_2$ and $t_0$ are integration constants. Clearly, the two solutions coincide provided we identify $s_0=(2\pi T\nu_{\tau})^2$, $a_1=\pi T\nu_{\tau}/a$ and $a_2=b/2a+{\tau}_0$.

The analysis of the regular static solutions above was based on the effective action \eqref{eq:IJTSchwarzian} for $\omega=1$. Interestingly, the same results follow for any $\omega\neq 0$. In particular, for generic $\omega$ we find
\begin{equation}\label{eq:IJTSchwarzian-1}
    \begin{aligned}
        \widetilde I_{\omega}[\alpha,\nu_{\tau}]= \frac{\ell_2}{\kappa_2^2}\int \dd t\,\nu_{\tau}\, {\cal S}_{\omega}[\nu_{\tau}]=\frac{\ell_2}{\kappa_2^2}\int \dd t\,\nu_{\tau}\,\left((1-\omega^2)\frac{s_0}{2\nu_{\tau}^2}-\omega^2\{{\tau},t\}\right)\,.
    \end{aligned}
\end{equation}
Varying this with respect to ${\tau}(t)$ keeping $\nu_{\tau}$ fixed leads to the same constraint, $\partial_{\tau}\{{\tau},t\}=0$. Moreover, on-shell $\{{\tau},t\}=-2\pi^2T^2$ and $s_0=(2\pi T\nu_{\tau})^2$, which implies that for any $\omega$ the on-shell value of the effective action is 
\begin{equation}\label{eq:Iw-final}
    \begin{aligned}
        \widetilde I_{\omega}[\alpha,\nu_{\tau}]=\frac{\ell_2}{\kappa_2^2}\int \dd t\,\nu_{\tau}\,2\pi^2 T^2\,.
    \end{aligned}
\end{equation}

It is worth pointing out that for $\omega =1$, which is the value appearing naturally in JT gravity, we have
\begin{equation}
    S_{\rm JT}\Big|_{\rm on-shell} = \widetilde I_{\omega=1}[\alpha,\nu_{\tau}]= -\frac{\ell_2}{\kappa_2^2}\int \dd t\,\nu_{\tau}\,\{{\tau},t\}\,.
\end{equation}
In this scheme, everything is encoded in the Schwarzian effective action. What our analysis here shows is that the local counterterm \eqref{eq:counterterm} modifies the normalization of this piece of the effective action, as reflected in \eqref{eq:IJTSchwarzian-1}, without tampering with the final answer in \eqref{eq:Iw-final} which is $\omega$ independent. From this perspective, the Schwarzian effective action is a scheme-dependent contribution.

\subsubsection{Thermodynamics} 

Since the value of the effective action \eqref{eq:IJTSchwarzian} on the static black hole solution is independent of $\omega$, its Euclidean version can be identified with the Helmholtz free energy of the 2D black hole. Wick rotating to Euclidean time, $t_E\sim t_E+T^{-1}$, the Euclidean on-shell action is given by 
\begin{equation}
        \widetilde I_{\omega}^E[\alpha,\nu_{\tau}]= \frac{\ell_2}{\kappa_2^2}\int \dd t_E\left(-\frac{s_0}{2\nu_{\tau}}\right)=-\frac{2\pi^2\ell_2}{\kappa_2^2}\nu_{\tau} T\,.
\end{equation}
From \eqref{eq:sol} we read off the location of the horizon, $r_h=2\pi \ell_2^2T$, as well as the corresponding value of the scalar field, $\phi(r_h)=\nu_{\tau} r_h/\ell_2$. It follows that the black hole entropy is given by
\begin{equation}\label{eq:entropy}
    S=\frac{2\pi}{\kappa_2^2}\phi(r_h)=\frac{4\pi^2\ell_2}{\kappa_2^2}\nu_{\tau} T\,,
\end{equation}
while the heat capacity is
\begin{equation}
    C=\frac{\partial S}{\partial T}=\frac{4\pi^2\ell_2}{\kappa_2^2}\nu_{\tau}\,.
\end{equation}
Identifying the energy with the conserved charge \eqref{eq:charge1D}, modified by the appropriate normalization of the timelike Killing vector for the static solution \cite{Castro:2018ffi}, we have 
\begin{equation}\label{eq:2d-energy}
    E=\frac{m}{\nu_{\tau}}=\frac{\ell_2}{2\kappa_2^2}\frac{s_0}{\nu_{\tau}}=\frac{2\pi^2\ell_2}{\kappa_2^2}\nu_{\tau} T^2\,.
\end{equation}
Putting everything together, we verify that the Euclidean on-shell action for any $\omega$ is indeed proportional to the Helmholtz free energy, i.e.,
\begin{equation}
        \widetilde I_{\omega}^E[\alpha,\nu_{\tau}]=T^{-1}\left(E-TS\right)\,.
\end{equation}

It is important to mention that in discussing the thermodynamic properties of JT gravity, we are overlooking a key effect. It has been established that, at finite temperature, quantum corrections in the Schwarzian theory become the leading effect \cite{Stanford:2017thb,Iliesiu:2020qvm}, surpassing the naive thermodynamics stated here. Our goal here is to establish the relations between the off-shell actions and functionals. Writing this thermodynamic contribution, which ignores quantum corrections, serves as a check that the coefficients and couplings agree when we compare them with the higher-dimensional analysis.

\section{Near-extremality at the boundary of  \texorpdfstring{AdS$_3$/CFT$_2$}{AdS3/CFT2}} \label{sec:ads3}

This section will describe the emergence of near-extremal dynamics in the context of AdS$_3$/CFT$_2$. This task has two aspects worth highlighting. 
First, we will work partially off-shell in a reduced phase space: concretely, the radial dependence of the metric component is fully solved, and the remaining conditions to be fully on-shell are equivalent to imposing the Ward identities in the CFT$_2$.   This approach will show how the phase space and dynamics of Sec.\,\ref{Sec:JTgravity} enter AdS$_3$ gravity in the appropriate limit without tampering with the radial direction and identifying the variables off-shell appropriately.  Second, the parametrization of the metric functions will make explicit how variables of the CFT$_2$ are related to those in the nCFT$_1$. With this, we will show the connection between the Ward identities of the two systems. 

 \subsection{\texorpdfstring{CFT$_2$}{CFT2} Ward identities on curved backgrounds}
\label{sec:CFT2}
Our starting point is to consider a CFT$_2$ on a non-trivial two-dimensional background. A useful characterization of this background will be \cite{Cvetic:2016eiv}
\begin{equation}\label{eq:2D}
    ds^2_{\scaleto{\rm 2D}{4pt}}=- \alpha^2\dd u^2+\nu^2(\dd\theta + \mu \dd u )^2~,
\end{equation}
where here $\alpha(u)$, $\nu(u)$ and $\mu(u)$ are three arbitrary functions.\footnote{We could also consider $\alpha$, $\nu$ and $\mu$ to be functions of $\theta$. However, it suffices to have time-dependent sources for our analysis. In order to remove clutter, we are abusing notation here: $\alpha$ and $\nu$ in \eqref{eq:2D} should not be equated with those in \eqref{eq:local-AdS2}-\eqref{eq:phiJT}. The variables will be related via a simple relation in Sec.\,\ref{sec:off-shell-decouple-3d}, and appropriate notation will be introduced there.}
With this parametrization of the background metric, it is convenient to decompose the stress tensor such that its components tie directly to the appropriate sources we have introduced: $(\alpha, \nu, \mu)$. For this reason, we write 
\begin{equation}
\begin{aligned}\label{eq:defO2D}
    T_{\theta\theta}&=\nu \,\mathcal{O}^{\scaleto{\rm 2D}{4pt}}_{\theta}~,\\
T_{u\theta}&=\alpha\, \mathcal{O}^{\scaleto{\rm 2D}{4pt}}_{j}+\mu \,T_{\theta\theta}~,\\
T_{uu}&=\mu^2 \nu \, \mathcal{O}^{\scaleto{\rm 2D}{4pt}}_{\theta}+2\mu\alpha\, \mathcal{O}^{\scaleto{\rm 2D}{4pt}}_{j}-\frac{\alpha^2}{\nu}\mathcal{O}^{\scaleto{\rm 2D}{4pt}}_{u}~,
\end{aligned}
\end{equation}
where we have traded the three components of the stress tensor for the operators ${\cal O}^{\scaleto{\rm 2D}{4pt}}_s\equiv({\cal O}^{\scaleto{\rm 2D}{4pt}}_{u},{\cal O}^{\scaleto{\rm 2D}{4pt}}_j,{\cal O}^{\scaleto{\rm 2D}{4pt}}_\theta)$. It is now simple to verify the pairing between sources and operators, which reads 
\begin{equation}\label{VariationI}
   -\frac{1}{2} \int d^2x \sqrt{-g_{\scaleto{\rm 2D}{4pt}}}\, T_{ab}\,\delta g_{\scaleto{\rm 2D}{4pt}}^{ab} =\int d^2x\big(-\mathcal{O}^{\scaleto{\rm 2D}{4pt}}_j\nu \delta \mu+\mathcal{O}^{\scaleto{\rm 2D}{4pt}}_{\theta}\frac{\alpha}{\nu}\delta\nu+\mathcal{O}^{\scaleto{\rm 2D}{4pt}}_{u}\delta\alpha\big)~.
\end{equation}
 
The Ward identities for a CFT$_2$ are 
\begin{equation}
\label{2DWard}
     \nabla_a T^{ab}=0~, \quad T^a_{~a}=\frac{c}{24\pi}R^{(2)}~,
\end{equation}
where $c$ is the central charge, and $R^{(2)}$ is the Ricci scalar of the background geometry. For the moment, we are considering CFTs with no gravitational anomalies, i.e., $c_L=c_R=c$, but we will generalise this discussion in Sec.\,\ref{sec:CFTgrav}. In terms of the operators introduced in \eqref{eq:defO2D}, and for the background in \eqref{eq:2D}, these equations take a simple form. Assuming that the sources in \eqref{eq:2D} have only $u$-dependence, the diffeomorphism Ward identity reads 
\begin{equation}\label{eq:2DWI}
\begin{aligned}
    (\partial_u-\mu \partial_{\theta}) {\cal O}^{\scaleto{\rm 2D}{4pt}}_{j}+{\cal O}^{\scaleto{\rm 2D}{4pt}}_{j}\frac{\partial_u\nu}{\nu}+\frac{\alpha}{\nu}\partial_{\theta}\mathcal{O}^{\scaleto{\rm 2D}{4pt}}_{u}&=0~,\\
    (\partial_u-\mu \partial_{\theta}) {\cal O}^{\scaleto{\rm 2D}{4pt}}_{u}-{\cal O}^{\scaleto{\rm 2D}{4pt}}_{\theta}\frac{\partial_u\nu}{\nu}+\frac{\alpha}{\nu}\partial_{\theta}\mathcal{O}^{\scaleto{\rm 2D}{4pt}}_{j}&=0~,
\end{aligned}
\end{equation}
while the trace Ward identity is
\begin{equation}\label{eq:2Dtrace}
    \mathcal{O}^{\scaleto{\rm 2D}{4pt}}_{\theta}+{\cal O}^{\scaleto{\rm 2D}{4pt}}_{u}=\frac{c}{24\pi}\frac{2}{\alpha}\partial_u\left(\frac{\partial_u\nu}{\alpha}\right)~,
\end{equation}
where we used that the Ricci scalar is given by
\begin{equation}
    R^{(2)}=\frac{2}{\alpha\nu}\partial_u\left(\frac{\partial_u\nu}{\alpha}\right)~.
\end{equation}

Finally, there are two charges that are important to highlight: the total energy $(E)$ and angular momentum $(J)$ of the system. In our conventions, they read
\begin{equation}\label{eq:charges2D}
    \begin{aligned}
        E &= -\int \dd \theta \left(\nu\,  {\cal O}^{\scaleto{\rm 2D}{4pt}}_{u}-\frac{c}{24\pi}\left(\frac{\partial_u\nu}{\alpha}\right)^2\right)~,\\
               J &= \int \dd \theta\, \nu\, {\cal O}^{\scaleto{\rm 2D}{4pt}}_{j}~.
    \end{aligned}
\end{equation}
From \eqref{eq:2DWI}, we can see that these charges are conserved when the operators are independent of $\theta$.  

\subsection{\texorpdfstring{AdS$_3$}{AdS3} reduced phase space}\label{sec:AdS3-phase-space}

In this part, guided by Sec.\,\ref{sec:CFT2}, we will construct a geometry whose boundary metric is \eqref{eq:2D} and supported by a boundary stress tensor \eqref{eq:defO2D}. We will be working with a three-dimensional Einstein-Hilbert action, supplemented with a negative cosmological constant, i.e.,
\begin{equation}
    I_{\scaleto{\rm AdS_3}{6pt}}=\frac{1}{2\kappa_3^2}\int_{\mathcal{M}} \dd^3x\sqrt{-g}\left(R^{(3)}+\frac{2}{\ell^2_3}\right)+\frac{1}{\kappa_3^2}\int_{\partial \mathcal{M}}\dd^2x\sqrt{-h}\left(K+\frac{1}{\ell_3}\right)~,
\end{equation}
where $\kappa_3^2=8\pi G_3$, $\ell_3$ is the AdS$_3$ radius, and $K$ is the trace of the extrinsic curvature. 

It is well known how to describe asymptotically locally AdS$_3$ backgrounds for pure AdS$_3$ gravity, with one key fact being that the Fefferman-Graham expansion terminates \cite{Skenderis:1999nb}. Here, we will describe these geometries in the Bondi-gauge,\footnote{The Bondi gauge has been used in AdS$_3$ to study various boundary conditions and their associated asymptotic symmetries \cite{Ruzziconi:2021Int, Geiller:2021Bondi, McNees:2023FCB}.} where we start by characterizing the metric as
\begin{equation}\label{eq:3DlineElement}
    ds^2_{\scaleto{\rm 3D}{4pt}}=-F(r,u,\theta)\dd u^2-2\alpha(u) \dd u \dd r+\Phi^2(u,r)\big(\dd\theta+A(r,u,\theta)\dd u\big)^2~.
\end{equation}
One can determine the radial dependence of these functions such that they comply with the equations of motion. Concretely, we have
\begin{equation}\label{eq:3dDilaton}
    \Phi(u,r)=\nu(u) \frac{r}{\ell_3}+\Phi_0 ~,
\end{equation} 
 where $\Phi_0$ is constant, and the remaining metric functions are 
\begin{equation} \label{eq:radial-func3d}
    \begin{aligned}
      F(r,u,\theta) &=  \frac{\alpha^2 }{\nu^2}\,\Phi^2 + F_1(u,\theta)\, \Phi + F_0(u,\theta) + \frac{F_{-2}(u,\theta)^2}{\Phi^2} ~, \\
A(r,u,\theta)&=\mu(u)+\frac{F_{-2}(u,\theta)}{\Phi^2}~.
    \end{aligned}
\end{equation}
In relation to the CFT$_2$ setup, the boundary metric of this spacetime is given by \eqref{eq:2D} as desired; note that the coordinate $u$ is null in the bulk, but on the boundary, it becomes time-like. $\Phi$ controls the size of $S^1$, and hence we can use this field as a dial to get close to the UV (asymptotic AdS$_3$ region) or close to the IR (near horizon region).
The functions $F_i(u,\theta)$ control the one-point functions of the stress tensor. In relation to the notation in \eqref{eq:defO2D}, we find
\begin{equation}\label{eq:map3D-2D}
    \begin{aligned}
        F_{1}(u,\theta)&=2\kappa_3^2\frac{\alpha^2}{\nu^2}\int \dd u\,  \alpha\left(\mathcal{O}^{\scaleto{\rm 2D}{4pt}}_{\theta}+{\cal O}^{\scaleto{\rm 2D}{4pt}}_{u}\right)\,,\\
        F_0(u,\theta)&=2\ell_3\kappa_3^2\frac{\alpha^2}{\nu} {\cal O}^{\scaleto{\rm 2D}{4pt}}_{u}-\ell_3^2\left(\frac{\partial_u\nu}{\nu}\right)^2 -\Phi_0 F_{1}(u,\theta)\,,\\
         F_{-2}(u,\theta)&= \ell_3\kappa_3^2 \,\alpha\,{\cal O}^{\scaleto{\rm 2D}{4pt}}_{j}\,.
    \end{aligned}
\end{equation}
Here the operators $({\cal O}^{\scaleto{\rm 2D}{4pt}}_{\theta},{\cal O}^{\scaleto{\rm 2D}{4pt}}_{u},{\cal O}^{\scaleto{\rm 2D}{4pt}}_{j})$ depend on the variables $u$ and $\theta$. 
As stated, the background is still not a solution to Einstein's equations.  Imposing that the metric \eqref{eq:3DlineElement} solves the equations of motion requires that
\begin{equation}\label{eq:EOM3D-WI}
\begin{aligned}
    (\partial_u-\mu \partial_{\theta}) {\cal O}^{\scaleto{\rm 2D}{4pt}}_{j}+{\cal O}^{\scaleto{\rm 2D}{4pt}}_{j}\frac{\partial_u\nu}{\nu}+\frac{\alpha}{\nu}\partial_{\theta}\mathcal{O}^{\scaleto{\rm 2D}{4pt}}_{u}&=0~,\\
    (\partial_u-\mu \partial_{\theta}) {\cal O}^{\scaleto{\rm 2D}{4pt}}_{u}-{\cal O}^{\scaleto{\rm 2D}{4pt}}_{\theta}\frac{\partial_u\nu}{\nu}+\frac{\alpha}{\nu}\partial_{\theta}\mathcal{O}^{\scaleto{\rm 2D}{4pt}}_{j}&=0~,\\
    {\cal O}^{\scaleto{\rm 2D}{4pt}}_{\theta}+{\cal O}^{\scaleto{\rm 2D}{4pt}}_{u}=\frac{c}{24\pi}\frac{2}{\alpha}\partial_u\left(\frac{\partial_u\nu}{\alpha}\right)~,
\end{aligned}
\end{equation}
where the central charge is given in terms of gravitational parameters by 
\begin{equation}\label{eq:3DCentralCharge}
    c=\frac{3\ell_3}{2G_3}~.
\end{equation}
These are exactly the Ward identities of the CFT$_2$: the conservation equation \eqref{eq:2DWI} and trace anomaly \eqref{eq:2Dtrace}. Hence, imposing that \eqref{eq:3DlineElement}-\eqref{eq:map3D-2D} is locally AdS$_3$ is equivalent to imposing the Ward identities of the CFT. 

It is important to note that \eqref{eq:3DlineElement}-\eqref{eq:map3D-2D} generalise the Ba\~nados phase space introduced in \cite{Banados:1998gg}. Here, we have parametrized the solution for a time-dependent boundary metric, and in contrast to \cite{Banados:1998gg}, the trace anomaly is not solved for.  Our parametrization of AdS$_3$  is an adaptation of \cite{Skenderis:1999nb,deHaro:2000vlm} where we are working in the Bondi gauge, rather than a Fefferman-Graham gauge. The advantage of working in the Bondi gauge, relative to Fefferman-Graham, is that $\Phi$ in \eqref{eq:3dDilaton} is independent of the operators $({\cal O}^{\scaleto{\rm 2D}{4pt}}_{\theta},{\cal O}^{\scaleto{\rm 2D}{4pt}}_{u},{\cal O}^{\scaleto{\rm 2D}{4pt}}_{j})$; for this reason, the decoupling limit will not tamper with a redefinition of the radial direction.

\subsubsection{BTZ black hole}
It is useful to describe the BTZ black hole \cite{Banados:1992wn,Banados:1992gq} in our parametrisation of the phase space (\ref{eq:3DlineElement}). First, we set $\nu=\ell_3$ and $\Phi_0=0$ so that $\Phi$  is simply the radial variable. The remaining sources are chosen as $\alpha=1$ and $\mu=0$. Then, the metric is given by
\begin{equation}\label{eq:BTZ}
    ds^2_{\scaleto{\rm BTZ}{4pt}}=-\left(\frac{r^2}{\ell_3^2}-m+\frac{j^2}{4r^2}\right)\dd u^2-2\dd u\dd r+r^2\left(\dd \theta+\frac{j}{2r^2}\dd u\right)^2~,
\end{equation}
where $m$ and $j$ are constants and are related to the inner ($r_-$) and outer ($r_+$) horizons via 
\begin{equation}
    m=\frac{r^2_{+}+r^2_{-}}{\ell_3^2}~,\quad j=\frac{2r_{+}r_{-}}{\ell_3}~.
\end{equation}
The energy and angular momentum of the black hole are given by $M_{\scaleto{\rm BTZ}{4pt}}=\frac{m}{8G_3}$ and  $J_{\scaleto{\rm BTZ}{4pt}}=\frac{j}{8G_3}$, respectively. In our parametrisation, this standard metric follows after making the following identifications with the operators $({\cal O}^{\scaleto{\rm 2D}{4pt}}_{\theta},{\cal O}^{\scaleto{\rm 2D}{4pt}}_{u},{\cal O}^{\scaleto{\rm 2D}{4pt}}_{j})$, consistent with the conserved charges \eqref{eq:charges2D}
\begin{equation}
\begin{aligned}\label{eq:map-BTZ-cft}
    {\cal O}^{\scaleto{\rm 2D}{4pt}}_{u}&=-{\cal O}^{\scaleto{\rm 2D}{4pt}}_{\theta}=-\frac{1}{2\pi}M_{\scaleto{\rm BTZ}{4pt}}~,\\
    {\cal O}^{\scaleto{\rm 2D}{4pt}}_{j}&=\frac{1}{2\pi \ell_3} J_{\scaleto{\rm BTZ}{4pt}}~.
\end{aligned}
\end{equation}
The Hawking temperature $T$, angular velocity $\Omega$, and black hole entropy $S$ are
\begin{equation} \label{eq:thermo-BTZ}
    T=\frac{r^2_{+}-r^2_{-}}{2\pi \ell_3^2 r_{+}}~,\quad \Omega=\frac{r_{-}}{\ell_3 r_{+}}~,\quad S=\frac{\pi r_{+}}{2G_3}~.
\end{equation}

\subsubsection{Near-extremal thermodynamics} For an extremal black hole, the temperature vanishes, and the energy is set by the angular momentum, satisfying the condition $\ell_3 M_{\scaleto{\rm BTZ}{4pt}}=J_{\scaleto{\rm BTZ}{4pt}}$, which is equivalent to the horizons coinciding, i.e., $r_{+}=r_{-}\equiv r_{0}$. To study the low-temperature regime, we expand around extremality as $r_{\pm}=r_0\pm\epsilon$, while keeping $J_{\scaleto{\rm BTZ}{4pt}}$ fixed. This gives at leading order in $\epsilon$: 
\begin{equation}
    T=\frac{2 \epsilon}{\pi \ell_3^2 }~,
\end{equation}
and the response in the energy $\Delta M_{\scaleto{\rm BTZ}{4pt}}$ and the entropy $\Delta S$ to leading order is given respectively by\footnote{For brevity, we are only stating the naive semi-classical expressions for entropy and mass in the near-extremal regime. Quantum corrections can be large, see \cite{Ghosh:2019rcj,Heydeman:2020hhw,Pal:2023cgk}, but will not impact the subsequent analysis. We will discuss briefly quantum corrections in Sec.\,\ref{sec:disc}. }
\begin{equation}\label{eq:near-btz-entropy}
    \Delta M_{\scaleto{\rm BTZ}{4pt}} = \pi^2 \frac{c}{12} \ell_3 T^2~,\qquad \Delta S = \pi^2 \frac{c}{6} \ell_3 T~.
\end{equation}
For future reference, we note that the extremality condition in terms of the operators $({\cal O}^{\scaleto{\rm 2D}{4pt}}_{\theta},{\cal O}^{\scaleto{\rm 2D}{4pt}}_{u},{\cal O}^{\scaleto{\rm 2D}{4pt}}_{j})$ can be expressed as 
\begin{equation}\label{eq:3DExtrCond1}
    {\cal O}^{\scaleto{\rm 2D}{4pt},*}_{\theta}+{\cal O}^{\scaleto{\rm 2D}{4pt},*}_{u}=0~, \quad {\cal O}^{\scaleto{\rm 2D}{4pt},*}_{u}+{\cal O}^{\scaleto{\rm 2D}{4pt},*}_{j}=0~.
\end{equation}
Here, the asterisk in the operators is to emphasise that these relations hold only at extremality. Strictly speaking, the first condition is the absence of an anomaly for the stationary solution. In addition, we notice that in terms of the parameter $r_0$, we can express ${\cal O}^{\scaleto{\rm 2D}{4pt},*}_{j}$ as
\begin{equation}\label{eq:3DExtrCond2}
    {\cal O}^{\scaleto{\rm 2D}{4pt},*}_{j}=\frac{r_0^2}{\kappa^2_3\ell^2_3}~. 
\end{equation}

\subsection{Off-shell decoupling limit}\label{sec:off-shell-decouple-3d}

We may now construct the decoupling limit that implements the near-extremal limit of the reduced phase space in \eqref{eq:3DlineElement}-\eqref{eq:map3D-2D}.  We refer to our approach as an off-shell decoupling limit, since it does not require imposing the equations of motion \eqref{eq:EOM3D-WI}. 

To gain physical insight into our construction, let us recall that the radial coordinate $r$ roughly corresponds to the energy scale along the holographic renormalization group flow. In this context, the linear form of the UV dilaton,
\begin{equation}\label{eq:energyscalePHI}
    \Phi(u,r)=\nu \frac{r}{\ell_3}+\Phi_0 ~,
\end{equation}
can be interpreted as follows. The source $\nu$ acts as a scaling factor for the energy scale $r$, while the constant $\Phi_0$ serves as a reference energy. By redefining $\nu$ as $\nu = \lambda \tilde{\nu}$ and taking the limit $\lambda \to 0$ while keeping $\tilde{\nu}$ fixed, we effectively zoom in on the reference energy $\Phi_0$.

Another important aspect of the decoupling limit is to keep the angular momentum $J$, defined in \eqref{eq:charges2D}, finite as $\lambda$ goes to zero. Concretely, this means
\begin{equation}\label{eq:decoupling-J}
    \mathcal{O}_{j}^{\scaleto{\rm 2D}{4pt}}=\frac{ \mathcal{O}_{j}^{(-1)}}{\lambda}~,
\end{equation}
where $\mathcal{O}_{j}^{(-1)}$ is fixed. 
Therefore, the defining feature of the decoupling limit is to take 
\begin{equation}\label{eq:decoupling-0}
   \nu = \lambda \tilde{\nu} ~, \qquad \lambda\to 0~, 
\end{equation}
while keeping $\tilde{\nu}$ fixed combined with \eqref{eq:decoupling-J}. By itself, this limit introduces divergences in the line element \eqref{eq:3DlineElement}-\eqref{eq:map3D-2D}. To make the limit well defined in the metric functions in \eqref{eq:radial-func3d}, we therefore expand our operators in terms of $\lambda$ as
\begin{equation}\label{eq:decoupling-1}
    \mathcal{O}_{i}^{\scaleto{\rm 2D}{4pt}}=\frac{ \mathcal{O}_{i}^{(-1)}}{\lambda}+ \mathcal{O}_{i}^{(0)}+ \mathcal{O}_{i}^{(1)}\lambda+\cdots~,
\end{equation}
for $\mathcal{O}_{i}^{\scaleto{\rm 2D}{4pt}} =\{ \mathcal{O}_{u}^{\scaleto{\rm 2D}{4pt}},\mathcal{O}_{\theta}^{\scaleto{\rm 2D}{4pt}}\}$. From a geometric perspective, the expansion of $\mathcal{O}_{i}^{\scaleto{\rm 2D}{4pt}}$ starts with $\lambda^{-1}$ because of the dependence on $\nu$ in \eqref{eq:radial-func3d};  from a CFT$_2$ perspective, this will lead to a finite energy and angular momentum as $\lambda \to 0$ in \eqref{eq:charges2D}. Demanding regularity of $F(r,u,\theta)$ in the limit $\lambda\to 0$ gives
\begin{equation}\label{eq:decoupling-2}
    \begin{aligned}
        \mathcal{O}_{j}^{(-1)}&=\mathcal{O}_{\theta}^{(-1)}=-\mathcal{O}_{u}^{(-1)}
        =\frac{c}{12\pi \ell_3^2}\frac{\Phi^2_0}{\tilde{\nu}}~,\\
        \mathcal{O}_{\theta}^{(0)}&=-\mathcal{O}_{u}^{(0)}=0~.
    \end{aligned}
\end{equation}
Next, the regularity of $A(r,u,\theta)$ requires, in addition to \eqref{eq:decoupling-2}, the implementation of the shift: 
\begin{equation}\label{eq:decoupling-3}
    \mu=\tilde{\mu}-\frac{\alpha}{\lambda \tilde{\nu}}~. 
\end{equation}
With the redefinitions \eqref{eq:decoupling-0}-\eqref{eq:decoupling-3}, the limit $\lambda\to0$, where  $\tilde \nu$, $\tilde \mu$, $\alpha$, $\Phi_0$ and $\mathcal{O}_{i}^{(n)}$ are held fixed, the line element \eqref{eq:3DlineElement} is smooth and well defined.\footnote{A closely related formulation of the same off-shell decoupling limit was discussed in \cite{Chaturvedi:2020jyy}. The difference there is that, besides $\tilde \nu$ and $\tilde \mu$, $\alpha \nu$ is kept fixed instead of $\alpha$, as $\nu\to 0$. As a consequence, the form \eqref{eq:2D} of the boundary metric is not preserved along the renormalization group flow discussed in \cite{Chaturvedi:2020jyy}, resulting in a different gauge in the IR.} In particular, we find
\begin{equation}\label{eq:3DlineElement-limit}
    ds^2_{\scaleto{\rm 3D}{4pt}}\underset{\lambda \to 0}{=}-F_{\rm IR}(r,u,\theta)\dd u^2-2\alpha(u) \dd u \dd r+\Phi_0^2\left(\dd\theta-\frac{2\alpha(u)}{\ell_3\kappa_3^2 \Phi_0} r\dd u\right)^2 + O(\lambda)~,
\end{equation}
where
\begin{equation}
   \begin{aligned}\label{eq:decoupling-4}
   F_{\rm IR}(r,u,\theta)&:= \lim_{\lambda \to 0} F(r,u,\theta) \\
   &=   \frac{\alpha^2 r^2}{(\ell_3/2)^2}+2r \frac{\alpha^2}{\tilde{\nu}}\frac{\kappa_3^2}{\ell_3}\int \dd u\, \alpha(\mathcal{O}_{\theta}^{(1)}+\mathcal{O}_{u}^{(1)}) +2\frac{\alpha^2}{\tilde{\nu}}\ell_3\kappa_3^2 \mathcal{O}_{u}^{(1)}-\ell_3^2\left(\frac{\partial_u\tilde{\nu}}{\tilde{\nu}}\right)^2~.
   \end{aligned}
\end{equation}
From here, we already recognise that the geometry, at fixed $\theta$, becomes locally AdS$_2$.   

It is important to stress that in \eqref{eq:decoupling-J}-\eqref{eq:decoupling-4} the equations of motion \eqref{eq:EOM3D-WI} are \textit{not} used. We now turn to the effect of this decoupling limit on the dynamics of the system. Taking the decoupling limit in \eqref{eq:EOM3D-WI}, to leading order in $\lambda$, we find  
\begin{equation}\label{eq:conditions-3D-decouple}
\begin{aligned}
    \partial_{\theta}\mathcal{O}_{i}^{(-1)}= \partial_{u}(\tilde{\nu} \mathcal{O}_{i}^{(-1)})&=0~,\quad i\in\{u,\theta,j\}~, \\
     \partial_{\theta}\mathcal{O}_{i}^{(1)}&=0~,\quad i\in\{u,\theta\}~.
\end{aligned}
\end{equation}
Hence, the decoupling limit drops all the $\theta$-dependence of our phase space. Note that the conditions for $\mathcal{O}_{i}^{(-1)}$ in \eqref{eq:conditions-3D-decouple} are compatible with the regularity condition in \eqref{eq:decoupling-2}. With this, the geometry \eqref{eq:decoupling-4} becomes an $S^1$ fibration over AdS$_2$ with $\ell_2 =\ell_3/2$. To order $\lambda$, which is the first backreaction of the system, the equations of motion reduce to
\begin{equation}\label{eq:3DIRWard}
\begin{aligned}
\mathcal{O}_{\theta}^{(1)}+\mathcal{O}_{u}^{(1)}&=\frac{c}{24\pi}\frac{2}{\alpha}\partial_u\left(\frac{\partial_u\tilde{\nu}}{\alpha}\right)~,\\
     \partial_u \mathcal{O}_{u}^{(1)}-\mathcal{O}_{\theta}^{(1)}\frac{\partial_u\tilde{\nu}}{\tilde{\nu}}&=0~.   
    \end{aligned}
\end{equation}
Hence, the Ward identities of the CFT$_2$ have a well-defined behaviour as we take $\lambda\to 0$.

Finally, applying the decoupling limit directly to the generating functional \eqref{VariationI} yields
\begin{multline}
    \delta I_{\text{IR}}=\int \dd^2 x\left(\left(\frac{\mathcal{O}_{\theta}^{(-1)}-\mathcal{O}_{j}^{(-1)}}{\lambda\tilde\nu}\right)\frac{\alpha}{\tilde\nu}\delta\tilde\nu+\left(\frac{\mathcal{O}_{u}^{(-1)}+\mathcal{O}_{j}^{(-1)}}{\lambda\tilde\nu}\right)\delta\alpha\right)\\
    -2\int \dd^2 x\,\mathcal{O}^{(-1)}_j\tilde\nu \delta \tilde\mu+\lambda\int \dd^2 x\left(\mathcal{O}_{\theta}^{(1)}\frac{\alpha}{\tilde\nu}\delta\tilde\nu+\mathcal{O}_{u}^{(1)}\delta\alpha\right)+O(\lambda^2)~.   
\end{multline}
Invoking the regularity conditions given in (\ref{eq:decoupling-2}), this variation remains finite as $\lambda \to 0$. Since  $\nu \mathcal{O}_j= \tilde\nu \mathcal{O}_j^{(-1)}$ is kept fixed in the decoupling limit, we also need to perform the Legendre transform:
\begin{equation}\label{eq:LegendreJ}
    I_{\text{IR}}\to I_{\text{IR}}+2\int \dd^2 x\,\mathcal{O}^{(-1)}_j\tilde\nu \tilde\mu~,
\end{equation}
so that the ${\cal O}(\lambda^0)$ term in the IR variation of the effective action is proportional to $\tilde{\mu}\delta (\tilde\nu \mathcal{O}_j^{(-1)})$, and hence vanishes in the decoupling limit. The variational principle resulting from this Legendre transformation is discussed extensively in \cite{Chaturvedi:2020jyy}. With a slight abuse of notation, therefore, the variation of the IR effective action after the Legendre transformation \eqref{eq:LegendreJ} takes the following form to leading order in $\lambda$:
\begin{equation}\label{eq:effectiveI3d}
    \delta I_{\text{IR}}=2\pi\lambda \int \dd u\left(\mathcal{O}_{\theta}^{(1)}\frac{\alpha}{\tilde\nu}\delta\tilde\nu+\mathcal{O}_{u}^{(1)}\delta\alpha\right)~.
\end{equation}

\subsubsection{Comparison with JT gravity} 

We can see now that our decoupling limit reproduces the features of JT gravity described in Sec.\,\ref{Sec:JTgravity}. In particular, the blackening factor \eqref{eq:decoupling-4} resembles \eqref{eq:fJT}, and the equations of motion \eqref{eq:3DIRWard} are compatible with \eqref{eq:traceJT}-\eqref{eq:WardIdJT}.  We will now make the map precise, taking into account the ambiguities that we encountered in Sec.\,\ref{Sec:JTgravity}. 

Our first step is to identify the leading term in $ r$ that dominates $f_{\rm IR}$ in \eqref{eq:decoupling-4} with that in \eqref{eq:fJT}. A natural identification is 
\begin{equation}\label{eq:relate2d3d}
      \ell_2=\frac{\ell_3}{2}~,\quad \alpha_{\scaleto{\rm 2D}{4pt}} = \alpha_{\scaleto{\rm 3D}{4pt}}~,
\end{equation}
where we are placing labels on the variables to stress their relation and context. In this identification, without loss of generality, the radial coordinate is the same in 2D and 3D.\footnote{Accounting for a radial redefinition between the two- and three-dimensional notations is simply a combination of the local symmetries ii)+iii) in Sec.\,\ref{sec:2Dphase-space}.} Next, the dilaton in JT gravity is related to  the 3D metric function $\Phi$ via 
\begin{equation}\label{eq:relate2d3d-1}
    \Phi-\Phi_0=\lambda\tilde{\nu} \frac{r}{\ell_3}\equiv L \phi~, 
\end{equation}
where we have introduced a length scale $L$ (which can be thought of as a Kaluza-Klein radius, but its specific value is not important).  With this, and in comparison to \eqref{eq:phiJT}, we have $\phi_0=0$ and
\begin{equation}
    \phi=\nu_{\scaleto{\rm 2D}{4pt}}\frac{r}{\ell_2}~.
\end{equation}
Here, $\nu_{\scaleto{\rm 2D}{4pt}}$ denotes the source introduced in the context of JT gravity. Via \eqref{eq:relate2d3d}-\eqref{eq:relate2d3d-1} we therefore have
\begin{equation}\label{eq:AdS3Map}
  \nu_{\scaleto{\rm 2D}{4pt}}= \frac{\lambda}{2L}\tilde{\nu}~.
\end{equation}
With these identifications, the IR lapse function in \eqref{eq:decoupling-4} can be rewritten as
\begin{equation}
   F_{\rm IR}(u,r)=  \frac{\alpha_{\scaleto{\rm 2D}{4pt}}^2 r^2}{\ell_2^2}+2\frac{\kappa_3^2}{L}\lambda \frac{\alpha_{\scaleto{\rm 2D}{4pt}}^2}{4\nu_{\scaleto{\rm 2D}{4pt}}}\int \dd u\, \alpha_{\scaleto{\rm 2D}{4pt}}(\mathcal{O}_{\theta}^{(1)}+\mathcal{O}_{u}^{(1)})\, \frac{r}{\ell_2}+2 \frac{\kappa_3^2}{L}\lambda\frac{\alpha_{\scaleto{\rm 2D}{4pt}}^2}{\nu_{\scaleto{\rm 2D}{4pt}}}\ell_2 \mathcal{O}_{u}^{(1)}-4\ell_2^2\left(\frac{\partial_u\nu_{\scaleto{\rm 2D}{4pt}}}{\nu_{\scaleto{\rm 2D}{4pt}}}\right)^2~.
\end{equation}
By comparing this expression with equations \eqref{eq:fJT} and \eqref{eq:JToperators}, we recognize that the decoupling limit in AdS$_3$ yields the identifications $\omega^2 =\gamma^{-2}= 4$, provided we relate the operators via
\begin{equation}
    L \kappa_2^2\mathcal{O}_u = \lambda \kappa_3^2\mathcal{O}_u^{(1)}~,\quad L \kappa_2^2\mathcal{O}_{\phi} =\lambda \kappa_3^2\mathcal{O}_{\theta}^{(1)}~.
\end{equation}
It is simple to check that with this relation, equations \eqref{eq:decoupling-4} reproduce \eqref{eq:traceJT} and \eqref{eq:WardIdJT} with $\omega=2$.
Applying this relation to the IR generating functional $I_{\text{IR}}$ defined in (\ref{eq:effectiveI3d}), we find
\begin{equation}
    \delta I_{\text{IR}}=2\pi L\frac{\kappa_2^2}{\kappa_3^2} \int \dd u \Big({\cal O}_u\, \delta \alpha_{\scaleto{\rm 2D}{4pt}} + \frac{\alpha_{\scaleto{\rm 2D}{4pt}}}{\nu_{\scaleto{\rm 2D}{4pt}}}{\cal O}_\phi \,\delta \nu_{\scaleto{\rm 2D}{4pt}} \Big)=\delta I_{\omega=2}~,
\end{equation}
 where $I_{\omega}$ is the functional in \eqref{eq:IJT}. In the last equality, we used that $2\pi L \kappa_2^2 = \kappa_3^2$, which comes from matching the generating functionals $I_{\text{IR}}$ with $I_{\omega=2}$.\footnote{For the example considered here, the relation between $\kappa_2$ and $\kappa_3$ follows directly from dimensional reduction.} It is worth relating $I_{\text{IR}}$ more explicitly to the generating functional of JT gravity. Following the relations in \eqref{eq:MapI} we have 
\begin{equation}
    I_{\text{JT}}\equiv I_{\omega=1}=I_{\text{IR}}-3\frac{\ell_2}{2\kappa_2^2}\int \dd u \sqrt{-h}\,\phi^{-1}h^{uu}(\partial_u\phi)^2~.
\end{equation}
The value $\omega=1$ is the natural choice that arises in JT gravity prior to the ambiguities due to $\omega$ (see \eqref{eq:on-shell-actionJT}). This relation shows the non-trivial interpolation of taking the low energy limit of the UV (AdS$_3$/CFT$_2$) and its effective description of the limit in terms of JT gravity. The difference is encoded by a local counterterm which modifies the value of $\omega$. 

To end, it is worth exploring if the interpolating value of $\omega$ we obtain in the near-extremal limit of AdS$_3$/CFT$_2$ and the analysis of JT gravity has an impact at tree-level thermodynamics. In particular, let us compare the entropy in 2D given by \eqref{eq:entropy} and the near-extremal BTZ entropy in \eqref{eq:near-btz-entropy}. Based on our dictionary in this section, we have
\begin{equation}
    \frac{c}{12}= \frac{\ell_2}{L} \frac{1}{\kappa_2^2}~,
\end{equation}
and, on a thermal state, the energies are related via 
\begin{equation}
    \ell_3 \Delta M_{\scaleto{\rm BTZ}{4pt}} = -2\pi \lambda^2\tilde{\nu}  \, {\cal O}_u^{(1)} = 2 L \nu_{\scaleto{\rm 2D}{4pt}} E_{\scaleto{\rm 2D}{4pt}} ~,
\end{equation}
where $E_{\scaleto{\rm 2D}{4pt}}$ is given by \eqref{eq:2d-energy}. With this, we see that 
\begin{equation}
    S_{\scaleto{\rm 2D}{4pt}} = 2\pi \sqrt{\frac{2\ell_2}{\kappa_2^2}\nu_{\scaleto{\rm 2D}{4pt}}E_{\scaleto{\rm 2D}{4pt}}} =  2\pi \sqrt{\frac{c}{12}\ell_3 \Delta M_{\scaleto{\rm BTZ}{4pt}} } = \Delta S_{\scaleto{\rm 3D}{4pt}}~.
\end{equation}
Therefore, the semi-classical entropy values match as expected from the analysis in, for example, \cite{Ghosh:2019rcj}. In the context of our analysis, it is important to highlight that this agreement is regardless of the value of $\omega$.

\section{Near-extremality at the boundary of TMG} \label{sec:TMG}
In this section, we generalize the analysis of Sec.\,\ref{sec:ads3} to situations where the CFT$_2$ has both a conformal and a gravitational anomaly, inducing independent left and right central charges, $c_{L}$ and $c_R$, respectively. From the gravitational perspective, we will introduce $c_{L}$ and $c_R$ by considering Topologically Massive Gravity (TMG) with boundary conditions suitable for AdS$_3$/CFT$_2$ as done in \cite{Deser:2002iw,Solodukhin:2005ns,Kraus:2005zm,Li:2008dq,Hotta:2008yq,Skenderis:2009nt}. 

The main appeal of this situation is to track further the imprint of the CFT$_2$ data in the near-extremal dynamics.  This will reinforce some aspects already encountered in Sec.\,\ref{sec:ads3}, related to how we implement the decoupling limit, and it will also illustrate the role of the left and right moving sectors of the CFT$_2$ in the near-extremal dynamics. With this in mind, we will only consider the minimal modifications to the analysis of Sec.\,\ref{sec:ads3} such that the effects of $c_L\neq c_R$ are manifest.  

It is important to mention that TMG, as a semi-classical gravitational theory, suffers instabilities and unitarity violations; see, for example,  \cite{Li:2008yz,Grumiller:2008qz,Maloney:2009ck}. Although these are important dynamical aspects of the theory, our reduced phase space does not contain these pathological modes, hence, instabilities will not influence the analysis here. 

\subsection{\texorpdfstring{CFT$_2$}{CFT2} Ward identities with a gravitational anomaly}
\label{sec:CFTgrav}
In analogy to Sec.\,\ref{sec:CFT2}, we want to place a two-dimensional CFT with unequal left and right central charges, i.e., $c_R \neq c_L$, on a curved background. However, the CFT$_2$ cannot be consistently coupled to dynamical gravity due to the presence of a gravitational anomaly. This anomaly can be interpreted either as a diffeomorphism anomaly, which leads to non-conservation of the stress tensor, or as a local Lorentz anomaly, which results in a non-symmetric stress tensor \cite{Kraus:2005zm,Solodukhin:2006chern}.\footnote{For a concise summary of the different notions of stress tensors, see \cite{Castro:2019vog}.}
For our purposes, we express the gravitational anomaly in terms of the diffeomorphism anomaly. This choice allows us to define the stress tensor as the variation of the action with respect to the metric,
\begin{equation}\label{eq:TMGgeneratingFunction}
    \delta I=-\frac{1}{2} \int d^2x \sqrt{-g_{\scaleto{\rm 2D}{4pt}}}\, T_{ab}\,\delta g_{\scaleto{\rm 2D}{4pt}}^{ab} ~.
\end{equation}
The Ward identities, with the gravitational and trace anomalies, are now\footnote{There is a typo in eq.~(2.9) of \cite{Castro:2019vog}. The correct expressions are obtained by replacing $\bar c\to\bar c/2$.}
\begin{equation}\label{eq:TMGward}
\begin{aligned}
    \nabla_aT^{ab}=-\frac{\bar{c}}{48\pi}g^{ab}\epsilon^{cd} \partial_e\partial_c \Gamma^e_{ad}~,\\
    T^a_{\phantom{a}a}=\frac{c}{24\pi}R^{(2)}+\frac{\bar{c}}{24\pi}\epsilon_c{}^a\nabla^{[b}\Gamma^{c]}_{ba}~.
\end{aligned}
\end{equation}
Here, $\epsilon_{ab}$ is the Levi-Civita tensor normalized such that $\sqrt{-g}\,\epsilon^{01}=-1$, while $R^{(2)}$ and $\Gamma^e_{ad}$ represent respectively the Ricci scalar and the Christoffel connection of the background geometry. This form of writing the Ward identities in the presence of the diffeomorphism anomaly is consistent with the Wess-Zumino consistency conditions. The violation of covariant conservation of the stress tensor is parametrised by the central charge $\bar{c}$, while the Weyl anomaly is parametrised by the central charge $c$. These are expressed in terms of the right and left central charges of the CFT$_2$ as follows: 
\begin{equation}\label{eq:centralBar}
    c=\frac{c_{L}+c_{R}}{2}~, \qquad \bar{c}=\frac{c_{L}-c_{R}}{2}~.
\end{equation}
We note that the terms accompanying the central charges $c$ and $\bar{c}$ in the Ward identities (\ref{eq:TMGward}) are local expressions of the background geometry. 

The Ward identities \eqref{eq:TMGward} can be alternatively written in the form  
\begin{equation}\label{eq:TMGward-2}
\begin{aligned}
    \nabla_aT^{ab}=\frac{\bar{c}}{48\pi}\epsilon^{bc}\nabla_c R^{(2)}-\nabla_a t^{ab}~,\\
    T^a_{\phantom{a}a}=\frac{c}{24\pi}R^{(2)}-t^a{}_a~,
\end{aligned}
\end{equation}
where $t^{ab}$ is the local and symmetric Bardeen-Zumino stress tensor \cite{Bardeen:1984pm,Skenderis:2009nt,Jensen:2012kj}
\begin{equation}\label{eq:BZStressTensor}
t^{ab}\equiv -\frac12\nabla_c(X^{cab}+X^{cba}-X^{abc})\,,
\end{equation}
with
\begin{equation}
    X^{ab}{}_c\equiv \frac{\bar{c}}{48\pi}(\epsilon^{ad}\Gamma^b_{cd}+\epsilon^{bd}\Gamma^a_{cd})~.
\end{equation}
It follows that the covariant tress tensor, $(T_{\rm cov})^{ab}\equiv T^{ab}+t^{ab}$ satisfies
\begin{equation}\label{eq:TMGwardCov}
\begin{aligned}
    \nabla_a(T_{\rm cov})^{ab}=\frac{\bar{c}}{48\pi}\epsilon^{bc}\nabla_c R^{(2)}~,\\
   (T_{\rm cov})^a{}_{a}=\frac{c}{24\pi}R^{(2)}~.
\end{aligned}
\end{equation}

Suppose that ${\cal T}^{ab}$ is the stress tensor of a CFT without a gravitational anomaly. We can then construct a stress tensor that satisfies the covariant Ward identities \eqref{eq:TMGwardCov} by defining
\begin{equation}\label{eq:SFlow}
    (T_{\rm cov})^{ab}={\cal T}^{ab}-\frac{\bar c}{2c}(\epsilon^a{}_c {\cal T}^{cb}+\epsilon^b{}_c {\cal T}^{ca})\,.
\end{equation}
The trace of this tensor is
\begin{equation}
    (T_{\rm cov})^a{}_{a}={\cal T}^a{}_{a}=\frac{c}{24\pi}R^{(2)}\,,
\end{equation}
while its covariant divergence takes the form
\begin{equation}
    \nabla_a(T_{\rm cov})^{ab}=-\frac{\bar c}{2c}\epsilon^a{}_c \nabla_a{\cal T}^{cb}=\frac{\bar c}{2c}\epsilon^{bc}\nabla_c{\cal T}^a_a=\frac{\bar c}{48\pi}\epsilon^{bc}\nabla_c R^{(2)}\,.
\end{equation}
The relation \eqref{eq:SFlow}, therefore, allows us to obtain the stress tensor of a CFT with a diffeomorphism anomaly from that of a diffeomorphism invariant CFT. This fact will play a key role in obtaining the decoupling limit of a CFT with a diffeomorphism anomaly.

As in Sec.\,\ref{sec:ads3}, we can now make some more explicit choices for the non-trivial two-dimensional background. We adopt the same background metric as in our previous discussion of CFT$_2$ in Sec.\,\ref{sec:CFT2}, as this choice remains unaffected by the gravitational anomaly, namely
\begin{equation}\label{eq:TMGboundaryds}
    ds^2_{\scaleto{\rm 2D}{4pt}}=- \alpha^2\dd u^2+\nu^2(\dd\theta + \mu \dd u )^2~, 
\end{equation}
where $\alpha(u)$, $\nu(u)$ and $\mu(u)$ are three arbitrary functions. A key advantage of choosing the stress tensor \eqref{eq:TMGgeneratingFunction} in a manner consistent with the variational principle in the presence of the gravitational anomaly is that it enables us to apply a decomposition similar to that used in Sec.\,\ref{sec:CFT2} to our current case:
\begin{equation}
\begin{aligned}
    T_{\theta\theta}&=\nu \,\mathcal{O}^{\scaleto{\rm TMG}{4pt}}_{\theta}~,\\
T_{u\theta}&=\alpha\, \mathcal{O}^{\scaleto{\rm TMG}{4pt}}_{j}+\mu \,T_{\theta\theta}~,\\
T_{uu}&=\mu^2 \nu \, \mathcal{O}^{\scaleto{\rm TMG}{4pt}}_{\theta}+2\mu\alpha\, \mathcal{O}^{\scaleto{\rm TMG}{4pt}}_{j}-\frac{\alpha^2}{\nu}\mathcal{O}^{\scaleto{\rm TMG}{4pt}}_{u}~.
\end{aligned}
\end{equation}
From the variation of the generating function (\ref{eq:TMGgeneratingFunction}), we obtain a pairing between the sources and operators given by
\begin{equation}\label{eq:TMGgenerating2}
   \delta I=\int \dd^2 x\Big(-\mathcal{O}^{\scaleto{\rm TMG}{4pt}}_j\nu \delta \mu+\mathcal{O}^{\scaleto{\rm TMG}{4pt}}_{\theta}\frac{\alpha}{\nu}\delta\nu+\mathcal{O}^{\scaleto{\rm TMG}{4pt}}_{u}\delta\alpha\Big)~.
\end{equation}
With this parametrization, the Ward identities \eqref{eq:TMGward} now read
\begin{equation}\label{eq:TMGequations}
    \begin{aligned}
   (\partial_u-\mu \partial_{\theta}) \mathcal{O}^{\scaleto{\rm TMG}{4pt}}_{u}-\mathcal{O}^{\scaleto{\rm TMG}{4pt}}_{\theta}\frac{\partial_u\nu}{\nu}+\frac{\alpha}{\nu}\partial_{\theta}\mathcal{O}^{\scaleto{\rm TMG}{4pt}}_{j}&=\frac{\bar{c}}{48\pi}\frac{\alpha}{\nu\partial_u\nu}\partial_u\left(\left(\frac{\nu\partial_u\nu}{\alpha^2}\right)^2\partial_u\mu\right)~, \\
    (\partial_u-\mu \partial_{\theta}) \mathcal{O}^{\scaleto{\rm TMG}{4pt}}_{j}+\mathcal{O}^{\scaleto{\rm TMG}{4pt}}_{j}\frac{\partial_u\nu}{\nu}+\frac{\alpha}{\nu}\partial_{\theta}\mathcal{O}_{u}^{\scaleto{\rm TMG}{4pt}}&=-\frac{\bar{c}}{48\pi} \frac{1}{\nu}\partial_u^2\left(\frac{\nu\partial_u\nu}{\alpha^2}\right)~, \\
    \mathcal{O}^{\scaleto{\rm TMG}{4pt}}_{\theta}+\mathcal{O}^{\scaleto{\rm TMG}{4pt}}_{u}&=\frac{c}{24\pi}\frac{2}{\alpha}\partial_u\left(\frac{\partial_u\nu}{\alpha}\right)+\frac{\bar{c}}{48\pi}\frac{1}{\alpha}\partial_u\left(\frac{\nu^2\partial_u\mu}{\alpha^2}\right)~.
    \end{aligned}
\end{equation}
The right-hand side of these equations consists of local functions of the sources, arising from the curvature terms in \eqref{eq:TMGward}.

The identification \eqref{eq:SFlow}, after adding the Bardeen-Zumino stress tensor \eqref{eq:BZStressTensor}, provides a map between the operators $(\mathcal{O}^{\scaleto{\rm TMG}{4pt}}_{u},\mathcal{O}^{\scaleto{\rm TMG}{4pt}}_{\theta},\mathcal{O}^{\scaleto{\rm TMG}{4pt}}_{j})$ and the operators $({\cal O}^{\scaleto{\rm 2D}{4pt}}_{u},{\cal O}^{\scaleto{\rm 2D}{4pt}}_{\theta},{\cal O}^{\scaleto{\rm 2D}{4pt}}_{j})$ of the CFT without a diffeomorphism anomaly, discussed in Sec.\,\ref{sec:ads3}. Namely,
\begin{equation}
    \begin{aligned}\label{eq:TMGrotation-0}
\mathcal{O}^{\scaleto{\rm TMG}{4pt}}_{u}&={\cal O}^{\scaleto{\rm 2D}{4pt}}_{u}+\frac{\bar c}{c}{\cal O}^{\scaleto{\rm 2D}{4pt}}_{j}+\frac{\bar{c}}{24\pi}\frac{\nu\partial_u\nu\partial_u\mu}{\alpha^3}~,\\
\mathcal{O}^{\scaleto{\rm TMG}{4pt}}_{\theta}&={\cal O}^{\scaleto{\rm 2D}{4pt}}_{\theta}-\frac{\bar c}{c}{\cal O}^{\scaleto{\rm 2D}{4pt}}_{j}+\frac{\bar{c}}{48\pi}\frac{\nu^2}{\alpha}\partial_u\left(\frac{\partial_u\mu}{\alpha^2}\right)~,\\
\mathcal{O}^{\scaleto{\rm TMG}{4pt}}_{j}&={\cal O}^{\scaleto{\rm 2D}{4pt}}_{j}+\frac{\bar c}{c}{\cal O}^{\scaleto{\rm 2D}{4pt}}_{u}-\frac{\bar{c}}{48\pi} \frac{1}{\nu}\partial_u\left(\frac{\nu\partial_u\nu}{\alpha^2}\right)-\frac{\bar{c}}{24\pi}\frac{(\partial_u\nu)^2}{\nu\alpha^2}~.\\
    \end{aligned}
\end{equation}

It is convenient to rewrite these relations in the more compact form
\begin{equation}\label{eq:TMGrotation}
    \begin{aligned}
\widehat{\mathcal{O}}^{\scaleto{\rm 2D}{4pt}}_{u}&={\cal O}^{\scaleto{\rm 2D}{4pt}}_{u}+\frac{\bar c}{c}{\cal O}^{\scaleto{\rm 2D}{4pt}}_{j}~,\\
\widehat{\mathcal{O}}^{\scaleto{\rm 2D}{4pt}}_{\theta}&={\cal O}^{\scaleto{\rm 2D}{4pt}}_{\theta}-\frac{\bar c}{c}{\cal O}^{\scaleto{\rm 2D}{4pt}}_{j}~,\\
\widehat{\mathcal{O}}^{\scaleto{\rm 2D}{4pt}}_{j}&={\cal O}^{\scaleto{\rm 2D}{4pt}}_{j}+\frac{\bar c}{c}{\cal O}^{\scaleto{\rm 2D}{4pt}}_{u}-\frac{\bar{c}}{24\pi}\frac{(\partial_u\nu)^2}{\nu\alpha^2}~,\\
    \end{aligned}
\end{equation}
where we have introduced the auxiliary operators
\begin{equation}\label{eq:hatOs}
    \begin{aligned}
\widehat{\mathcal{O}}^{\scaleto{\rm 2D}{4pt}}_{u}&\equiv\mathcal{O}^{\scaleto{\rm TMG}{4pt}}_{u}-\frac{\bar{c}}{24\pi}\frac{\nu\partial_u\nu\partial_u\mu}{\alpha^3}~,\\
\widehat{\mathcal{O}}^{\scaleto{\rm 2D}{4pt}}_{\theta}&\equiv\mathcal{O}^{\scaleto{\rm TMG}{4pt}}_{\theta}-\frac{\bar{c}}{48\pi}\frac{\nu^2}{\alpha}\partial_u\left(\frac{\partial_u\mu}{\alpha^2}\right)~,\\
\widehat{\mathcal{O}}^{\scaleto{\rm 2D}{4pt}}_{j}&\equiv\mathcal{O}^{\scaleto{\rm TMG}{4pt}}_{j}+\frac{\bar{c}}{48\pi} \frac{1}{\nu}\partial_u\left(\frac{\nu\partial_u\nu}{\alpha^2}\right)~.\\
    \end{aligned}
\end{equation}
Notice that the transformations \eqref{eq:TMGrotation} become completely homogeneous in terms of the shifted operators ${\cal O}^{\scaleto{\rm 2D}{4pt}}_{u}-\frac{c}{24\pi}\frac{(\partial_u\nu)^2}{\nu\alpha^2}$ and $\widehat{\cal O}^{\scaleto{\rm 2D}{4pt}}_{u}-\frac{c}{24\pi}\frac{(\partial_u\nu)^2}{\nu\alpha^2}$. The first of these is what defines the energy in \eqref{eq:charges2D}, while, as we will see momentarily, the latter defines the energy in TMG. Moreover, the auxiliary operators \eqref{eq:hatOs} are obtained by adding a local -- but not covariant -- counterterm to the generating function of TMG. In particular, from \eqref{eq:TMGgenerating2} follows that 
\begin{equation}\label{eq:OhatVar}
   \delta (I+I_{\rm ct\,, \scaleto{\rm TMG}{4pt}})=\int \dd^2 x\Big(-\widehat{\mathcal{O}}^{\scaleto{\rm 2D}{4pt}}_j\nu \delta \mu+\widehat{\mathcal{O}}^{\scaleto{\rm 2D}{4pt}}_{\theta}\frac{\alpha}{\nu}\delta\nu+\widehat{\mathcal{O}}^{\scaleto{\rm 2D}{4pt}}_{u}\delta\alpha\Big)~,
\end{equation}
where
\begin{equation}\label{eq:OhatCT}
    I_{\rm ct\,, \scaleto{\rm TMG}{4pt}}=-\frac{\bar{c}}{48\pi}\int \dd^2 x\, \mu\,\partial_u\left(\frac{\nu\partial_u\nu}{\alpha^2}\right)~.
\end{equation}

One may verify that both sets of operators $({\cal O}^{\scaleto{\rm 2D}{4pt}}_{u},{\cal O}^{\scaleto{\rm 2D}{4pt}}_{\theta},{\cal O}^{\scaleto{\rm 2D}{4pt}}_{j})$ and $(\widehat{\cal O}^{\scaleto{\rm 2D}{4pt}}_{u},\widehat{\cal O}^{\scaleto{\rm 2D}{4pt}}_{\theta},\widehat{\cal O}^{\scaleto{\rm 2D}{4pt}}_{j})$ satisfy the Ward identities \eqref{eq:2DWI}-\eqref{eq:2Dtrace}, where the diffeomorphism anomaly is absent. This means that, as is the case with the Weyl anomaly, within the effectively one-dimensional phase space we are working in, the diffeomorphism anomaly can be eliminated by a local counterterm. Notice, however, that this would not be the case if the sources were functions of both $u$ and $\theta$. Moreover, it cannot be overemphasized that the relations between the operators $({\cal O}^{\scaleto{\rm 2D}{4pt}}_{u},{\cal O}^{\scaleto{\rm 2D}{4pt}}_{\theta},{\cal O}^{\scaleto{\rm 2D}{4pt}}_{j})$ and $(\mathcal{O}^{\scaleto{\rm TMG}{4pt}}_{u},\mathcal{O}^{\scaleto{\rm TMG}{4pt}}_{\theta},\mathcal{O}^{\scaleto{\rm TMG}{4pt}}_{j})$, and that between $(\widehat{\cal O}^{\scaleto{\rm 2D}{4pt}}_{u},\widehat{\cal O}^{\scaleto{\rm 2D}{4pt}}_{\theta},\widehat{\cal O}^{\scaleto{\rm 2D}{4pt}}_{j})$ and $(\mathcal{O}^{\scaleto{\rm TMG}{4pt}}_{u},\mathcal{O}^{\scaleto{\rm TMG}{4pt}}_{\theta},\mathcal{O}^{\scaleto{\rm TMG}{4pt}}_{j})$, are entirely different. The latter corresponds to a local counterterm that exists only within the effectively one-dimensional phase space. On the contrary, the former is a general non-local deformation of any relativistic CFT. 

Finally, it is useful to construct explicitly conserved charges in this case. Due to the presence of the diffeomorphism anomaly, the generating function is not invariant under diffeomorphisms, i.e., $\delta_{\xi}I\neq0$ for an arbitrary diffeomorphism $\xi^a$.  As a consequence, the stress tensor is not covariantly conserved, as expressed in the Ward identity (\ref{eq:TMGward}). Nevertheless, conserved charges can still be defined under certain conditions. In particular, let us consider diffeomorphisms that preserve the gauge choice of the background metric (\ref{eq:TMGboundaryds}). In this case, the transformation of the sources is:
\begin{equation}
    \delta \alpha=\partial_u(\epsilon\alpha)+\alpha\sigma~,\quad \delta\nu=\epsilon\partial_u\nu+ \nu\sigma~,\quad \delta \mu=\partial_u(\epsilon\mu+\varphi)~.
\end{equation}
Here, $\epsilon$, $\varphi$ and $\sigma$ represent time-reparametrisations, $\theta$-reparametrisations, and a Weyl transformation, respectively. The conformal killing vectors (CKVs) are determined by imposing $\delta_{\scaleto{\rm CKV}{4pt}} \alpha=\delta_{\scaleto{\rm CKV}{4pt}}\mu=\delta_{\scaleto{\rm CKV}{4pt}}\nu=0$. This leads to the following solutions for the infinitesimal transformations:
\begin{equation}
    \epsilon=\xi_1\frac{\nu}{\alpha}~,\quad \sigma=-\xi_1\frac{\partial_u\nu}{\alpha}~,\quad \varphi=\xi_2-\xi_1\frac{\nu}{\alpha}\mu~,
\end{equation}
where $\xi_1$ and $\xi_2$ are arbitrary constants. These transformations trivially leave the generating function invariant, since 
\begin{equation}
   \delta_{\scaleto{\rm CKV}{4pt}} I=\int \dd^2 x(-\mathcal{O}^{\scaleto{\rm TMG}{4pt}}_j\nu \delta_{\scaleto{\rm CKV}{4pt}} \mu+\mathcal{O}^{\scaleto{\rm TMG}{4pt}}_{\theta}\frac{\alpha}{\nu}\delta_{\scaleto{\rm CKV}{4pt}}\nu+\mathcal{O}^{\scaleto{\rm TMG}{4pt}}_{u}\delta_{\scaleto{\rm CKV}{4pt}}\alpha)=0~.
\end{equation}

If the operators $({\cal O}^{\scaleto{\rm TMG}{4pt}}_u,{\cal O}^{\scaleto{\rm TMG}{4pt}}_j,{\cal O}^{\scaleto{\rm TMG}{4pt}}_\theta)$ are independent of $\theta$, these CKVs together with the Ward identities (\ref{eq:TMGward}) imply the existence of conserved charges. In particular, integrating by parts this variation of the generating function and keeping total derivative terms we derive
\begin{equation}
\begin{aligned}
    0=\int  \dd^2 x\Bigg[-&\xi_2\partial_u\left(\nu{\cal O}^{\scaleto{\rm TMG}{4pt}}_j+\frac{\bar{c}}{48\pi} \partial_u\left(\frac{\nu\partial_u\nu}{\alpha^2}\right)\right)
    \\&+\xi_1\partial_u\left(\nu{\cal O}^{\scaleto{\rm TMG}{4pt}}_{u}-\frac{c}{24\pi}\left(\frac{\partial_u\nu}{\alpha}\right)^2-\frac{\bar{c}}{24\pi}\frac{\nu^2\partial_u\nu\partial_u\mu}{\alpha^3}\right)\Bigg]~.
\end{aligned}
\end{equation}
Since the constants $\xi_1$ and $\xi_2$ are arbitrary, we identify two conserved charges, given by 
\begin{equation}\label{eq:TMGCharges}
\begin{aligned}
    E_{\scaleto{\rm TMG}{4pt}}&
    =-\int \dd \theta \left(\nu\widehat{\cal O}^{\scaleto{\rm 2D}{4pt}}_{u}-\frac{c}{24\pi}\left(\frac{\partial_u\nu}{\alpha}\right)^2\right)~,\\
    J_{\scaleto{\rm TMG}{4pt}}&
    =\int \dd \theta\, \nu\widehat{{\cal O}}^{\scaleto{\rm 2D}{4pt}}_j~,
\end{aligned}
\end{equation}
where we have used the relations \eqref{eq:hatOs}. $J_{\scaleto{\rm TMG}{4pt}}$ represents the angular momentum, while $E_{\scaleto{\rm TMG}{4pt}}$ denotes the energy of the system. These charges agree with \eqref{eq:charges2D} when $\bar c= 0$ ($c_L=c_R$). However, the relations \eqref{eq:TMGrotation} imply that they are related to those in \eqref{eq:charges2D} even when $\bar{c}\neq 0$. In particular, 
\begin{equation}\label{eq:TMGto2DCharges}
    E_{\scaleto{\rm TMG}{4pt}}=E-\frac{\bar c}{c}J~,\qquad
    J_{\scaleto{\rm TMG}{4pt}}=J-\frac{\bar c}{c}E~.
\end{equation}
Notice that this transformation of the charges preserves the extremality condition since, 
\begin{equation}\label{eq:TMGto2Dextremality}
    E_{\scaleto{\rm TMG}{4pt}}-J_{\scaleto{\rm TMG}{4pt}}=\left(1+\frac{\bar c}{c}\right)(E-J)~.
\end{equation}

\subsection{TMG reduced phase space}

In three dimensions, the Einstein-Hilbert action supplemented by a gravitational Chern-Simons term defines the topological massive gravity (TMG) theory \cite{Deser:1982TMG,Deser:1982TMG2,Deser:1991TMG}, and we will mostly follow the conventions of \cite{Castro:2019vog}. The action is given by
\begin{equation}\label{eq:TMGaction}
    I_{\scaleto{\rm CS}{4pt}}=\frac{1}{2\kappa^2_3}\int \dd^3x \sqrt{-g}\Big(R+\frac{2}{\ell_3}\Big)+\frac{1}{4\kappa^2_3\bm{\mu}}\int \dd^3x \sqrt{-g}\, \epsilon^{\alpha\beta \gamma}\left(\Gamma^{\mu}_{\alpha \nu}\partial_{\beta} \Gamma_{\gamma\mu}^{\nu}+\frac{2}{3}\Gamma_{\alpha \nu}^{\mu}\Gamma_{\beta \sigma}^{\nu}\Gamma_{\gamma\mu}^{\sigma}\right)~,
\end{equation}
where $\bm{\mu}$ is a real coupling constant with the dimensions of mass;  we adopt the convention that the three-dimensional Levi-Civita tensor satisfies $\sqrt{-g}\, \epsilon^{012}=-1$. The equations of motion of TMG are
\begin{equation}\label{eq:TMGeom}
    R_{\mu\nu}-\frac{1}{2}g_{\mu\nu}R-\frac{1}{\ell_3^2}g_{\mu\nu}=-\frac{1}{\bm \mu}C_{\mu\nu}~,
\end{equation}
where $C_{\mu\nu}$ is the Cotton-York tensor,
\begin{equation}
    C_{\mu\nu}\equiv\epsilon^{\phantom{\mu}\gamma \lambda}_{\mu}\,\nabla_{\gamma }\left(R_{\lambda\nu}-\frac{1}{4}g_{\lambda\nu}R\right)~.
\end{equation}
Despite the presence of the Christoffel connection $\Gamma_{\alpha \nu}^{\gamma}$ in the action, the equations of motion remain covariant. Moreover, all locally AdS$_3$ solutions have a vanishing Cotton-York tensor $C_{\mu\nu} = 0$, making them solutions of TMG, as seen in (\ref{eq:TMGeom}).
For a nonzero Cotton-York tensor, $C_{\mu\nu} \neq 0$, TMG admits new solutions, a subclass of which corresponds to warped AdS$_3$ spacetimes, including warped black holes \cite{Anninos:2003bhTMG,Clement:2007bhTMG,Anninos:2009bhTMG}. These asymptotically warped AdS$_3$ geometries do not satisfy the Brown-Henneaux boundary conditions and will not be considered here. 

In what follows, we focus exclusively on locally AdS$_3$ solutions with a vanishing Cotton-York tensor, $C_{\mu\nu}=0$. In this case, it is well-known that the dictionary with the dual CFT$_2$ leads to \cite{Kraus:2005zm}
\begin{equation}\label{eq:centralRL}
    c_{L}=\frac{3\ell_3}{2G_3}\left(1+\frac{1}{\bm{\mu} \ell_3}\right)~, \quad c_{R}=\frac{3\ell_3}{2G_3}\left(1-\frac{1}{\bm{\mu} \ell_3}\right)~.
\end{equation}
Alternatively, via (\ref{eq:centralBar}), we have
\begin{equation}
    c=\frac{3\ell_3}{2G_3}~,\quad \bar{c}=\frac{1}{\bm{\mu}\ell_3}\frac{3\ell_3}{2G_3}~.
\end{equation}

The reduced phase space is again given by \eqref{eq:3DlineElement}-\eqref{eq:radial-func3d}. However, we now need to relate the metric functions  \eqref{eq:radial-func3d} to the CFT$_2$ variables in Sec.\,\ref{sec:CFTgrav} in a way compatible with the Ward identities \eqref{eq:TMGequations}. This is achieved by using the same map between the metric functions and the CFT$_2$ variables $({\cal O}^{\scaleto{\rm 2D}{4pt}}_{u},{\cal O}^{\scaleto{\rm 2D}{4pt}}_{\theta},{\cal O}^{\scaleto{\rm 2D}{4pt}}_{j})$ given in \eqref{eq:map3D-2D}, but expressing these operators in terms of the TMG variables $(\mathcal{O}^{\scaleto{\rm TMG}{4pt}}_{u},\mathcal{O}^{\scaleto{\rm TMG}{4pt}}_{\theta},\mathcal{O}^{\scaleto{\rm TMG}{4pt}}_{j})$ by inverting the transformation \eqref{eq:TMGrotation}. This gives,
\begin{equation}\label{eq:TMGinverserotation}
    \begin{aligned}
\mathcal{O}^{\scaleto{\rm 2D}{4pt}}_{u}&=\frac{c}{c^2-\bar{c}^2}\left(c\,\widehat{\cal O}^{\scaleto{\rm 2D}{4pt}}_{u}-\bar c\,\widehat{\cal O}^{\scaleto{\rm 2D}{4pt}}_{j}-\frac{\bar{c}^2}{24\pi}\frac{(\partial_u\nu)^2}{\nu\alpha^2}\right)~,\\
\mathcal{O}^{\scaleto{\rm 2D}{4pt}}_{\theta}&=\frac{c}{c^2-\bar{c}^2}\left(c\,\widehat{\cal O}^{\scaleto{\rm 2D}{4pt}}_{\theta}+\bar c\,\widehat{\cal O}^{\scaleto{\rm 2D}{4pt}}_{j}\right)-\frac{\bar{c}^2}{c^2-\bar{c}^2}\left(\widehat{\cal O}^{\scaleto{\rm 2D}{4pt}}_{\theta}+\widehat{\cal O}^{\scaleto{\rm 2D}{4pt}}_{u}-\frac{c}{24\pi}\frac{(\partial_u\nu)^2}{\nu\alpha^2}\right)~,\\
\mathcal{O}^{\scaleto{\rm 2D}{4pt}}_{j}&=\frac{c}{c^2-\bar{c}^2}\left(c\,\widehat{\cal O}^{\scaleto{\rm 2D}{4pt}}_{j}-\bar c\,\widehat{\cal O}^{\scaleto{\rm 2D}{4pt}}_{u}+\frac{c\bar{c}}{24\pi}\frac{(\partial_u\nu)^2}{\nu\alpha^2}\right)~,\\
    \end{aligned}
\end{equation}
where $(\widehat{\cal O}^{\scaleto{\rm 2D}{4pt}}_{u},\widehat{\cal O}^{\scaleto{\rm 2D}{4pt}}_{\theta},\widehat{\cal O}^{\scaleto{\rm 2D}{4pt}}_{j})$ are expressed in terms of the TMG variables through \eqref{eq:hatOs}.

\subsubsection{BTZ black hole in TMG}

The line element of the BTZ black hole is still given by \eqref{eq:BTZ}. However, the gravitational Chern-Simons term in the action \eqref{eq:TMGaction} modifies the definition of the holographic stress tensor \cite{Skenderis:2009nt}. 
Accordingly, the TMG mass, $M_{\scaleto{\rm TMG}{4pt}}$, and angular momentum, $J_{\scaleto{\rm TMG}{4pt}}$, are now related to the horizon radii $r_{\pm}$ as
\begin{equation}
    \begin{aligned}
        \ell_3M_{\scaleto{\rm TMG}{4pt}}+J_{\scaleto{\rm TMG}{4pt}} &=\left(1-\frac{1}{\bm{\mu} \ell_3}\right) \frac{(r_++r_-)^2}{8G_3\ell_3} ~,\\
        \ell_3M_{\scaleto{\rm TMG}{4pt}}-J_{\scaleto{\rm TMG}{4pt}} &=\left(1+\frac{1}{\bm{\mu} \ell_3}\right) \frac{(r_+-r_-)^2}{8G_3\ell_3}~.
    \end{aligned}
\end{equation}
These expressions can be easily understood by noticing that the holographic stress tensor of TMG realizes precisely the stress tensor deformation \eqref{eq:SFlow}. 

The TMG mass and angular momentum are related with the operators $({\cal O}^{\scaleto{\rm TMG}{4pt}}_{\theta},{\cal O}^{\scaleto{\rm TMG}{4pt}}_{u},{\cal O}^{\scaleto{\rm TMG}{4pt}}_{j})$ according to the conserved charges \eqref{eq:TMGCharges}, namely 
\begin{equation}
\begin{aligned}\label{eq:map-O-TMG}
     \mathcal{O}^{\scaleto{\rm TMG}{4pt}}_{u}=- \mathcal{O}^{\scaleto{\rm TMG}{4pt}}_{\theta}&=-\frac{M_{\scaleto{\rm TMG}{4pt}}}{2\pi}~,\\
     \mathcal{O}^{\scaleto{\rm TMG}{4pt}}_{j}&=\frac{J_{\scaleto{\rm TMG}{4pt}}}{2\pi \ell_3}~.
\end{aligned}
\end{equation}
These operators are related to those of the original CFT$_2$, $({\cal O}^{\scaleto{\rm 2D}{4pt}}_{u},{\cal O}^{\scaleto{\rm 2D}{4pt}}_{\theta},{\cal O}^{\scaleto{\rm 2D}{4pt}}_{j})$, through the transformation \eqref{eq:TMGrotation}, which leads to the transformation \eqref{eq:TMGto2DCharges} of the conserved charges. This gives
\begin{equation}
    \ell_3M_{\scaleto{\rm TMG}{4pt}}=\ell_3M_{\scaleto{\rm BTZ}{4pt}}-\frac{\bar c}{c}J_{\scaleto{\rm BTZ}{4pt}}~,\qquad
    J_{\scaleto{\rm TMG}{4pt}}=J_{\scaleto{\rm BTZ}{4pt}}-\frac{\bar c}{c}\ell_3 M_{\scaleto{\rm BTZ}{4pt}}~,
\end{equation} 
or equivalently 
\begin{equation}
    \ell_3M_{\scaleto{\rm TMG}{4pt}}\pm J_{\scaleto{\rm TMG}{4pt}}=\left(1\mp \frac{\bar c}{c}\right)(\ell_3M_{\scaleto{\rm BTZ}{4pt}}\pm J_{\scaleto{\rm BTZ}{4pt}})~.
\end{equation} 

The Hawking temperature $T$ and angular velocity $\Omega$ remain unchanged relative to \eqref{eq:thermo-BTZ}, namely
\begin{equation}
    T=\frac{r^2_{+}-r^2_{-}}{2\pi \ell_3^2 r_{+}}~,\qquad \Omega=\frac{r_{-}}{\ell_3 r_{+}}~.
\end{equation}
However, the semi-classical Wald entropy of the black hole receives a contribution from the gravitational Chern-Simons term, in addition to the area-law \cite{Saida:1999ec, Kraus:2005vz, Solodukhin:2005ns,Sahoo:2006vz,Tachikawa:2006sz}, and it is given by 
\begin{equation}
    S_{\scaleto{\rm TMG}{4pt}}=\frac{\pi}{6\ell_3}\left(c_L(r_{+}-r_{-})+c_R(r_{+}+r_{-})\right)~.
\end{equation}

\subsubsection{Near-extremal thermodynamics} 

At extremality, we again have $r_{+}=r_{-}=r_{0}$ and zero temperature. In this limit, the relation between the TMG mass and angular momentum is
\begin{equation}\label{eq:TMGextremality}
    J_{\scaleto{\rm TMG}{4pt}}=\ell_3 M_{\scaleto{\rm TMG}{4pt}}=\frac{c_R}{6\ell^2_3}r_0^2~,
\end{equation}
while the extremal entropy is
\begin{equation}\label{eq:TMGextremalEntropy}
    S_{\scaleto{\rm TMG}{4pt}}=\frac{c_R\pi r_0}{3\ell_3}~.
\end{equation}
The linear response in temperature of the energy $\Delta M_{\scaleto{\rm TMG}{4pt}}$ and entropy $\Delta S_{\scaleto{\rm TMG}{4pt}}$, at fixed $J_{\scaleto{\rm TMG}{4pt}}$, is
\begin{equation}
\begin{aligned}\label{eq:near-BTZ-TMG}
    \Delta M_{\scaleto{\rm TMG}{4pt}}= \pi^2 \frac{c_L}{12} \ell_3 T^2~,\qquad
    \Delta S_{\scaleto{\rm TMG}{4pt}}=\pi^2\frac{c_L}{6} \ell_3 T~.
\end{aligned}
\end{equation}

The right sector controls the notion of extremality, as we can see from (\ref{eq:TMGextremality}) and \eqref{eq:TMGextremalEntropy}. On the other hand, the linear response around extremality is controlled only by the left sector, as seen from \eqref{eq:near-BTZ-TMG}. The presence of the gravitational anomaly in TMG disentangles the role of the right and left degrees of freedom of the CFT$_2$ \cite{Castro:2019vog}. 

\subsection{Off-shell decoupling limit}\label{sec:off-shell-decouple-TMG}

We will now implement the decoupling limit for the TMG phase space described by \eqref{eq:map3D-2D} and \eqref{eq:TMGinverserotation}. The procedure closely parallels the AdS$_3$/CFT$_2$ case discussed in Sec.\,\ref{sec:off-shell-decouple-3d}, so the presentation here will be rather brief. The starting point is again to scale the sources $\nu$ and $\mu$ as  
\begin{equation}\label{eq:decoupling-0b}
   \nu = \lambda \tilde{\nu} ~, \qquad \mu=\tilde{\mu}-\frac{\alpha}{\lambda\tilde{\nu}}~,\qquad \lambda\to 0~, 
\end{equation}
while keeping $\tilde{\nu}$ and $\tilde\mu$ fixed. A key difference, however, is that while the decoupling limit in Sec.\,\ref{sec:off-shell-decouple-3d} is taken while keeping the angular momentum $J_{\scaleto{\rm BTZ}{4pt}}$ fixed, the decoupling limit of TMG involves keeping $J_{\scaleto{\rm TMG}{4pt}}$ fixed instead. As we will see, this affects crucially the asymptotic form of the remaining phase space variables.

Since the parametrisation of the metric functions for TMG is again given by \eqref{eq:map3D-2D} in terms of the operators  $({\cal O}^{\scaleto{\rm 2D}{4pt}}_{u},{\cal O}^{\scaleto{\rm 2D}{4pt}}_{\theta},{\cal O}^{\scaleto{\rm 2D}{4pt}}_{j})$ of Sec.\,\ref{sec:off-shell-decouple-3d}, we can directly apply the operator expansions \eqref{eq:decoupling-J} and \eqref{eq:decoupling-1}. The corresponding expansions for the TMG variables $(\mathcal{O}^{\scaleto{\rm TMG}{4pt}}_{u},\mathcal{O}^{\scaleto{\rm TMG}{4pt}}_{\theta},\mathcal{O}^{\scaleto{\rm TMG}{4pt}}_{j})$ can then be deduced through the relations \eqref{eq:TMGrotation-0}. Let us parametrise the expansions \eqref{eq:decoupling-J} and \eqref{eq:decoupling-1} as
\begin{equation}\label{eq:decoupling-2D}
    \mathcal{O}_{i}^{\scaleto{\rm 2D}{4pt}}=\frac{\mathcal{O}_{i}^{\scaleto{\rm 2D}{4pt}(-1)}}{\lambda}+ \mathcal{O}_{i}^{\scaleto{\rm 2D}{4pt}(0)}+\mathcal{O}_{i}^{\scaleto{\rm 2D}{4pt}(1)}\lambda+\cdots~,
\end{equation}
where now $i=(u,\theta,j)$. Inserting these expansions in the expressions \eqref{eq:TMGrotation-0} for the TMG variables we find that these too admit such expansions as $\lambda\to 0$, namely
\begin{equation}\label{eq:decoupling-TMG}
    \mathcal{O}_{i}^{\scaleto{\rm TMG}{4pt}}=\frac{\mathcal{O}_{i}^{\scaleto{\rm TMG}{4pt}(-1)}}{\lambda}+ \mathcal{O}_{i}^{\scaleto{\rm TMG}{4pt}(0)}+ \mathcal{O}_{i}^{\scaleto{\rm TMG}{4pt}(1)}\lambda+\cdots~.
\end{equation}
Similarly, the operators $(\widehat{\cal O}^{\scaleto{\rm 2D}{4pt}}_{u},\widehat{\cal O}^{\scaleto{\rm 2D}{4pt}}_{\theta},\widehat{\cal O}^{\scaleto{\rm 2D}{4pt}}_{j})$ in \eqref{eq:hatOs} can be expanded as
\begin{equation}\label{eq:decoupling-hatOs}
    \widehat{\mathcal{O}}_{i}^{\scaleto{\rm 2D}{4pt}}=\frac{\widehat{\mathcal{O}}_{i}^{\scaleto{\rm 2D}{4pt}(-1)}}{\lambda}+ \widehat{\mathcal{O}}_{i}^{\scaleto{\rm 2D}{4pt}(0)}+\widehat{\mathcal{O}}_{i}^{\scaleto{\rm 2D}{4pt}(1)}\lambda+\cdots~.
\end{equation}

From the expressions \eqref{eq:TMGCharges} for the TMG charges follows that keeping $J_{\scaleto{\rm TMG}{4pt}}$ fixed in the decoupling limit requires that the expansion of $\widehat{\mathcal{O}}_{j}^{\scaleto{\rm 2D}{4pt}}$ takes the form
\begin{equation}\label{eq:decoupling-J-TMG}
    \widehat{\mathcal{O}}_{j}^{\scaleto{\rm 2D}{4pt}}=\frac{\widehat{\mathcal{O}}_{j}^{\scaleto{\rm 2D}{4pt}(-1)}}{\lambda}~,
\end{equation}
without any subleading corrections. However, this implies that we cannot anymore impose \eqref{eq:decoupling-J}. In particular, we need to allow for a nonzero $\mathcal{O}_{j}^{\scaleto{\rm 2D}{4pt}(1)}$. Using the values for the order $(-1)$ and $(0)$ expansion coefficients in \eqref{eq:3DIRWard}, together with $\mathcal{O}_{j}^{\scaleto{\rm 2D}{4pt}(0)}=0$, we find that the corresponding coefficients of the TMG expansions \eqref{eq:decoupling-TMG} are given by
\begin{equation}\label{eq:TMGorder-10}
\begin{aligned}
\mathcal{O}_{j}^{\scaleto{\rm TMG}{4pt}(-1)}=\mathcal{O}_{\theta}^{\scaleto{\rm TMG}{4pt}(-1)}=-\mathcal{O}_{u}^{\scaleto{\rm TMG}{4pt}(-1)}&=\frac{c_R}{12\pi\ell_3^2}\frac{\Phi_0^2}{\tilde{\nu}}~,\\
\mathcal{O}_{j}^{\scaleto{\rm TMG}{4pt}(0)}=\mathcal{O}_{u}^{\scaleto{\rm TMG}{4pt}(0)}=\mathcal{O}_{\theta}^{\scaleto{\rm TMG}{4pt}(0)}&=0~.\\
\end{aligned}
\end{equation}
The same values apply to the corresponding coefficients of the operators $(\widehat{\cal O}^{\scaleto{\rm 2D}{4pt}}_{u},\widehat{\cal O}^{\scaleto{\rm 2D}{4pt}}_{\theta},\widehat{\cal O}^{\scaleto{\rm 2D}{4pt}}_{j})$. Note that the expressions for $\mathcal{O}_{i}^{\scaleto{\rm TMG}{4pt}(-1)}$ are equivalent to the extremality conditions~\eqref{eq:TMGextremality} derived from the thermodynamic analysis, upon identifying $\tilde{\nu} = \ell_3$ and $\Phi_0 = r_0$.

At order $(1)$ as $\lambda\to 0$ we obtain the TMG coefficients
\begin{equation}\label{eq:TMG2Ddecoupling}
\begin{aligned}
\mathcal{O}_{u}^{\scaleto{\rm TMG}{4pt}(1)}&=\mathcal{O}_{u}^{\scaleto{\rm 2D}{4pt}(1)}+\frac{\bar c}{c}{\cal O}^{\scaleto{\rm 2D}{4pt}(1)}_{j}-\frac{\bar c}{24\pi}\frac{\tilde\nu\partial_u\tilde\nu}{\alpha^3}\partial_u\left(\frac{\alpha}{\tilde\nu}\right)~,\\
\mathcal{O}_{\theta}^{\scaleto{\rm TMG}{4pt}(1)}&=\mathcal{O}_{\theta}^{\scaleto{\rm 2D}{4pt}(1)}-\frac{\bar c}{c}{\cal O}^{\scaleto{\rm 2D}{4pt}(1)}_{j}-\frac{\bar c}{48\pi}\frac{\tilde\nu^2}{\alpha}\partial_u\left(\frac{1}{\alpha^2}\partial_u\left(\frac{\alpha}{\tilde\nu}\right)\right)~,\\
\mathcal{O}_{j}^{\scaleto{\rm TMG}{4pt}(1)}&={\cal O}^{\scaleto{\rm 2D}{4pt}(1)}_{j}+\frac{\bar c}{c}\left(\mathcal{O}_{u}^{\scaleto{\rm 2D}{4pt}(1)}-\frac{c}{24\pi}\frac{1}{\tilde\nu}\left(\frac{\partial_u\tilde\nu}{\alpha}\right)^2\right)-\frac{\bar c}{48\pi}\frac{1}{\tilde\nu}\partial_u\left(\frac{\tilde\nu\partial_u\tilde\nu}{\alpha^2}\right)~,
\end{aligned}
\end{equation}
or equivalently, in terms of the variables defined in \eqref{eq:hatOs},
\begin{equation}\label{eq:hatO2Ddecoupling}
\begin{aligned}
\widehat{\mathcal{O}}_{u}^{\scaleto{\rm 2D}{4pt}(1)}&=\mathcal{O}_{u}^{\scaleto{\rm 2D}{4pt}(1)}+\frac{\bar c}{c}{\cal O}^{\scaleto{\rm 2D}{4pt}(1)}_{j}~,\\
\widehat{\mathcal{O}}_{\theta}^{\scaleto{\rm 2D}{4pt}(1)}&=\mathcal{O}_{\theta}^{\scaleto{\rm 2D}{4pt}(1)}-\frac{\bar c}{c}{\cal O}^{\scaleto{\rm 2D}{4pt}(1)}_{j}~,\\
\widehat{\mathcal{O}}_{j}^{\scaleto{\rm 2D}{4pt}(1)}&={\cal O}^{\scaleto{\rm 2D}{4pt}(1)}_{j}+\frac{\bar c}{c}\left(\mathcal{O}_{u}^{\scaleto{\rm 2D}{4pt}(1)}-\frac{c}{24\pi}\frac{1}{\tilde\nu}\left(\frac{\partial_u\tilde\nu}{\alpha}\right)^2\right)~.
\end{aligned}
\end{equation}
Requiring that $\widehat{\mathcal{O}}_{j}^{\scaleto{\rm 2D}{4pt}}$ is exactly of the form \eqref{eq:decoupling-J-TMG} implies that we must set 
\begin{equation}\label{eq:TMGdecouplingOj}
    \widehat{\mathcal{O}}_{j}^{\scaleto{\rm 2D}{4pt}(1)}=0\quad \Leftrightarrow\quad {\cal O}^{\scaleto{\rm 2D}{4pt}(1)}_{j}=-\frac{\bar c}{c}\left(\mathcal{O}_{u}^{\scaleto{\rm 2D}{4pt}(1)}-\frac{c}{24\pi}\frac{1}{\tilde\nu}\left(\frac{\partial_u\tilde\nu}{\alpha}\right)^2\right)~.
\end{equation}

The Ward identities that the order $(1)$ TMG expansion coefficients satisfy can be determined from the Ward identities \eqref{eq:TMGequations}, by dropping all $\theta$-dependence. In particular, the non-trivial Ward identities take the form 
\begin{equation}\label{eq:TMGWardExpansion}
\begin{aligned}
    \mathcal{O}_{\theta}^{\scaleto{\rm TMG}{4pt}(1)}+\mathcal{O}_{u}^{\scaleto{\rm TMG}{4pt}(1)}&=\frac{c}{24\pi}\frac{2}{\alpha}\partial_u\left(\frac{\partial_u\tilde{\nu}}{\alpha}\right)-\frac{\bar{c}}{48\pi}\frac{1}{\alpha}\partial_u\left(\frac{\tilde{\nu}^2}{\alpha^2}\partial_u\left(\frac{\alpha}{\tilde{\nu}}\right)\right)~,\\
    \partial_u \mathcal{O}_{u}^{\scaleto{\rm TMG}{4pt}(1)}-\mathcal{O}_{\theta}^{\scaleto{\rm TMG}{4pt}(1)}\frac{\partial_u\tilde{\nu}}{\tilde{\nu}}&=-\frac{\bar{c}}{48\pi}\frac{\alpha}{\tilde{\nu}\partial_u\tilde{\nu}}\partial_u\left(\left(\frac{\tilde{\nu}\partial_u\tilde{\nu}}{\alpha^2}\right)^2\partial_u\left(\frac{\alpha}{\tilde{\nu}}\right)\right)~.
\end{aligned}
\end{equation}
The equation for $\mathcal{O}_{j}^{\scaleto{\rm TMG}{4pt}(1)}$ is trivially satisfied due to the TMG decoupling condition $\widehat{\mathcal{O}}_{j}^{\scaleto{\rm 2D}{4pt}(1)}=0$.

We are now ready to examine the form of the line element in the decoupling limit. As we have seen, the parameterization of the metric functions in terms of the operators $({\cal O}^{\scaleto{\rm 2D}{4pt}}_{u},{\cal O}^{\scaleto{\rm 2D}{4pt}}_{\theta},{\cal O}^{\scaleto{\rm 2D}{4pt}}_{j})$ is still given by \eqref{eq:map3D-2D}. However, we need to revisit the limit \eqref{eq:decoupling-4} in order to account for the fact that now $\mathcal{O}_{j}^{\scaleto{\rm 2D}{4pt}(1)}\neq 0$. For a nonzero $\mathcal{O}_{j}^{\scaleto{\rm 2D}{4pt}(1)}$ we obtain 
\begin{equation}
   \begin{aligned}\label{eq:decoupling-4-TMG-0}
   &F_{\rm IR}(r,u,\theta):= \lim_{\lambda \to 0} F(r,u,\theta) \\
   &=   \frac{\alpha^2 r^2}{(\ell_3/2)^2}+2r \frac{\alpha^2}{\tilde{\nu}}\frac{\kappa_3^2}{\ell_3}\int \dd u\, \alpha\big(\mathcal{O}_{\theta}^{\scaleto{\rm 2D}{4pt}(1)}+\mathcal{O}_{u}^{\scaleto{\rm 2D}{4pt}(1)}\big)+2\frac{\alpha^2}{\tilde{\nu}}\ell_3\kappa_3^2 \big(\mathcal{O}_{u}^{\scaleto{\rm 2D}{4pt}(1)}+\mathcal{O}_{j}^{\scaleto{\rm 2D}{4pt}(1)}\big)-\ell_3^2\left(\frac{\partial_u\tilde{\nu}}{\tilde{\nu}}\right)^2~.
   \end{aligned}
\end{equation}
We can now express this in terms of the TMG variables $(\mathcal{O}^{\scaleto{\rm TMG}{4pt}(1)}_{u},\mathcal{O}^{\scaleto{\rm TMG}{4pt}(1)}_{\theta},\mathcal{O}^{\scaleto{\rm TMG}{4pt}(1)}_{j})$, or equivalently $(\widehat{\cal O}^{\scaleto{\rm 2D}{4pt}(1)}_{u},\widehat{\cal O}^{\scaleto{\rm 2D}{4pt}(1)}_{\theta},\widehat{\cal O}^{\scaleto{\rm 2D}{4pt}(1)}_{j})$, using the relations \eqref{eq:TMGinverserotation} and the fact that $\widehat{\mathcal{O}}_{j}^{\scaleto{\rm 2D}{4pt}(1)}=0$. The result is 
\begin{equation}
   \label{eq:decoupling-4-TMG}
   F_{\rm IR}(r,u,\theta) =  \frac{\alpha^2 r^2}{(\ell_3/2)^2}+2r \frac{\alpha^2}{\tilde{\nu}}\frac{\kappa_3^2}{\ell_3}\int \dd u\, \alpha\big(\widehat{\mathcal{O}}_{\theta}^{\scaleto{\rm 2D}{4pt}(1)}+\widehat{\mathcal{O}}_{u}^{\scaleto{\rm 2D}{4pt}(1)}\big)+2\frac{\alpha^2}{\tilde{\nu}}\ell_3\kappa_3^2 \frac{c}{c_L}\widehat{\mathcal{O}}_{u}^{\scaleto{\rm 2D}{4pt}(1)}-\frac{c}{c_L}\ell_3^2\left(\frac{\partial_u\tilde{\nu}}{\tilde{\nu}}\right)^2~.
\end{equation}
The IR vector potential $A_{\text{IR}}$ remains identical to the case with $c_L=c_R$ ($\bar c =0$) in Sec.\,\ref{sec:off-shell-decouple-3d}, namely
\begin{equation}
    A_{\text{IR}}=\tilde{\mu}-\frac{2r\alpha}{\ell_3 \Phi_0}~.
\end{equation}

From the IR limit of the lapse function \eqref{eq:decoupling-4-TMG} we identify the IR variables 
\begin{equation}\label{eq:OIRTMG}
    \mathcal{O}_{u}^{\text{IR}}=\widehat{\mathcal{O}}_{j}^{\scaleto{\rm 2D}{4pt}(1)}~,\qquad
\mathcal{O}_{\theta}^{\text{IR}}=\widehat{\mathcal{O}}_{\theta}^{\scaleto{\rm 2D}{4pt}(1)}~,
\end{equation}
so that 
\begin{equation}\label{eq:TMGIRlapsef2}
    F_{\text{IR}}=\frac{\alpha^2 r^2}{(\ell_3/2)^2}+2r \frac{\alpha^2}{\tilde{\nu}}\frac{\kappa_3^2}{\ell_3}\int \dd u\, \alpha(\mathcal{O}_{u}^{\text{IR}}+\mathcal{O}_{\theta}^{\text{IR}}) +2\frac{\alpha^2}{\tilde{\nu}}\ell_3\kappa_3^2\frac{c}{c_L} \mathcal{O}_{ u}^{\text{IR}}-\ell_3^2\frac{c}{c_L}\left(\frac{\partial_u\tilde{\nu}}{\tilde{\nu}}\right)^2~.
\end{equation}
As we have seen, these operators satisfy the IR limit of the Ward identities \eqref{eq:2DWI}-\eqref{eq:2Dtrace} without a diffeomorphism anomaly, namely
\begin{equation}\label{eq:TMGWardExpansion2}
\begin{aligned}
&\mathcal{O}_{\theta}^{\text{IR}}+\mathcal{O}_{u}^{\text{IR}}=\frac{c}{24\pi}\frac{2}{\alpha}\partial_u\left(\frac{\partial_u\tilde{\nu}}{\alpha}\right)~,\\
 &   \partial_u \mathcal{O}_{u}^{\text{IR}}-\mathcal{O}_{\theta}^{\text{IR}}\frac{\partial_u\tilde{\nu}}{\tilde{\nu}}=0~.
\end{aligned}
\end{equation}

Recall that the operators $(\widehat{\cal O}^{\scaleto{\rm 2D}{4pt}}_{u},\widehat{\cal O}^{\scaleto{\rm 2D}{4pt}}_{\theta},\widehat{\cal O}^{\scaleto{\rm 2D}{4pt}}_{j})$, and hence $\mathcal{O}_{u}^{\text{IR}}$ and $\mathcal{O}_{\theta}^{\text{IR}}$, arise by adding to the TMG effective action the local but non-covariant counterterm \eqref{eq:OhatCT}, whose IR limit is
\begin{equation}\label{eq:OhatCT-IR}
    I_{\rm ct\,, \scaleto{\rm TMG}{4pt}}=2\pi\lambda\frac{\bar{c}}{48\pi}\int \dd u \frac{\alpha}{\tilde\nu}\,\partial_u\left(\frac{\tilde\nu\partial_u\tilde\nu}{\alpha^2}\right)~.
\end{equation}
Indeed, up to a total derivative, this local counterterm can be expressed in terms of the one-dimensional boundary metric $h_{uu}$ and the boundary value of the scalar $\phi$ as 
\begin{equation}\label{eq:OhatCT-IR-cov}
    S_{\rm ct\,, \scaleto{\rm TMG}{4pt}}=2\pi\lambda\frac{\bar{c}}{48\pi}\int \dd u\sqrt{-h}\,h^{uu}\phi^{-1}\partial_u\phi\big(\partial_u\phi-\Gamma^u_{uu}\phi\big)~,
\end{equation}
where $\Gamma^u_{uu}$ is the Christoffel connection of the one-dimensional boundary metric $h_{uu}$. The appearance of this quantity renders manifest that the local counterterm \eqref{eq:OhatCT-IR} is not covariant.

Using the values \eqref{eq:TMGorder-10} for the operator expansion coefficients, the fact that $\widehat{\mathcal{O}}^{\scaleto{\rm 2D}{4pt}(1)}_j=0$, and implicitly implementing the Legendre transform \eqref{eq:LegendreJ}, the IR limit of the variational principle \eqref{eq:OhatVar} becomes
\begin{equation}\label{eq:OhatVar-IR}
   \delta (I_{\rm IR}+I_{\rm ct\,, \scaleto{\rm TMG}{4pt}})=2\pi\lambda\int \dd u\Big(\mathcal{O}_{\theta}^{\text{IR}}\frac{\alpha}{\tilde\nu}\delta\tilde\nu+\mathcal{O}_{u}^{\text{IR}}\delta\alpha\Big)+ O(\lambda^2)~.
\end{equation}
As we have already pointed out, this demonstrates that the gravitational anomaly can be eliminated by a local but non-covariant counterterm in the effective one-dimensional theory. This is a generic feature of the description of anomalies when the effective theory changes dimension \cite{Achucarro:1993fd,Sahoo:2006vz}.

\subsubsection{Comparison with JT gravity} 

As expected, many aspects of the decoupling limit in TMG reproduce features of JT gravity discussed in Sec.\,\ref{Sec:JTgravity}. In particular, the lapse function in~\eqref{eq:TMGIRlapsef2} resembles that of~\eqref{eq:fJT}, and the Ward identities~\eqref{eq:TMGWardExpansion2} are compatible with~\eqref{eq:traceJT}–\eqref{eq:WardIdJT}. We will now make this relation precise, taking into account the ambiguities encountered in Sec.\,\ref{Sec:JTgravity}.

The initial steps of the comparison are unmodified from Sec.\,\ref{sec:off-shell-decouple-3d}. More concretely, the relations in \eqref{eq:relate2d3d}-\eqref{eq:AdS3Map} hold for TMG, since there is no use of dynamics in those equations.
With these identifications, the IR lapse function $F_{\rm IR}$ in (\ref{eq:TMGIRlapsef2}) now reads
\begin{equation}
   F_{\rm IR}=  \frac{\alpha^2_{\scaleto{\rm 2D}{4pt}} r^2}{\ell_2^2}+2\frac{\kappa_3^2}{L} \lambda \frac{\alpha^2_{\scaleto{\rm 2D}{4pt}}}{4\nu_{\scaleto{\rm 2D}{4pt}}}\int \dd u\, \alpha_{\scaleto{\rm 2D}{4pt}}(\mathcal{O}_{\theta}^{\text{IR}}+\mathcal{O}_{u}^{\text{IR}}) \frac{r}{\ell_2} +2\frac{\kappa_3^2}{L}\lambda \frac{c}{c_L}\frac{\alpha^2_{\scaleto{\rm 2D}{4pt}}}{\nu_{\scaleto{\rm 2D}{4pt}}} \ell_2\mathcal{O}_{u}^{\text{IR}}-4\ell_2^2\frac{c}{c_L}\left(\frac{\partial_u\nu_{\scaleto{\rm 2D}{4pt}}}{\nu_{\scaleto{\rm 2D}{4pt}}}\right)^2~.
\end{equation}
By comparing this expression with equations \eqref{eq:JToperators} in Sec.\,\ref{Sec:JTgravity}, we recognize that the decoupling limit in TMG yields the identifications $\omega^2 =\gamma^{-2}= 4\frac{c}{c_L}$, provided we relate the operators as
\begin{equation}\label{eq:map-O-TMG-2D}
    \kappa_2^2\mathcal{O}_u =\lambda \frac{\kappa_3^2}{L}\frac{c}{c_L}\mathcal{O}_u^{\text{IR}}~,\quad \kappa_2^2\mathcal{O}_{\phi} =\lambda \frac{\kappa_3^2}{L}\frac{c}{c_L}\mathcal{O}_{\theta}^{\text{IR}}~.
\end{equation}
Applying this relation to the IR generating functional $I^{\text{IR}}_{\text{3D}}$ defined in \eqref{eq:OhatVar-IR}, we find
\begin{equation}
    \delta( I_{\text{IR}}+ I_{\rm ct\,, \scaleto{\rm TMG}{4pt}}) =2\pi L\frac{c_L}{c}\frac{\kappa_2^2}{\kappa_3^2} \int \dd u \Big({\cal O}_u\, \delta \alpha_{\scaleto{\rm 2D}{4pt}} + \frac{\alpha_{\scaleto{\rm 2D}{4pt}}}{\nu_{\scaleto{\rm 2D}{4pt}}}{\cal O}_\phi \,\delta \nu_{\scaleto{\rm 2D}{4pt}} \Big)=\delta I_{\omega}~, \quad \omega^2=4\frac{c}{c_L}.
\end{equation}
In the last equality, we are setting 
\begin{equation}
    2\pi L  \kappa_2^2= \frac{c}{c_L}\kappa_3^2~,
\end{equation}
to identify the functionals. We note that this relation does not follow from dimensional reduction, which would instead yield the naive relation $2\pi L \kappa_2^2 = \kappa_3^2$. The reason behind this discrepancy is that TMG is a higher-derivative theory. For instance, to compute the black hole entropy in TMG, one must use the Wald entropy formula rather than the standard Bekenstein–Hawking formula, which holds only for two-derivative theories. Our off-shell construction of the decoupling limit properly captures this subtlety by working in terms of the CFT operators, whose Ward identities encode this information.

Thus, for TMG, we obtain the following mapping:
\begin{equation}
    I_{\omega=1}=I_{\omega}+(1-\omega^2)\frac{\ell_2}{2\kappa_2^2}\int \dd u \sqrt{-h}\,\phi^{-1}h^{uu}(\partial_u\phi)^2~, \quad \omega^2= 4\frac{c}{c_L}~,
\end{equation}
where we now have
\begin{equation}
    \frac{c_L}{12}= \frac{\ell_2}{L} \frac{1}{\kappa_2^2}~.
\end{equation}
Therefore, the dynamics of JT gravity is controlled solely by the left-movers of the CFT$_2$, while the right-movers are frozen in the near-extremal regime~\cite{Castro:2019vog}. 

Finally, it is instructive to check that the near-extremal thermodynamics is correctly captured by JT gravity. Similar to the case in Sec.\,\ref{sec:off-shell-decouple-3d}, we have 
\begin{equation}
    \ell_3 \Delta M_{\scaleto{\rm TMG}{4pt}} = -2\pi \lambda^2\tilde{\nu}  \, {\cal O}_u^{\text{IR}} = 2 L \nu_{\scaleto{\rm 2D}{4pt}} E_{\scaleto{\rm 2D}{4pt}} ~,
\end{equation}
where we used \eqref{eq:map-O-TMG}, \eqref{eq:near-BTZ-TMG}, and \eqref{eq:map-O-TMG-2D}. It is then simple to check that 
\begin{equation}
    S_{\scaleto{\rm 2D}{4pt}} = 2\pi \sqrt{\frac{2\ell_2}{\kappa_2^2}\nu_{\scaleto{\rm 2D}{4pt}}E_{\scaleto{\rm 2D}{4pt}}} =  2\pi \sqrt{\frac{c_L}{12}\ell_3 \Delta M_{\scaleto{\rm TMG}{4pt}} } = \Delta S_{\scaleto{\rm TMG}{4pt}}~,
\end{equation}
where $S_{\scaleto{\rm 2D}{4pt}}$ is given by \eqref{eq:entropy} and  $\Delta S_{\scaleto{\rm TMG}{4pt}}$ by \eqref{eq:near-BTZ-TMG}. As in previous sections, although we encounter a non-trivial value of  $\omega$, which controls the Schwarzian action, the match of the thermodynamic entropy does not rely on $\omega$.

\section{Near-extremality at the boundary of \texorpdfstring{AdS$_4$/CFT$_3$}{AdS4/CFT3}} \label{sec:ads5}

In this section, we introduce a novel construction of the decoupling limit within the AdS$_4$/CFT$_3$ correspondence, where we can track how the near-extremal dynamics is imprinted on the UV data on both the gravitational and CFT sides by matching the appropriate phase spaces in the IR and UV. The gravitational analysis will be done for Einstein-Maxwell theory with a negative cosmological constant, and the reduced phase space we construct is a non-trivial gravitational dressing around the Reissner-Nordstr\"om AdS$_4$ black hole. On the CFT$_3$, this corresponds to a configuration with a non-trivial stress tensor and a $U(1)$ current placed on a curved manifold of topology $\mathbb{R}\times M_2$. 

However, the situation in this section is substantially more intricate relative to AdS$_3$/CFT$_2$. First, unlike in two-dimensional CFTs, in three dimensions there are no anomalies. This makes a possible connection with the Schwarzian dynamics more elusive. We will address this by making contact with the counterterm ambiguity discussed in Sec.\,\ref{sec:Schw}. Second, on the gravity side, the presence of local degrees of freedom prevents the complete characterisation of solutions in the Einstein equations with a fixed boundary metric. To address this challenge, we focus on a specific subsector of Einstein-Maxwell theory with a negative cosmological constant: a generalisation of the Robinson-Trautman (RT) family of algebraically special solutions  \cite{RT:1962}.

\subsection{\texorpdfstring{CFT$_3$}{CFT3} Ward identities on curved backgrounds}

We will again begin with a brief discussion of the Ward identities of generic three-dimensional CFTs that possess a $U(1)$ global symmetry. The relevant currents in this case are the stress tensor, $T^{ab}$, and a $U(1)$ current, $J^a$. If we place the theory on a curved background and turn on a background gauge field, we can read off these currents through the variation of the generating function $I[g,{\cal A}]$ with respect to the corresponding sources, namely
\begin{equation}\label{eq:3DvarI}
    \delta I=\int \dd^3x \sqrt{-g} \Big(\frac{1}{2}T^{ab}\delta g_{ab}+ J^b\delta {\cal A}_{b}\Big)~.
\end{equation}
The Ward identities that these operators satisfy follow from the invariance of $I[g,{\cal A}]$ under diffeomorphisms, U(1) gauge, and local Weyl transformations of the background sources. Respectively, they take the form
\begin{equation}
\label{eq:3DWard}
    \nabla_a T^{ab}+J_a {\cal F}^{ab}=0~, \quad \nabla_bJ^b=0~, \quad T^a_{~a}=0~,
\end{equation}
where ${\cal F}=\dd{\cal A}$. A key difference with the two-dimensional CFTs discussed above is that there is no possibility of a Weyl or gravitational anomaly in three dimensions. 

Given a conformal Killing vector $\xi^a$ of the background $g_{ab}$, ${\cal A}_a$, the Ward identities \eqref{eq:3DWard} imply that the following quantity is conserved \cite{Papadimitriou:2005ii}
\begin{equation}\label{eq:CKCharge3D}
    {\cal Q}[\xi]=-\int_{\cal C} \dd \sigma_a\, \xi^b(T^a{}_b+{\cal A}_b J^a)~,
\end{equation}
i.e. it is independent of the choice of Cauchy surface ${\cal C}$. Here, $\dd \sigma_a$ is the directional volume element on the Cauchy surface ${\cal C}$. Moreover, the conservation of the U(1) current in  \eqref{eq:3DWard} and the Bianchi identity ${\rm d}{\cal F}=0$ imply respectively the conservation of the electric and magnetic charges 
\begin{equation}\label{eq:EMCharges3D}
    {\cal Q}_e=\int_{\cal C} \dd\sigma_a\, J^a~,\quad {\cal Q}_m=\int_{\cal C} {\cal F}~.
\end{equation}

We will place the CFT on a curved background $g$ of the form $\mathbb{R}\times M_2$ and turn on a background gauge field ${\cal A}_a$. We will parametrise the metric as
\begin{equation}
\label{eq:3Dmetric}
    ds^2_{\scaleto{\rm 3D}{4pt}}=-\alpha^2 \dd u^2+2 \nu^2 e^{2\chi}\dd z\dd\bar{z}~,
\end{equation}
where $\alpha=\alpha(u)$, $\nu=\nu(u)$ depend only on the time $u$, while $\chi=\chi(u,z,\bar{z})$ depends on all three coordinates. Moreover, we take the background gauge field to be of the form 
\begin{equation}
\label{eq:3Dpotential}
    {\cal A}=\mu\, \dd u+{\cal A}_z\dd z+{\cal A}_{\bar z}\dd \bar z~,
\end{equation}
where $\mu=\mu(u)$ and the spatial components are arbitrary functions of all coordinates for now. This shares several features with the two-dimensional background in \eqref{eq:2D}. The components $\alpha$ and $\nu$ will play the same role as in two dimensions: $\alpha$ acts as time dilation, while $\nu$ is a Weyl factor for $M_2$. However, the angular velocity in the earlier sections is now traded for an electric chemical potential, $\mu$, which enters as a component of the background gauge field. The geometry of $M_2$, i.e., $2e^{2\chi} \dd z \dd\bar{z}$, will be held fixed.

On this background, we decompose the stress tensor and current such that their components are tied directly to the appropriate sources we have introduced: $(\alpha, \nu, \mu)$. Thus, we have
\begin{equation}
\label{eq:3DStresstensor}
\begin{aligned}
    T_{uu}&=-\frac{\alpha^2}{\nu^2}\mathcal{O}_{u}^{\scaleto{\rm 3D}{4pt}}~, \quad T_{z\bar{z}}=\frac{1}{2}\mathcal{O}_{z\bar z}^{\scaleto{\rm 3D}{4pt}} e^{2\chi}~,\\
     T_{uz}&=\frac{\alpha}{\nu^2}t_{uz}~, \quad  T_{zz}=\frac{1}{\alpha}t_{zz}~,
\end{aligned}
\end{equation}
plus the corresponding barred components. In particular, the trace of the stress tensor is given by
\begin{equation}\label{eq:3D2Dtrace}
    T^a{}_a=\frac{1}{\nu^2}({\cal O}^{\scaleto{\rm 3D}{4pt}}_u+{\cal O}^{\scaleto{\rm 3D}{4pt}}_{z\bar z})~.
\end{equation}
Here, $\mathcal{O}_{u}^{\scaleto{\rm 3D}{4pt}}=\mathcal{O}_{u}^{\scaleto{\rm 3D}{4pt}}(u,z,\bar{z})$, and $\mathcal{O}^{\scaleto{\rm 3D}{4pt}}_{z\bar z}=\mathcal{O}^{\scaleto{\rm 3D}{4pt}}_{z \bar z}(u,z,\bar{z})$. The components $t_{uz}$ and $t_{zz}$ (and their barred versions) also depend on the local coordinates, but they will not be treated as dynamical operators. As will be the case in the subsequent AdS$_4$ construction, we consider a sector of the CFT$_3$ where these components are locally determined by the curvature of the background \eqref{eq:3Dmetric}.   

The $U(1)$ current is decomposed as 
\begin{equation}
\label{eq:u(1)current}
    J=-\frac{\alpha}{\nu}\mathcal{O}^{\scaleto{\rm 3D}{4pt}}_{e}\dd u+\frac{j_z}{\alpha}\dd z+\frac{j_{\bar{z}}}{\alpha}\dd\bar{z}~,
\end{equation}
where  $\mathcal{O}_{e}^{\scaleto{\rm 3D}{4pt}}=\mathcal{O}_{e}^{\scaleto{\rm 3D}{4pt}}(u,z,\bar{z})$ controls the electric charge carried by the current, and the spatial components $j_z$ also depend on  $(u,z,\bar{z})$. In this notation, turning on a magnetic charge corresponds to setting
\begin{equation}\label{eq:magnetic}
    {\cal F}_{z\bar z}=\nu e^{2\chi} \mathcal{O}^{\scaleto{\rm 3D}{4pt}}_{m}~.
\end{equation}

In this way, the variation \eqref{eq:3DvarI} of the generating function is given by paring the sources with their respective operators\footnote{The components $ T_{uz}$ and $ T_{zz}$ are inert in our discussion. In particular, we are not turning on sources in the background metric for those components. This is motivated by the setup we will have in AdS$_4$. However, appropriate modifications to turn on those sources are straightforward.} 
\begin{equation}
\label{eq:3DvariationI}
    \delta I=\int \dd^3x\left(e^{2\chi}\left(\mathcal{O}_{z\bar z}^{\scaleto{\rm 3D}{4pt}}\frac{\alpha}{\nu}\delta\nu+\mathcal{O}_{u}^{\scaleto{\rm 3D}{4pt}}\delta\alpha+\mathcal{O}_{e}^{\scaleto{\rm 3D}{4pt}}\nu\delta \mu\right)+j_z\delta {\cal A}_{\bar{z}}+j_{\bar{z}}\delta {\cal A}_z\right)~.
\end{equation}
Finally, with this notation, the Ward identities \eqref{eq:3DWard} take the form
\begin{equation}\label{eq:3dWardequations}
    \begin{aligned}
\partial_u(e^{2\chi}\mathcal{O}_{u}^{\scaleto{\rm 3D}{4pt}})-\frac{1}{2\nu^2}\partial_u(\nu^2 e^{2\chi})\mathcal{O}_{z\bar z}^{\scaleto{\rm 3D}{4pt}}&=\frac{1}{\alpha}( {\cal F}_{uz}j_{\bar z} +{\cal F}_{u\bar z}j_z)-\frac{\alpha}{\nu^2} (\partial_{\bar z} t_{uz} + \partial_zt_{u\bar z})~,\\
    \partial_z \mathcal{O}_{z\bar z}^{\scaleto{\rm 3D}{4pt}}+2\frac{\nu}{\alpha} {\cal F}_{uz}\mathcal{O}^{\scaleto{\rm 3D}{4pt}}_{e} &=\frac{2}{\alpha}e^{-2\chi}\left({\cal F}_{z\bar z}j_z-\partial_{\bar z}t_{zz}+\partial_u(e^{2\chi}t_{uz})\right) ~,\\
     \partial_{\bar z} \mathcal{O}_{z\bar z}^{\scaleto{\rm 3D}{4pt}}+2\frac{\nu}{\alpha} {\cal F}_{u\bar z}\mathcal{O}^{\scaleto{\rm 3D}{4pt}}_{e} &=\frac{2}{\alpha}e^{-2\chi}\left({\cal F}_{\bar z z}j_{\bar z}-\partial_{z}t_{\bar z\bar z}+\partial_u(e^{2\chi}t_{u\bar z})\right) ~,
    \end{aligned}
\end{equation}
which correspond to the conservation equation of the stress tensor, while the $U(1)$ conservation and trace conditions give respectively
\begin{equation}\label{eq:3dWardequations-1}
    \begin{aligned}
    \partial_u(\nu e^{2\chi}\mathcal{O}^{\scaleto{\rm 3D}{4pt}}_{e})&= -\partial_z j_{\bar z}-\partial_{\bar z}j_z~,
    \end{aligned}
\end{equation}
and
\begin{equation}\label{eq:3dWardequations-2}
    \begin{aligned}
      \mathcal{O}_{u}^{\scaleto{\rm 3D}{4pt}}+\mathcal{O}_{z\bar z}^{\scaleto{\rm 3D}{4pt}}&=0~.  
    \end{aligned}
\end{equation}

Given a conformal Killing vector of the background \eqref{eq:3Dmetric}-\eqref{eq:3Dpotential}, the associated conserved charge \eqref{eq:CKCharge3D} takes the form
\begin{equation}\label{eq:CKCharge3D0}
    {\cal Q}[\xi]=-\int_{M_2}2 e^{2\chi}\dd z\dd\bar{z}\, \big(\xi^u\alpha\,\mathcal{O}_{u}^{\scaleto{\rm 3D}{4pt}}-\xi^{z}t_{uz}-\xi^{\bar z}t_{u\bar z}+(\xi^u\mu+\xi^z{\cal A}_z+\xi^{\bar z}{\cal A}_{\bar z}) \nu\, \mathcal{O}^{\scaleto{\rm 3D}{4pt}}_{e}\big)~.
\end{equation}
In particular, if $\chi(u,z,\bar z)$, ${\cal A}_z(u,z,\bar z)$ and ${\cal A}_{\bar z}(u,z,\bar z)$ are independent of $u$ and we work in a gauge where $\nu\mu/\alpha$ is constant, the background \eqref{eq:3Dmetric}-\eqref{eq:3Dpotential} admits the timelike conformal Killing vector $\xi=(\nu/\alpha)\partial_u$, which leads to the conserved energy \begin{equation}\label{eq:Energy3D}
    E=-\int_{M_2}2 e^{2\chi}\dd z\dd\bar{z}\, \nu\left(\mathcal{O}_{u}^{\scaleto{\rm 3D}{4pt}}+\frac{\nu\mu}{\alpha}\,\mathcal{O}^{\scaleto{\rm 3D}{4pt}}_{e}\right)~.
\end{equation} 
Moreover, the electric and magnetic charges \eqref{eq:EMCharges3D0} become
\begin{equation}\label{eq:EMCharges3D0}
    {\cal Q}_e=\int_{M_2} 2 e^{2\chi}\dd z\dd\bar{z}\, \nu\, \mathcal{O}^{\scaleto{\rm 3D}{4pt}}_{e}~,\quad {\cal Q}_m=\int_{M_2} 2 e^{2\chi}\dd z\dd\bar{z}\,\nu\, \mathcal{O}^{\scaleto{\rm 3D}{4pt}}_{m}~.
\end{equation}
In particular, the second term in the energy \eqref{eq:Energy3D} is proportional to the electric charge ${\cal Q}_e$. It follows that for backgrounds of this form we can define the simplified conserved energy
\begin{equation}\label{eq:EnergyPrime3D}
    E'=-\int_{M_2}2 e^{2\chi}\dd z\dd\bar{z}\,\nu\,\mathcal{O}_{u}^{\scaleto{\rm 3D}{4pt}}~.
\end{equation}

\subsection{Generalised Robinson-Trautman phase space}\label{sec:4.5.2}

In this section we construct a reduced phase space in AdS$_4$ that has the operators $\mathcal{O}_{u}^{\scaleto{\rm 3D}{4pt}}$, $\mathcal{O}_{z\bar z}^{\scaleto{\rm 3D}{4pt}}$, $\mathcal{O}_{e}^{\scaleto{\rm 3D}{4pt}}$ turned on, and carries also the sources $(\alpha, \nu,\mu)$. The simplest gravitational theory that contains these ingredients is a four-dimensional Einstein-Maxwell theory with a negative cosmological constant,
\begin{equation}\label{eq:MEAction}
    S=\frac{1}{2\kappa^2_4}\int \dd^4x \sqrt{-g}\left(R-2\Lambda\right)-\frac{1}{4}\int \dd^4x\sqrt{-g}\,F^2~,
\end{equation}
where $\Lambda=-3/\ell_4^2$ and $\kappa_4^2=8\pi G_4$.
A useful way to construct the desired phase space is to consider a generalisation of the charged Robinson-Trautman (RT) solutions of this theory. The RT family of algebraically special vacuum Einstein solutions describes spacetimes with a null geodesic congruence that has vanishing shear and rotation but nonzero expansion \cite{RT:1962} (see \cite{Stephani:2003tm} for a comprehensive review). These generally time-dependent geometries exactly solve the radial part of the Einstein equations, reducing them to a single transverse equation, known as the RT equation. This equation can be interpreted holographically as a Ward identity in the dual CFT$_3$ \cite{Skenderis:2014RT}.

Einstein-Maxwell theory admits electrically and magnetically charged versions of such algebraically special solutions. Here, we generalise the charged RT spacetimes with a negative cosmological constant in two ways. Firstly, we turn on suitable sources such that the three-dimensional metric (\ref{eq:3Dmetric}) matches the boundary metric of the bulk spacetime. This ensures that the transverse equations of motion of the four-dimensional Einstein-Maxwell theory will be identified with the CFT$_3$ Ward identities (\ref{eq:3dWardequations}). However, since the conformal anomaly is identically zero in this case, the holographic stress tensor obtained from the charged RT class of solutions has a vanishing trace. Therefore, in order to obtain the trace in terms of the operators $\mathcal{O}_{u}^{\scaleto{\rm 3D}{4pt}}$ and $\mathcal{O}^{\scaleto{\rm 3D}{4pt}}_{z\bar z}$ as in \eqref{eq:3D2Dtrace}, we need to further generalise the RT ansatz so that the trace of the holographic stress tensor is given by \eqref{eq:3D2Dtrace}. This is then set to zero once the equations of motion are imposed.

We will work in a Bondi-like gauge with a bulk metric of the form 
\begin{equation}\label{eq:3dbulkmetric}
    ds^2_{\scaleto{\rm 4D}{4pt}}=-\Omega^{2}f\dd u^2+\Omega^{-2}\left(-2\alpha(u) \dd u\dd r+2\Phi^2(r,u)e^{2\chi}\dd z\dd\bar{z}\right)~,
\end{equation}
where $f$  and $\Omega$ are functions of $r,u,z$, and $\bar z$. Similar to the AdS$_3$ cases in the previous sections, the scalar $\Phi$ controls the size of the base manifold $M_2$ with metric $2e^{2\chi}\dd z\dd\bar{z}$ and is parametrised as
\begin{equation}\label{eq:dilaton-4d}
    \Phi(r,u)=\nu(u) \frac{r}{\ell_4} +\Phi_0~,
\end{equation}
where $\Phi_0$ is constant. The fieldstrength components of the dyonic class of solutions are given by
\begin{equation}
\label{eq:Fbulk}
\begin{aligned}
   F_{ur}&=-\frac{\alpha\nu(\mathcal{O}^{\scaleto{\rm 3D}{4pt}}_{q}+\mathcal{O}^{\scaleto{\rm 3D}{4pt}}_{\bar{q}})}{\Phi^2}~, \quad F_{z\bar{z}}=\nu e^{2\chi}(\mathcal{O}^{\scaleto{\rm 3D}{4pt}}_{q}-\mathcal{O}^{\scaleto{\rm 3D}{4pt}}_{\bar{q}})~,\\
    F_{uz}&=j_z+\frac{\ell_4\alpha}{\Phi}\partial_z\mathcal{O}^{\scaleto{\rm 3D}{4pt}}_{q}~, \quad F_{u\bar{z}}=j_{\bar{z}}+\frac{\ell_4\alpha}{\Phi}\partial_{\bar{z}}\mathcal{O}^{\scaleto{\rm 3D}{4pt}}_{\bar{q}}~.
\end{aligned}
\end{equation}
with  $\mathcal{O}^{\scaleto{\rm 3D}{4pt}}_{q}=\mathcal{O}^{\scaleto{\rm 3D}{4pt}}_{q}(u,z)$, $\mathcal{O}^{\scaleto{\rm 3D}{4pt}}_{\bar q}=\mathcal{O}^{\scaleto{\rm 3D}{4pt}}_{\bar q}(u,\bar z)$, while $j_{z}$ and $j_{\bar{z}}$ are functions of $u,z,\bar z$. The combinations 
\begin{equation}\label{eq:map-charges}
    \mathcal{O}^{\scaleto{\rm 3D}{4pt}}_{e} = \mathcal{O}^{\scaleto{\rm 3D}{4pt}}_{q}+\mathcal{O}^{\scaleto{\rm 3D}{4pt}}_{\bar q} ~,\qquad \mathcal{O}^{\scaleto{\rm 3D}{4pt}}_{m} = \mathcal{O}^{\scaleto{\rm 3D}{4pt}}_{q}-\mathcal{O}^{\scaleto{\rm 3D}{4pt}}_{\bar q}~,
\end{equation}
can be identified respectively with electric and magnetic currents in the dual CFT$_3$. In particular, the boundary background fieldstrength is 
\begin{equation}\label{eq:Fbulk-boundary}
{\cal F}_{z\bar{z}}=\nu e^{2\chi}(\mathcal{O}^{\scaleto{\rm 3D}{4pt}}_{q}-\mathcal{O}^{\scaleto{\rm 3D}{4pt}}_{\bar{q}})~,\quad
    {\cal F}_{uz}=j_z~, \quad {\cal F}_{u\bar{z}}=j_{\bar{z}}~.
\end{equation}
In the following, we will show how to place these variables on-shell, and how the transverse bulk equations of motion translate to Ward identities in the dual CFT$_3$.

Let us first consider the dynamics of the gauge field, which it is the simplest. Maxwell's equations, together with the Bianchi identity, imply that the variables that parametrise the fieldstrength \eqref{eq:Fbulk} satisfy the transverse constraints 
\begin{equation}
    \begin{aligned}\label{eq:BM-4}
       \partial_u \left(\nu e^{2\chi}\mathcal{O}^{\scaleto{\rm 3D}{4pt}}_{q}\right) &=- \partial_{\bar z} j_z~, \\
             \partial_u \left(\nu e^{2\chi}\mathcal{O}^{\scaleto{\rm 3D}{4pt}}_{\bar q}\right) &=- \partial_{z} j_{\bar z}~. \\
    \end{aligned}
\end{equation}
We recognize the sum of these constraints as the conservation of the electric current in the dual CFT$_3$, given in \eqref{eq:3dWardequations-1}. The difference of the two equations in \eqref{eq:BM-4} that is imposed by the Bianchi identity corresponds to the conservation of the topological current in the dual CFT$_3$. Provided these constraints are satisfied, the fieldstrength expressions \eqref{eq:Fbulk} can be integrated to determine the corresponding gauge field, namely
\begin{equation}\label{eq:vectorRT}
    A(r,u,z,\bar z)=\left(-\frac{(\mathcal{O}^{\scaleto{\rm 3D}{4pt}}_{q}+\mathcal{O}^{\scaleto{\rm 3D}{4pt}}_{\bar{q}})\ell_4\alpha}{\Phi}+\mu\right)\dd u- \left(\int \dd \bar{z}\,\nu e^{2\chi}{\cal O}_{q}^{\scaleto{\rm 3D}{4pt}}\right)\dd z- \left(\int \dd z\, \nu e^{2\chi}{\cal O}_{\bar q}^{\scaleto{\rm 3D}{4pt}}\right)\dd \bar z~,
\end{equation}
where the function $\mu=\mu(u)$ plays the role of a source for the operator $\mathcal{O}^{\scaleto{\rm 3D}{4pt}}_{e}$ appearing in \eqref{eq:3Dpotential}. 

Imposing the Einstein equations is more involved. Based on the form of the dyonic RT solutions, an appropriate ansatz for the  metric function $f(r,u,z,\bar z)$ and $\Omega(r,u,z,\bar z)$ in \eqref{eq:3dbulkmetric} is
\begin{equation}\label{eq:FRT}
\begin{aligned}
    f(r,u,z,\bar z)&=\frac{\alpha^2}{\nu^2}\Phi^2+2\ell_4\frac{\alpha}{\nu}\left(\partial_u\Phi +\Phi\partial_u\chi\right)-\frac{\ell^2_4\alpha^2}{\nu^2}\Delta\chi+\frac{f_{-1}(u,z,\bar{z})}{\Phi}+\frac{f_{-2}(u,z,\bar{z})}{\Phi^2}~,\\
    \Omega(r,u,z,\bar z)&=1+\frac{f_{-3}(u,z,\bar{z})}{\Phi^3}~,
\end{aligned}
\end{equation}
where $\Delta = 2 e^{-2\chi}\partial_z\partial_{\bar z}$ is the Laplace-Beltrami operator on the base two-manifold. With hindsight, we parametrise the functions $f_i(u,z,\bar{z})$ in terms of the dual CFT$_3$ variables in (\ref{eq:3DvariationI}) as
\begin{equation}
\begin{aligned}
     f_{-1}(u,z,\bar{z})&=\frac{1}{2}\ell_4\kappa^2_4\frac{\alpha^2}{\nu} (\mathcal{O}^{\scaleto{\rm 3D}{4pt}}_{u}-\mathcal{O}^{\scaleto{\rm 3D}{4pt}}_{z\bar z})~,\\
        f_{-2}(u,z,\bar{z})&=2\ell_4^2\kappa^2_4\alpha^2\mathcal{O}^{\scaleto{\rm 3D}{4pt}}_{q}\mathcal{O}^{\scaleto{\rm 3D}{4pt}}_{\bar{q}}~,\\
        f_{-3}(u,z,\bar{z})&=\frac{1}{24}\ell_4\kappa_4^2\nu(\mathcal{O}^{\scaleto{\rm 3D}{4pt}}_{u}+\mathcal{O}^{\scaleto{\rm 3D}{4pt}}_{z\bar z})~.
    \end{aligned}
\end{equation}

Inserting this ansatz in the Einstein equations following from \eqref{eq:MEAction} we find that they are satisfied provided the following transverse constraints hold, in addition to \eqref{eq:BM-4}:
\begin{equation}\label{eq:EoM3D}
    \begin{aligned}
e^{-3\chi}\partial_u\left(e^{3\chi}\mathcal{O}^{\scaleto{\rm 3D}{4pt}}_{u}\right)+\mathcal{O}^{\scaleto{\rm 3D}{4pt}}_{u}\frac{\partial_u\nu}{\nu}&=\frac{\alpha}{\nu^2}\frac{\ell^2_4}{2\kappa^2_4}\Delta^2\chi+\frac{2}{\alpha} e^{-2\chi}j_zj_{\bar{z}}~,\\
     \partial_z\mathcal{O}^{\scaleto{\rm 3D}{4pt}}_{u}-4 \frac{\nu}{\alpha}\mathcal{O}^{\scaleto{\rm 3D}{4pt}}_{\bar{q}}j_z&=0~,\\
       \partial_{\bar z}\mathcal{O}^{\scaleto{\rm 3D}{4pt}}_{u}-4 \frac{\nu}{\alpha}\mathcal{O}^{\scaleto{\rm 3D}{4pt}}_{{q}}j_{\bar z}&=0~,\\
      \mathcal{O}^{\scaleto{\rm 3D}{4pt}}_{u}+\mathcal{O}^{\scaleto{\rm 3D}{4pt}}_{z\bar z}&=0~.
    \end{aligned}
\end{equation}
Notice that these equations also imply that 
\begin{equation}\label{eq:extra-condition}
    \partial_{\bar z}\left(\mathcal{O}^{\scaleto{\rm 3D}{4pt}}_{\bar{q}}j_z\right)= \partial_{ z}\left(\mathcal{O}^{\scaleto{\rm 3D}{4pt}}_{{q}}j_{\bar z}\right)~,
\end{equation}
which corresponds to an integrability condition for the second and third equations in \eqref{eq:EoM3D}.

The transverse equations \eqref{eq:BM-4}, \eqref{eq:EoM3D} and \eqref{eq:extra-condition} are a generalisation of the RT equations for the dyonic algebraically special solutions of Einstein-Maxwell theory with a negative cosmological constant. They are equivalent to the Ward identities \eqref{eq:3dWardequations}, \eqref{eq:3dWardequations-1} and \eqref{eq:3dWardequations-2} provided $t_{uz}$ and $t_{zz}$ defined in \eqref{eq:3DStresstensor}, take the special form 
\begin{equation}\label{eq:map-tuz}
    t_{uz}=-\frac{\ell_4^2}{2\kappa^2_4}\partial_z\Delta\chi~,\quad t_{zz}=\frac{\ell_4^2}{\kappa^2_4}\partial_u(e^{\chi}\partial_z^2 e^{-\chi})~,
\end{equation}
which corresponds to specific components of the Cotton-York tensor of the boundary metric \eqref{eq:3Dmetric} \cite{HR:2014RT,Skenderis:2014RT}. Indeed, evaluating the holographic electric U(1) current we find that it takes the form \eqref{eq:u(1)current}, with $ \mathcal{O}^{\scaleto{\rm 3D}{4pt}}_{e}$ given by the sum in \eqref{eq:map-charges}. Moreover, the computing the holographic stress tensor corresponding to these configurations using holographic renormalization we obtain
\begin{equation}\label{eq:holT3D}
    \begin{aligned}
        T_{uu}&=-\frac{\alpha^2}{\nu^2}{\cal O}^{\scaleto{\rm 3D}{4pt}}_u~,\quad
        T_{z\bar{z}}=\frac{1}{2}e^{2\chi}{\cal O}^{\scaleto{\rm 3D}{4pt}}_{z\bar z}~,\\
        T_{uz}&=\frac{\alpha}{\nu^2}t_{uz}~,\quad \
        T_{zz}=\frac{1}{\alpha}t_{zz}~,
    \end{aligned}
\end{equation}
which coincides with the general stress tensor decomposition \eqref{eq:3DStresstensor}, except that now $t_{uz}$ and $t_{zz}$ take the specific form \eqref{eq:map-tuz}. The fact that these components of the holographic stress tensor are determined locally in terms of the background geometry on the boundary is a general property of asymptotically locally AdS$_4$ algebraically special solutions \cite{HR:2014RT,Skenderis:2014RT}. 

The family of solutions we consider here represents a subset of the most general phase space of a CFT$_3$, as is clear from the special form of the stress tensor components \eqref{eq:map-tuz} and the vanishing of the corresponding sources. As a result, the stress tensor of the dual CFT$_3$ coincides with the stress tensor of a conformal fluid on a non-flat background \cite{HR:2014RT}. In particular, the boundary stress tensor dual to the RT spacetimes can be decomposed as 
\begin{equation}\label{eq:PF}
    T_{ab}=T_{ab}^{\scaleto{\rm PF}{4pt}}+Z_{ab}~,
\end{equation}
where $T_{ab}^{\scaleto{\rm PF}{4pt}}$ is the perfect fluid stress tensor
\begin{equation}
    T_{ab}^{\scaleto{\rm PF}{4pt}}=p(g_{ab}+U_aU_b)+\epsilon U_aU_b~,
\end{equation}
where $U=\alpha\, \dd u$ is a time-like vector field that satisfies $U_aU^a=-1$, while the energy density $\epsilon$ and pressure $p$ are respectively
\begin{equation}\label{eq:hydro}
    \epsilon=-\frac{\mathcal{O}^{\scaleto{\rm 3D}{4pt}}_{u}}{\nu^2}~,\qquad p=\frac{\mathcal{O}^{\scaleto{\rm 3D}{4pt}}_{z\bar z}}{2\nu^2}~.
\end{equation}
The conformal equation of state, $\epsilon=2p$, is satisfied, reflecting the traceless stress tensor of a CFT$_3$. However, the perfect fluid stress tensor \eqref{eq:PF} is not covariantly conserved on the background \eqref{eq:3Dmetric}. The conservation of $T_{ab}$ on this background is ensured by the presence of the symmetric tensor $Z_{ab}$, which is defined as
\begin{equation}
    Z_{ab}\equiv \frac{\ell^2_4}{\kappa^2_4} (\delta^d_{a}-U_aU^d)C_{dbc}U^c~.
\end{equation}
In this expression, $C_{abc}$ is the Cotton tensor associated with the boundary metric $g_{ab}$. This reproduces the components \eqref{eq:map-tuz}. Recall that in three dimensions the Cotton tensor plays the role of the Weyl tensor in higher dimensions. Namely, a three-dimensional manifold is conformally flat if and only if it possesses a vanishing Cotton tensor. The presence of a nonzero Cotton tensor has been argued to encode gravitational radiation in the CFT$_3$ \cite{Cotton:2024rad}, in addition to the electromagnetic radiation captured by the U(1) current $J$ given in (\ref{eq:u(1)current}).

\subsubsection{AdS Reissner-Nordstr\"om black hole}

The dyonic AdS$_4$ Reissner–Nordstr\"om black hole belongs to the phase space described by the generalised charged RT solution. In the Bondi-like gauge introduced above, the metric is given by 
\begin{equation}
    ds^2=-f(r)\dd u^2-2\dd u\dd r+\frac{4r^2}{(1+z\bar{z})^2}\dd z\dd\bar{z}~, 
\end{equation}
where the blackening factor is
\begin{equation}
    \quad f(r)=\frac{ r^2}{\ell^2_{4}}+1-\frac{2MG_4}{r}+\frac{(Q^2+P^2)G_4}{4\pi r^2}~,
\end{equation}
and the gauge field reads
\begin{equation}\label{eq:gauge-field-RN}
    A=\left(-\frac{1}{4\pi}\frac{Q}{r}+\mu_0\right)\dd u+\frac{P}{4\pi}\Big(\frac{\dd z}{z}-\frac{\dd\bar{z}}{\bar{z}}\Big)\frac{i}{(1+z\bar{z})}~,
\end{equation}
with $\mu_0$ a constant. The factor of $1/4\pi$ in front of the mass parameter, $M$, the electric charge, $Q$, and the magnetic charge, $P$, ensures that these coincide with the corresponding conserved charges \eqref{eq:EMCharges3D0} and \eqref{eq:EnergyPrime3D}. 

In relation to the above notation, the Reissner–Nordstr\"om solution corresponds to setting $\alpha=1$, $\nu =\ell_4$, $\mu=0$ and $\Phi_0=0$, which gives $\Phi=r$.\footnote{In going from the gauge potential \eqref{eq:vectorRT} to the expression \eqref{eq:gauge-field-RN} for the Reissner–Nordstr\"om solution we have dropped a pure gauge term proportional to the electric charge $Q$.} Moreover, the metric on the base 2-manifold $M_2$ is
\begin{equation}
   2 e^{2\chi}\dd z\dd\bar{z}= \frac{4}{(1+z\bar{z})^2}\dd z\dd\bar{z}~,  
\end{equation}
which describes a round 2-sphere. In this case, the boundary metric \eqref{eq:3Dmetric} is conformally flat, i.e. it possesses a vanishing Cotton tensor. As a result, the components $t_{uz}$, $t_{zz}$ of the stress tensor in \eqref{eq:map-tuz}, as well as their complex conjugates, vanish. Similarly, $j_z=j_{\bar z}=0$ for the Reissner–Nordstr\"om black hole solution. The remaining components of the stress tensor and U(1) current are determined by the one-point functions
\begin{equation}
    \begin{aligned}
        \mathcal{O}^{\scaleto{\rm 3D}{4pt}}_{u}=-\mathcal{O}^{\scaleto{\rm 3D}{4pt}}_{z\bar z} =-\frac{M}{4\pi}~,\\
        \mathcal{O}^{\scaleto{\rm 3D}{4pt}}_{q}= \frac{1}{8\pi \ell_4}(Q+iP)~, \\ 
        \mathcal{O}^{\scaleto{\rm 3D}{4pt}}_{\bar q}=\frac{1}{8\pi \ell_4}(Q-iP)~.
    \end{aligned}
\end{equation}  

In order to describe the thermodynamics in a simple manner, we rewrite the blackening factor in terms of its roots, i.e.,
\begin{equation}
    f(r)=\frac{1}{r^2 \ell^2_{4} }(r-r_{+})(r-r_{-})(r+r_1)(r+r_2)~,
\end{equation}
where $r_1$ and $r_2$ are defined through the following relations
\begin{equation}
    r_1+r_2=r_{+}+r_{-}~, \quad r_1 r_2=\ell^2_{4}+r^2_{+}+r^2_{-}+r_{-}r_{+}~.
\end{equation}
Here, $r_{+}$ is the outer horizon, and $r_{-}$ the inner horizon; $r_{1,2}$ are unphysical roots of $f(r)$. 
In terms of $r_{\pm}$, the mass $M$, and charges $Q$, $P$ are
\begin{equation}
\begin{aligned}
    MG_4&=\frac{(r_{+}+r_{-})}{2}\left(1+\frac{r^2_{+}+r^2_{-}}{ \ell^2_{4}}\right)~, \\ 
    (Q^2+P^2)\frac{G_4}{4\pi}&=r_{+}r_{-}\left(1+\frac{r^2_{+}+r^2_{-}+r_{-}r_{+}}{\ell^2_{4}}\right)~.
\end{aligned}
\end{equation}

The Hawking temperature $T$ and black hole entropy $S$ are given by
\begin{equation}
    \begin{aligned}
T&=\frac{(r_{+}-r_{-})}{4\pi}\frac{(\ell_4^2+3r^2_{+}+r^2_{-}+2r_{-}r_{+})}{\ell^2_4r^2_{+}}~,\\
    S&=\frac{\pi r_{+}^2}{G_4}~.
    \end{aligned}
\end{equation}
The electric potential $ \mu_e$ and magnetic potential $ \mu_m$ are determined by the gauge field \eqref{eq:gauge-field-RN}. In particular, demanding regularity at the horizon sets $\mu_0=Q/4\pi r_+$, and hence
 \begin{equation}
    \mu_e\equiv A_u\big|_{r=\infty}-A_u\big|_{r=r_+}
    =\frac{Q}{4\pi r_{+}}~.
\end{equation}
Moreover, regularity at the horizon of the magnetic dual potential determines
\begin{equation}
\mu_m=\frac{P}{4\pi r_{+}}~.    
\end{equation}

Using these expressions for the thermodynamic variables of the Reissner–Nordstr\"om black hole, the first law takes the form
\begin{equation}
\dd M = T \dd S + \mu_e \dd Q + \mu_m \dd P~,    
\end{equation}
while the corresponding Smarr relation is given by
\begin{equation}\label{eq:SmarrRelation}
    pV+M=\mu_eQ+\mu_m P+ST+{\frac{r_+}{2G_4}}~,
\end{equation}
where the volume of the 2-sphere on the boundary is $V=4\pi\nu^2=4\pi \ell_4^2$, since we have set $\nu=\ell_4$. Moreover, the pressure $p$ is given in \eqref{eq:hydro}. The Smarr formula \eqref{eq:SmarrRelation} is a consequence of the generalized Euler relation $p + \epsilon = s T + \mu_e q_e + \mu_m q_m + y \partial_y p$, where $y$ encodes the effect of spatial curvature in the thermal system \cite{RM:2024Smarr}.  

\subsubsection{Near-extremal thermodynamics} 

At extremality, $r_{+}=r_{-}=r_0$, the mass  and charges satisfy the constraint
\begin{equation}\label{eq:ExtremalityMQ}
   G_4M_0=r_{0}\left(1+\frac{2 r^2_0}{\ell^2_{4}}\right)~, \quad (\mu_{e,0}Q_0+\mu_{m,0}P_0)G_4=r_0\left(1+\frac{3r^2_{0}}{\ell^2_{4}}\right)~.
\end{equation}
At zero temperature, the Smarr relation \eqref{eq:SmarrRelation} relates the extremal mass $M_0$ with charges $Q_0$, $P_0$ as
\begin{equation}\label{eq:SmarrlawT=0}
    M_0-(\mu_{e,0}Q_0+\mu_{m,0}P_0)=-pV+\frac{r_0}{2G_4}=-\frac{r_0^3}{\ell^2_4G_4}~.
\end{equation}
This expression will play an important role in the subsequent discussion. Finally, the extremality conditions (\ref{eq:ExtremalityMQ}) written in terms of CFT$_3$ variables are
\begin{equation}\label{3D:extremalitycond1}
    \mathcal{O}^{\scaleto{\rm 3D}{4pt},*}_{u}=-\frac{2r_{0}}{\kappa^2_4}\left(1+\frac{2 r^2_0}{\ell^2_{4}}\right)~, \quad \mathcal{O}^{\scaleto{\rm 3D}{4pt},*}_{q}\mathcal{O}^{\scaleto{\rm 3D}{4pt},*}_{\bar{q}}=\frac{r_{0}}{2\kappa^2_4\ell_4^2}\left(1+\frac{3r^2_{0}}{\ell^2_{4}}\right)~.
\end{equation}

To study the low-temperature regime, we expand around the horizon radii near their extremal value as $r_{\pm}=r_0\pm\epsilon+{\cal O}(\epsilon^2)$, keeping $P_0$ and $Q_0$ fixed. This gives \cite{Almheiri:2016fws}
\begin{equation}
    T=\frac{\epsilon}{2\pi \ell_2^2}~, \quad \ell_2^2=\frac{\ell_4^2r_0^2}{6r_0^2+\ell^2_4}~,
\end{equation}
where $\ell_2$ is the effective AdS$_2$ radius. The linear responses in energy $\Delta M$ and entropy $\Delta S$ are\footnote{The expression for $\Delta M$ is obtained by keeping $P^2+Q^2$ fixed up to ${\cal O}(\epsilon^2)$, which requires the ${\cal O}(\epsilon^2)$ terms in the expansion of $r_\pm$ near $r_0$. Namely,  $r_\pm=r_0\pm\epsilon+ r_\pm^{(2)}\epsilon^2+{\cal O}(\epsilon^3)$, where $r_+^{(2)}+r_-^{(2)}=(\ell_4^2+2r_0^2)/(\ell_4^2+6r_0^2)r_0$.} 
\begin{equation}\label{eq:4d-energy}
    \Delta M=2\pi^2\ell^2_2\left(\frac{8\pi r_0}{\kappa_4^2}\right)T^2~, \quad \Delta S=4\pi^2\ell_2^2\left(\frac{8\pi r_0}{\kappa_4^2}\right)T~.
\end{equation}

\subsection{Off-shell decoupling limit}

We now proceed to implement the decoupling limit for the generalisation of the Robinson–Trautman phase space described in Sec.\,\ref{sec:4.5.2}. The procedure closely parallels that in Sec.\,\ref{sec:off-shell-decouple-3d} and Sec.\,\ref{sec:off-shell-decouple-TMG}.  

We define again the decoupling limit via the scaling
\begin{equation}\label{eq:decoupling-0c}
   \nu = \lambda \tilde{\nu} ~, \qquad \lambda\to 0~, 
\end{equation}
while keeping $\tilde{\nu}$ fixed. This is motivated by thinking of the radial direction as a holographic energy scale: taking $\nu$ small or large effectively changes the radial profile of $\Phi(u,r)$ in \eqref{eq:dilaton-4d}. 
In addition, we require that the electric and magnetic charges \eqref{eq:EMCharges3D0} remain finite in the limit $\lambda \to 0$. From   \eqref{eq:map-charges} then follows that the operators $\mathcal{O}^{\scaleto{\rm 3D}{4pt}}_{q}$ and $\mathcal{O}^{\scaleto{\rm 3D}{4pt}}_{\bar{q}}$ must scale as
\begin{equation}\label{eq:fixedQQ}
    \mathcal{O}^{\scaleto{\rm 3D}{4pt}}_q=\frac{ \mathcal{O}_q^{(-1)}}{\lambda}~, \quad \mathcal{O}^{\scaleto{\rm 3D}{4pt}}_{\bar{q}}=\frac{ \mathcal{O}_{\bar{q}}^{(-1)}}{\lambda}~, 
\end{equation}
with $\mathcal{O}_{q}^{(-1)}$ and $\mathcal{O}_{\bar{q}}^{(-1)}$ held fixed. 

As in the earlier cases, this scaling introduces divergences in the geometry. In order to account for these divergences and eliminate them we expand the remaining operators as
\begin{equation}
    \mathcal{O}^{\scaleto{\rm 3D}{4pt}}_{i}=\frac{ \mathcal{O}_{i}^{(-1)}}{\lambda}+ \mathcal{O}_{i}^{(0)}+ \mathcal{O}_{i}^{(1)}\lambda+ \mathcal{O}_{i}^{(2)}\lambda^2+\cdots\,,\quad \quad i\in (u,z\bar{z})~.
\end{equation}
In particular, the expansion of $\mathcal{O}^{\scaleto{\rm 3D}{4pt}}_{i}$ starts at order $1/\lambda$, as inferred from dimensional analysis of the line element (\ref{eq:3dbulkmetric})–(\ref{eq:FRT}). Imposing regularity of the lapse function $f(r,u,z,\bar{z})$ in equation~(\ref{eq:FRT}) as $\lambda \to 0$ yields the constraints
\begin{equation}
\label{eq:RTregularity}
\begin{aligned} \mathcal{O}_{q}^{(-1)}\mathcal{O}_{\bar{q}}^{(-1)}=\frac{3\Phi_0^4}{2\ell_4^2\kappa^2_4\tilde{\nu}^2}-\frac{\Phi_0^2}{2\kappa^2_4\tilde{\nu}^2}\Delta\chi~, \\
    \mathcal{O}_{u}^{(-1)}=-\mathcal{O}_{z\bar{z}}^{(-1)}=-\frac{4\Phi_0^3}{\ell_4\kappa^2_4\tilde{\nu}}+\frac{2
    \ell_4\Phi_0}{\kappa^2_4\tilde{\nu}}\Delta\chi~,\\
    \mathcal{O}_{u}^{(0)}=-\mathcal{O}_{z\bar{z}}^{(0)}=-\frac{2\Phi^2_0}{\kappa^2_4\alpha}\partial_u\chi~.
\end{aligned}
\end{equation}
Notice that since $\mathcal{O}^{\scaleto{\rm 3D}{4pt}}_q$ is a function of $u$, $z$ only, while $\mathcal{O}^{\scaleto{\rm 3D}{4pt}}_{\bar q}$ is a function of $u$, $\bar z$ only, the first equation in \eqref{eq:RTregularity} imposes an implicit constraint on $\Delta\chi$. We also observe that setting $\Phi_0=r_0$, $\tilde\nu=\ell_4$ and $\Delta\chi=-1$ in the order $(-1)$ coefficients reproduces the extremal values in \eqref{3D:extremalitycond1}.

Imposing the asymptotic conditions \eqref{eq:decoupling-0c}–\eqref{eq:RTregularity} in the limit $\lambda \to 0$, while keeping $\tilde{\nu}$, $\tilde{\mu}$, $\alpha$, $\Phi_0$, and $\chi$ fixed, the line element \eqref{eq:3dbulkmetric} remains smooth and well-defined. Namely, we find the following IR expression for the lapse function:
\begin{equation}\label{eq:lapseIR4d}
   \begin{aligned}
   f_{\rm IR}(r,u,z,\bar{z})&:= \lim_{\lambda \to 0} f(r,u,z,\bar{z}) \\
   &=  \frac{\alpha^2 r^2}{\left(\frac{\ell_4^2\Phi_0^2}{6\Phi_0^2-\ell^2_4\Delta\chi}\right)}+2r\left(2\alpha\partial_u\chi+\frac{\alpha\partial_u\tilde{\nu}}{\tilde{\nu}}\right)+\frac{\ell_4\kappa^2_4\alpha^2}{2\Phi_0\tilde{\nu}}(\mathcal{O}^{(1)}_{u}-\mathcal{O}^{(1)}_{z\bar{z}})~.
   \end{aligned}
\end{equation}
In contrast to the lower-dimensional examples discussed previously, the resulting geometry exhibits a more intricate dependence on the transverse coordinates $z$ and $\bar{z}$. We emphasise once again that this decoupling limit is entirely off-shell in directions transverse to the radial coordinate, as we have not imposed the equations of motion~(\ref{eq:EoM3D}).

Applying the decoupling limit to the equations of motion (\ref{eq:EoM3D}) simplifies the system further. The lowest order conditions from \eqref{eq:EoM3D} are 
\begin{equation}
\begin{aligned}
    \Delta^2\chi&=0~, \\
    \partial_u \big(e^{3\chi}\tilde\nu\mathcal{O}_{u}^{(-1)}\big)&=0~,\\
     \partial_z \mathcal{O}^{(-1)}_{u}= \partial_{\bar z} \mathcal{O}^{(-1)}_{u}&=0~. 
\end{aligned}
\end{equation}
Requiring consistency between these equations and the regularity conditions \eqref{eq:RTregularity} imposes 
\begin{equation}\label{eq:constant-chi}
    \partial_u \chi = 0 ~, \qquad  \Delta\chi= R_0~,
\end{equation}
where $R_0$ is a constant. This  also implies that $\mathcal{O}_{u}^{(0)}=\mathcal{O}_{z\bar z}^{(0)}=0$.  Notice that $R_0$ determines the topology of the horizon: $R_0<0$ corresponds to $S^2$, $R_0=0$ to $\mathbb{R}^2$, and $R_0>0$ to $\mathbb{H}_2$. We will focus on the case of the round $S^2$ with $R_0=-1$, which corresponds to the base 2-manifold of the Reissner–Nordstr\"om black hole. In this case,
\begin{equation}
       \chi=\log\left(\frac{\sqrt{2}}{1+z\bar{z}}\right)~.
\end{equation}

At the next order, implementing the decoupling limit on \eqref{eq:EoM3D} imposes
\begin{equation}
\begin{aligned}
    e^{-3\chi}\partial_u ( e^{3\chi}\mathcal{O}_{u}^{(0)})-\mathcal{O}_{z\bar{z}}^{(0)}\frac{\partial_u\tilde{\nu}}{\tilde{\nu}}=e^{-2\chi}\frac{2}{\alpha} j_zj_{\bar{z}}~,\\
\partial_z\mathcal{O}^{(0)}_{u}=4\mathcal{O}^{(-1)}_{\bar{q}}j_z\frac{\tilde{\nu}}{\alpha}~,\\
\partial_{\bar z}\mathcal{O}^{(0)}_{u}=4\mathcal{O}^{(-1)}_{q}j_{\bar z}\frac{\tilde{\nu}}{\alpha}~.
\end{aligned}
\end{equation}
The regularity conditions \eqref{eq:RTregularity}, combined with \eqref{eq:constant-chi}, then further imply that the sources must vanish, i.e. 
\begin{equation}
\begin{aligned}
j_z=j_{\bar{z}}=0~.
\end{aligned}
\end{equation}
With these simplifications, the regularity conditions (\ref{eq:RTregularity}) reduce to
\begin{equation}\label{3D:ExtremalRN}
\begin{aligned}
\mathcal{O}_{q}^{(-1)}\mathcal{O}_{\bar{q}}^{(-1)}=\frac{\Phi_0^2}{2\kappa^2_4\tilde{\nu}^2}\Big(1+\frac{3\Phi_0^2}{\ell_4^2}\Big)~, \\
     \mathcal{O}_{u}^{(-1)}=-\frac{2\Phi_0}{\kappa^2_4}\frac{\ell_4}{\tilde{\nu}}\Big(1+\frac{2\Phi_0^2}{\ell_4^2}\Big)~,
\end{aligned} 
\end{equation}
which, as we already pointed out, match the extremality conditions obtained from thermodynamics in (\ref{3D:extremalitycond1}) upon identifying $\tilde{\nu} = \ell_4$ and $\Phi_0 = r_0$. Finally, the Ward identities \eqref{eq:EoM3D} also imply that
\begin{equation}\label{3D:WrongWard}
    \begin{aligned}
        \partial_u \mathcal{O}_{u}^{(1)}-\mathcal{O}_{z\bar{z}}^{(1)}\frac{\partial_u\tilde{\nu}}{\tilde{\nu}}=0~,\\ 
    \mathcal{O}_{z\bar{z}}^{(1)}+\mathcal{O}_{u}^{(1)}=0~.
    \end{aligned}
\end{equation}

It is interesting to observe that the decoupling limit of the charged Robinson–Trautman solution drives the geometry towards the fixed point of the Calabi flow (see \cite{Skenderis:2014RT} and references therein), which is related to a conformal fluid in equilibrium \cite{HR:2014RT,Skenderis:2014RT,Petropoulos:2015RT,Bakas:2016RT,Skenderis:2017RT,Petropoulos:2017RT}. This is reminiscent of the mechanism discussed in \cite{Bobev:2020jlb} in a supersymmetric context.  

Besides regularity of the line element and of the Ward identities, we must also demand that the gauge potential $A(r,u,z,\bar{z})$ remains finite in the decoupling limit. From equation~(\ref{eq:vectorRT}) follows that, in addition to the line element regularity conditions \eqref{eq:RTregularity}, to ensure the regularity of the gauge potential we need to implement the shift 
\begin{equation}\label{eq:decoupling-3c}
    \mu(u)=\tilde{\mu}(u)+\frac{1}{\lambda }\frac{\ell_4\alpha}{\Phi_0}(\mathcal{O}_{q}^{(-1)}+\mathcal{O}_{\bar{q}}^{(-1)})~,
\end{equation}
which results in the following expression for the gauge field
\begin{equation}\label{eq:AIR-RT}
\begin{aligned}
    A_{\text{IR}}&:= \lim_{\lambda \to 0} A(r,u,z,\bar{z}) \\
&=\left(\frac{\alpha\tilde{\nu}}{\Phi_0^2}\left(\mathcal{O}_{q}^{(-1)}+\mathcal{O}_{\bar{q}}^{(-1)}\right)  r +\tilde{\mu}\right)\dd u+\frac{2\tilde{\nu}}{(1+z\bar{z})}\left(\mathcal{O}_{q}^{(-1)}\frac{\dd z}{z}+\mathcal{O}_{\bar{q}}^{(-1)}\frac{\dd\bar{z}}{\bar{z}}\right)~.
\end{aligned}
\end{equation}

To simplify the discussion, in the following we turn off the magnetic charge of the solution by setting $\mathcal{O}_{q}^{(-1)}=\mathcal{O}_{\bar{q}}^{(-1)}$. The electric charge, therefore, reduces to $\mathcal{O}_{e}^{(-1)}=2\mathcal{O}_{q}^{(-1)}$. Moreover, the spatial components of the gauge potential \eqref{eq:AIR-RT} become pure gauge, which we drop. With all of the above ingredients taken into account,  the lapse function $f_{\text{IR}}$ and vector potential $A_{\text{IR}}$ simplify to:
\begin{equation}\label{eq:3DIRf}
\begin{aligned}
    f_{\text{IR}}&=\frac{\alpha^2 r^2}{\left(\frac{\ell_4^2\Phi_0^2}{6\Phi_0^2+\ell^2_4}\right)}+2r\frac{\alpha\partial_u\tilde{\nu}}{\tilde{\nu}}+\frac{\ell_4\kappa^2_4\alpha^2}{2\Phi_0\tilde{\nu}}(\mathcal{O}^{(1)}_{u}-\mathcal{O}^{(1)}_{z\bar{z}})~,\\
A_{\text{IR}}&=\left(\frac{\mathcal{O}_{e}^{(-1)}\alpha\tilde{\nu} r}{\Phi_0^2}+\tilde{\mu}\right)\dd u~.
\end{aligned}
\end{equation}

Finally, applying the decoupling limit directly to the generating functional \eqref{eq:3DvariationI} yields:
\begin{equation}
    \delta I_{\text{IR}}=\int d^3x\,e^{2\chi}\Big(\lambda\left[\mathcal{O}_{z\bar{z}}^{(1)}\frac{\alpha}{\tilde{\nu}}\delta\tilde{\nu}+\mathcal{O}_{u}^{(1)}\delta\alpha+\mathcal{O}_{e}^{(-1)}\tilde{\nu}\delta \tilde{\mu}\right]+\frac{1}{\lambda}\delta\left(\frac{2\Phi_0^3}{\ell_4\kappa^2_4}\frac{\alpha}{\tilde{\nu}}\right)\Big)+O(\lambda^2)~,
\end{equation}
where we have used the regularity conditions \eqref{3D:ExtremalRN}. At first glance, the $\lambda \to 0$ limit appears singular due to the last term, which arises from the Smarr relation at zero temperature~\eqref{eq:SmarrlawT=0}. However, this term is a total variation and, therefore, can be removed by adding a local counterterm.

To proceed, as in the earlier sections, we perform a Legendre transformation that replaces the source $\tilde{\mu}$ with its conjugate variable $\mathcal{O}_{e}^{(-1)} \tilde{\nu}$, which is a conserved charge that is kept fixed in the decoupling limit. As a result, the term  $\tilde{\mu}\delta(\mathcal{O}_{e}^{(-1)} \tilde{\nu})$ drops out. Hence, the generating functional simplifies to
\begin{equation}\label{eq:IRgenerating4d}
    \delta I_{\text{IR}}=4\pi\lambda\int du\left(\mathcal{O}_{z\bar{z}}^{(1)}\frac{\alpha}{\tilde{\nu}}\delta\tilde{\nu}+\mathcal{O}_{u}^{(1)}\delta\alpha\right)~, 
\end{equation}
where we have integrated over the transverse space, yielding the factor of $4\pi$. 

\subsubsection{Comparison with JT gravity} 

We can see now that many aspects of our decoupling limit reproduce the features of JT gravity described in Sec.\,\ref{Sec:JTgravity}. In particular, the blackening factor \eqref{eq:3DIRf} resembles \eqref{eq:fJT}, and the equations of motion \eqref{3D:WrongWard} are similar to \eqref{eq:traceJT}–\eqref{eq:WardIdJT}, though the system appears to be free of anomalies, as would be expected from its parent CFT$_3$ theory. We will now explain how the anomaly of Schwarzian dynamics shows up, taking into account the ambiguities that we encountered in Sec.\,\ref{Sec:JTgravity}.

Our first step is to compare the leading term in $r$ that dominates $f_{\text{IR}}$ in \eqref{eq:lapseIR4d} with that in \eqref{eq:fJT}. From this comparison we naturally identify \begin{equation}\label{eq:relation2d4da}
    \ell_2=\ell_4\sqrt{\frac{\Phi_0^2}{6\Phi_0^2+\ell^2_4}}~,\quad \alpha_{\scaleto{\rm 2D}{4pt}}=\alpha_{\scaleto{\rm 4D}{4pt}}~,
\end{equation}
where we inserted labels on the variables to emphasize their relation and context. In this identification, the radial coordinate is taken to be the same in 2D and 4D. In this way, the JT dilaton is related to the 4D metric function $\Phi$ as
\begin{equation}\label{eq:relation2d4db}
    \Phi-\Phi_0=\lambda\frac{\tilde{\nu}}{\ell_4}r\equiv L\phi~,
\end{equation}
where we introduced a length scale $L$ that plays the role of a Kaluza-Klein radius. Its specific value is not relevant for the current discussion. With this identification at hand, and in comparison to \eqref{eq:phiJT}, we have that
\begin{equation}
    \phi=\nu_{\scaleto{\rm 2D}{4pt}}\frac{r}{\ell_2}~.
\end{equation}
Here, with respect to \eqref{eq:phiJT}, the 2D radial gauge is such that $\phi_0=0$. In addition, from equations \eqref{eq:relation2d4da} and \eqref{eq:relation2d4db}, we deduce that
\begin{equation}
    \nu_{\scaleto{\rm 2D}{4pt}}=\frac{\lambda}{L}\frac{\ell_2}{\ell_4}\tilde{\nu}.
\end{equation}
With this identification, the lapse function $f_{\rm IR}$ from equation (\ref{eq:3DIRf}) can be rewritten as:
\begin{equation}
    f_{\text{IR}}(u,r)=\frac{\alpha^2_{\scaleto{\rm 2D}{4pt}} r^2}{\ell_2^2}+\Big(2\ell_2\alpha_{\scaleto{\rm 2D}{4pt}}\frac{\partial_u\nu_{\scaleto{\rm 2D}{4pt}}}{\nu_{\scaleto{\rm 2D}{4pt}}}\Big)\frac{r}{\ell_2}+\frac{\kappa^2_4}{L}\lambda\frac{\alpha^2_{\scaleto{\rm 2D}{4pt}}}{2\Phi_0\nu_{\scaleto{\rm 2D}{4pt}}}\ell_2(\mathcal{O}^{(1)}_{u}-\mathcal{O}^{(1)}_{z\bar{z}})~.
\end{equation}
By comparing this expression with equations \eqref{eq:fJT} and \eqref{eq:JToperators} from Sec.\,\ref{Sec:JTgravity}, we recognize that the decoupling limit in the generalised Robinson-Trautman solutions yields the identifications $\omega=\gamma = 0$, provided we relate the operators as:
\begin{equation}
    L\kappa_2^2\mathcal{O}_u =\lambda\frac{\kappa_4^2}{2\Phi_0}\mathcal{O}_u^{(1)}~,\quad L\kappa_2^2\mathcal{O}_{\phi} =\lambda\frac{\kappa_4^2}{2\Phi_0}\mathcal{O}_{z\bar{z}}^{(1)}~.
\end{equation}
Applying this relation to the IR generating functional $I_{\text{IR}}$ defined in (\ref{eq:IRgenerating4d}), we find:
\begin{equation}
    \delta I_{\text{IR}}=4\pi(2\Phi_0 L)\frac{\kappa_2^2}{\kappa_4^2} \int \dd u \Big({\cal O}_u\, \delta \alpha + \frac{\alpha}{\nu}{\cal O}_\phi \,\delta \nu \Big)=\delta I_{\omega=0}~,
\end{equation}
where $I_{\omega}$ is the functional in \eqref{eq:IJT}. In the last equality, we used that $8\pi \Phi_0 L\kappa_2^2 = \kappa_4^2 $ which comes from matching the functionals $I_{\text{IR}}$ and $I_{\omega}$, but also from simple dimensional reduction. Finally, the relation between $I_{\text{IR}}$ and the generating functional of JT follows from the equation \eqref{eq:MapI}, which for this particular case is 
\begin{equation}
    I_{\rm JT}=I_{\omega=1}=I_{\omega=0}+\frac{\ell_2}{2\kappa_2}\int \dd u \sqrt{-h}\,\phi^{-1}h_{uu}(\partial_u\phi)^2~.
\end{equation}
The value $\omega = 1$ emerges naturally in JT gravity, prior to introducing the ambiguities associated with $\omega$. This relation highlights the non-trivial interpolation between the low-energy limit of the UV theory described by AdS$_4$/CFT$_3$ and its effective infrared description in terms of JT gravity. In summary, we started with a CFT$_3$ that is free of Weyl anomalies, as captured by the generating functional $I_{\omega = 0}$. Then,  through our off-shell construction of the decoupling limit, we derived the Schwarzian dynamics, which exhibits a Weyl anomaly. The key mechanism underlying this result is that the anomaly originates from a local counterterm that exists only in one dimension. In higher dimensions, such a covariant counterterm does not exist, and thus the Weyl anomaly cannot be modified in this way.

Finally, we verify that, regardless of the value of $\omega$, the thermodynamics in the near-extremal limit of AdS$_4$/CFT$_3$ matches the tree-level thermodynamics of JT gravity. According to the dictionary established in this section, the energies are related as follows:
\begin{equation}
    \ell_4\Delta M_{\scaleto{\rm 4D}{4pt}}=-4\pi \lambda^2\tilde{\nu}  \mathcal{O}_u^{(1)}=\frac{L}{\ell_2}\ell_4\nu_{\rm \scaleto{\rm 2D}{4pt}}E_{\rm \scaleto{\rm 2D}{4pt}}~,
\end{equation}
where $\Delta M_{\scaleto{\rm 4D}{4pt}}$ is the energy of the near-extremal AdS Reissner-Nordstr\"om black hole given in \eqref{eq:4d-energy} and $E_{\rm \scaleto{\rm 2D}{4pt}}$ is given by \eqref{eq:2d-energy}. With this, we see that
\begin{equation}
    S_{\rm \scaleto{\rm 2D}{4pt}} = 2\pi \sqrt{\frac{2\ell_2}{\kappa_2^2}\nu_{\scaleto{\rm 2D}{4pt}}E_{\scaleto{\rm 2D}{4pt}}} =  2\pi \sqrt{2\ell_2^2\frac{8\pi \Phi_0}{\kappa_4^2}\Delta M_{\scaleto{\rm 4D}{4pt}}}=\Delta S_{\scaleto{\rm 4D}{4pt}}~,
\end{equation}
Therefore, the semi-classical entropy in both descriptions matches as expected from the analysis of -- for instance -- \cite{Nayak:2018qej}. It is important to emphasise that the agreement at the tree-level in our thermodynamic analysis holds independently of the value of $\omega$. 

\section{Discussion} \label{sec:disc}

We presented a novel decoupling limit, designed to capture the low-temperature, near-extremal limit of black holes within the context of AdS/CFT. There are some important aspects of our construction worth highlighting:
\begin{enumerate}
    \item In dealing with the gravitational dynamics, we presented three examples, in Sec.\,\ref{sec:ads3}-\ref{sec:ads5}, where it is consistent to input the radial dependence of the bulk fields and leave the transverse dependence arbitrary. This defines a reduced phase space in AdS$_{d+1}$, which we relate to appropriate one-point functions ${\cal O}^{\scaleto{d}{4pt}}_i$ and sources in CFT$_{d}$. Imposing the remaining Einstein's and Maxwell's equations is equivalent to imposing Ward identities on the CFT$_d$. 
    \item On this reduced phase space, we introduced a decoupling limit which acts on the one-point functions and sources. The well-known near-horizon limits inspire the decoupling limit: we fix the charge (angular momentum or electric/magnetic charge), and lower the mass down to an extreme. It also acts on the sources that control the boundary geometry: in the limit, the size of the spatial directions grows relative to the time direction. And, very importantly, the coordinates in AdS$_{d+1}$ remain unchanged in the decoupling limit.
\item The limit is implemented such that the metric and fields in AdS$_{d+1}$ are well-defined and finite. This fixes the behaviour of the operators under the limit. We also make sure that the boundary action in AdS$_{d+1}$, whose variation gives the one-point functions of the CFT$_d$, is finite in the limit. 
\item The result of the decoupling limit are IR values for CFT variables: for the one-point functions ${\cal O}^{\scaleto{d}{4pt}}_i \to {\cal O}^{\scaleto{\rm IR}{4pt}}_i$ and similarly for the sources.  These quantities are then matched with their counterparts in JT gravity, with one-point functions ${\cal O}^{\scaleto{\rm JT}{4pt}}_i$ and sources, which allows us to connect the reduced phase spaces of each theory.  In doing this, we do not rely on a dimensional reduction of the gravitational theory in $d+1$-dimensions; the decoupling limit naturally reduces the phase space to something that can be matched to JT gravity. 
\end{enumerate}

The connection between the reduced phase space resulting from the decoupling limit and that in JT gravity is subtle. For this reason, we revisited in Sec.\,\ref{Sec:JTgravity} several aspects of the phase space of JT gravity. One important finding in this discussion is the inherent ambiguities we identify in the two-dimensional theory. The local, and covariant, boundary counterterm we found is      
\begin{equation}
\begin{aligned}\label{eq:counterterm-1}
    S_{\rm ct}&=\frac{\ell_2}{2\kappa_2^2}\int \dd u\sqrt{-h}\,\phi^{-1}h^{uu}(\partial_u\phi)^2\,,
\end{aligned}
\end{equation}
where $\phi$ is the dilaton in JT gravity and $h_{uu}$ is the boundary metric. This term can be added with any coefficient $\omega^2$, and its effect is to redefine the one-point functions of the theory, thereby affecting the Ward identities of the theory. More explicitly, if we denote ${\cal O}_{u}^{\scaleto{\rm JT}{4pt}}$ and ${\cal O}_{\phi}^{\scaleto{\rm JT}{4pt}}$ the one-point functions naturally obtained in JT gravity, with sources $\alpha$ and $\nu$ respectively, then adding the counterterm $(\omega^2-1) S_{\rm ct}$ to the JT action induces the field redefinition
\begin{equation}
\label{eq:MapII-1}
\begin{aligned}
     {\cal O}_{u}^{\scaleto{\rm JT}{4pt}}&={\cal O}_{u}+\frac{(1-\omega^2)}{2}\frac{\ell_2}{\kappa_2^2}\frac{(\partial_u\nu)^2}{\nu\alpha^2}~,\\
     {\cal O}_{\phi}^{\scaleto{\rm JT}{4pt}}&={\cal O}_{\phi}-\frac{(1-\omega^2)}{2}\frac{\ell_2}{\kappa_2^2}\frac{(\partial_u\nu)^2}{\nu\alpha^2}+(1-\omega^2)\frac{\ell_2}{\kappa_2^2\alpha}\partial_u\left(\frac{\partial_u\nu}{\alpha}\right)~,\\
\end{aligned}
\end{equation}
where ${\cal O}_{u}$ and ${\cal O}_{\phi}$ satisfy the $\omega^2$-dependent Ward identities
\begin{equation}\label{eq:ward2d-1}
   \begin{aligned}
    &{\cal O}_{u}+\mathcal{O}_{\phi}= \frac{\omega^2\ell_2}{\kappa_2^2\alpha}\partial_u\left(\frac{\partial_u\nu}{\alpha}\right)\,,\\
     &\partial_u {\cal O}_u-{\cal O}_{\phi}\frac{\partial_u\nu}{\nu}=0\,,
    \end{aligned} 
\end{equation}
whereas ${\cal O}_{u}^{\scaleto{\rm JT}{4pt}}$ and ${\cal O}_{\phi}^{\scaleto{\rm JT}{4pt}}$ satisfy these same equations with $\omega^2=1$. This makes explicit that the concept of a ``conformal anomaly'' is inherently ambiguous because a counterterm can be added to remove it. The addition of $S_{\rm ct}$ also affects the derivation of the Schwarzian effective action, as shown explicitly in Sec.\,\ref{sec:schw-dyn}, see \eqref{eq:IJTSchwarzian-1}. Still, the classical on-shell value of the action, i.e. the free energy, is independent of $\omega$, which we show explicitly. 

As mentioned above, including this counterterm  $S_{\rm ct}$ is key when connecting the phase space of AdS$_{d+1}$ with the phase space of JT gravity. For the three examples studied here, we find:
\begin{description}
    \item[AdS$_3$/CFT$_2$ (Sec.\,\ref{sec:ads3})] In the absence of gravitational anomalies, i.e., $c_L=c_R=c$, we find a relation of the form \eqref{eq:MapII-1} with $\omega^2=4$. This value can be intuitively inferred by the fact that the AdS$_3$ and AdS$_2$ radii are related via $\ell_3=2\ell_2$.
    \item[AdS$_3$/CFT$_2$ (Sec.\,\ref{sec:TMG})] When $c_L\neq c_R$, and extremality sets the left-moving sector to its ground state, we find $\omega^2=4\frac{c}{c_L}$. In addition, we can also remove the 1D ``gravitational anomaly.'' The appropriate counterterm, written in terms of the dilaton in JT gravity, is  
    \begin{equation}\label{eq:OhatCT-IR-cov-1}
    S_{\rm ct\,,\scaleto{\rm TMG}{4pt}}=\frac{\bar{c}}{24}\int \dd u\sqrt{-h}\,h^{uu}\phi^{-1}\partial_u\phi\big(\partial_u\phi-\Gamma^u_{uu}\phi\big)~,
\end{equation}
which is not covariant and therefore affects the Ward identity associated with boundary diffeomorphisms. Recall that $\bar c=(c_L-c_R)/2$. This counterterm induces the field redefinitions in \eqref{eq:hatO2Ddecoupling}, which transforms the Ward identities \eqref{eq:TMGWardExpansion} into \eqref{eq:TMGWardExpansion2}. 
    \item[AdS$_4$/CFT$_3$ (Sec.\,\ref{sec:ads5})] In this case, the IR limit of  ${\cal O}^{\scaleto{\rm 3D}{4pt}}_i$ gives that $\omega^2=0$. This is the most striking case conceptually. Since there is no conformal anomaly in the CFT$_3$, when flowing to the IR, it is perhaps odd to change this feature. However, this affects the lore that the Schwarzian dynamics in the IR is universal. In our analysis, the Schwarzian appears if we add the local counterterm $S_{\rm ct}$ in \eqref{eq:counterterm-1}, which can be viewed as a UV/IR matching condition. 
\end{description}

The counterterm $S_{\rm ct}$ reflects a renormalisation scheme ambiguity in either IR or UV theories, but it is unambiguously determined when both theories are taken into account. In the above, we are taking an IR point of view, where we are adding the counterterm $S_{\rm ct}$ to match with the UV phase space, and hence tampering with the UV physics. In this sense, we are taking the Schwarzian dynamics in JT gravity, with $\omega=1$, as the fundamental EFT that describes the low-temperature dynamics. It is also important to address this result from the UV perspective. One way to interpret our results is to read off the EFT coefficients at low temperatures from the RG flow of the UV, and our off-shell decoupling limit provides one version of this flow. From this UV perspective, we are finding that the Schwarzian dynamics will only appear depending on the number of dimensions: it will play a role in AdS$_3$/CFT$_2$ where there is just a modification of the value of $\omega$, analogous to, for example, a $c$-theorem; in contrast, in AdS$_4$/CFT$_3$ our analysis indicates that the Schwarzian effective action plays no role in describing the dynamics and quantum effects at low temperatures, since $\omega=0$. Another possibility is that there is reason to add counterterms to the flow, and one should end up with $\omega\neq0$.  It would be interesting to investigate further these possibilities and understand in which sense the Schwarzian dynamics is universal at low energies or theory-dependent.

There are several future directions to explore. Our construction of the phase space was limited to three examples; however, this can easily be extended to other cases. The construction of RT geometries in flat space is very well known (and developed before the AdS cases). Hence, it would be interesting to quantify the role of the counterterm in \eqref{eq:counterterm-1}, and how it could guide the imprint of the Schwarzian dynamics far away from the horizon in asymptotically flat cases. The addition of rotation is also within reach: the generalisation of algebraically special backgrounds which include rotation is known as the Timofeev solutions \cite{Timofeev1996,Stephani:2003tm,HR:2014RT}. And it should also be possible to extend the analysis to higher dimensions, where it would be interesting to work out the case of AdS$_5$. 

 We have addressed the problem of extracting near-extremal dynamics by studying the phase space and identifying effective actions that describe this sector. These effective actions, when inserted in the Euclidean path integral, will capture both classical and quantum corrections to the black hole. From this perspective, it is important to understand what fixes $\omega^2$. From the direct evaluation of the gravitational path integral, such as the analysis in \cite{Iliesiu:2022onk,Kapec:2024zdj,Arnaudo:2024bbd}, we expect that there should be a physical criterion that fixes the value of $\omega^2$ or at least predicts why it varies from the UV to the IR.\footnote{The analysis done in \cite{Kolanowski:2024zrq} has a drawback in their construction of the modes outside the near horizon geometry for BTZ. The modes constructed there have a temperature cutoff (for a fixed temperature $T$, only a finite number of them are normalizable in AdS$_3$), hence we cannot make a comparison or draw conclusions relative to our reduced phase space in AdS$_3$.} It would be interesting to see if there is evidence of this counterterm from a Euclidean path integral perspective.     

Our analysis sits on the gravitational side of AdS/CFT; still, the way we cast the decoupling limits and analysed the phase space is designed towards a CFT understanding. From this CFT perspective, we are introducing a novel RG flow that makes the $d$-dimensional system flow to a one-dimensional quantum mechanics. Key aspects of the decoupling limit are based on the AdS side: the regularity conditions of the bulk fields, and landing on a JT description of the system. It would be very interesting to have a purely field-theoretical derivation of this flow across dimensions, and in particular, have a better understanding of the counterterms appearing in 1D. This is a difficult problem in field theory, see e.g., \cite{Bobev:2017uzs,Sacchi:2023omn} for a recent discussion. Recent attempts to characterise the one-dimensional model include \cite{Benini:2022bwa,Cabo-Bizet:2024gny,Heydeman:2024fgk}. We hope this can be addressed in future work.  

\section*{Acknowledgments}
We would like to thank Nikolay Bobev, Sameer Murthy and Achilleas Porfyriadis for helpful discussions. IP would like to thank the Isaac Newton Institute for Mathematical Sciences, Cambridge, for support and hospitality during the programme ``Black holes: bridges between number theory and holographic quantum information'', where work on this paper was undertaken. This work was supported by EPSRC grant no EP/R014604/1. RM would like to especially thank David Berenstein for valuable discussions related to this project and the surrounding topics. RM's research was supported by the Department of Energy under grant DE-SC 0011702, and by the Chilean Fulbright Commission and ANID through the Beca Igualdad de Oportunidades $\#$ 56180016. AC has been partially supported by STFC consolidated grant ST/X000664/1.

\appendix

\bibliography{all}{}
\bibliographystyle{ytphys}

\end{document}